\documentclass{jfm}

\RequirePackage{packages}
\RequirePackage{newcommands}

\newcommand{\arev}[1]{#1}
\newcommand{\adl}[1]{}

\ifCUPmtlplainloaded \else
\checkfont{eurm10}
\iffontfound
\IfFileExists{upmath.sty}
{\typeout{^^JFound AMS Euler Roman fonts on the system,
using the 'upmath' package.^^J}%
\usepackage{upmath}}
{\typeout{^^JFound AMS Euler Roman fonts on the system, but you
dont seem to have the}%
\typeout{'upmath' package installed. JFM.cls can take advantage
of these fonts,^^Jif you use 'upmath' package.^^J}%
\providecommand\upi{\upi}%
}
\else
\providecommand\upi{\upi}%
\fi
\fi

\ifCUPmtlplainloaded \else
\checkfont{msam10}
\iffontfound
\IfFileExists{amssymb.sty}
{\typeout{^^JFound AMS Symbol fonts on the system, using the
'amssymb' package.^^J}%
\usepackage{amssymb}%
  \let\leq=\leqslant
  \let\geq=\geqslant
}{}
\fi
\fi

\ifCUPmtlplainloaded \else
\IfFileExists{amsbsy.sty}
{\typeout{^^JFound the 'amsbsy' package on the system, using it.^^J}%
\usepackage{amsbsy}}
{\providecommand\boldsymbol[1]{\mbox{\boldmath $##1$}}}
\fi

\title[]{
Sheared free-surface flow over three-dimensional obstructions of finite amplitude
}
\author[A.~H. Akselsen and S.~\AA. Ellingsen]{Andreas H.\ Akselsen$^1$\thanks{Email address for correspondence: andreas.h.akselsen@ntnu.no} and Simen \AA. Ellingsen$^1$}
\affiliation{$^1$Department of Energy and Process Engineering, Norwegian University of Science and Technology, N-7491 Trondheim, Norway}
\date{\today}           

\begin{document}

\maketitle

\begin{abstract}

When shallow water flows over uneven bathymetry, the water surface is modulated. 
This type of problem has been revisited numerous times since it was first studied by Lord Kelvin in 1886.
Our study analytically examines currents whose unperturbed velocity profile $U(z)$ follows a power-law $z^q$, flowing over
a three-dimensional uneven bed. This particular form of $U$, which can model a miscellany of realistic flows, allows explicit analytical solutions. 
Arbitrary bed shapes can readily be imposed via Fourier's theorem provided their steepness is moderate.

Three-dimensional vorticity-bathymetry interaction effects are evident when the flow makes an oblique angle with, a sinusoidally corrugated bed. Streamlines are found to twist and the fluid particle drift is redirected away from the direction of the unperturbed current.

Furthermore, a perturbation technique is developed which satisfies the bottom boundary condition to arbitrary order also for large-amplitude obstructions which penetrate well into the current profile. This introduces higher-order harmonics of the bathymetry amplitude. States of resonance for first and higher order harmonics are readily calculated.
Although the method is theoretically restricted to bathymetries of moderate inclination, a wide variety of steeper obstructions are satisfactorily represented by the method, even provoking occurrences of recirculation.

All expressions are analytically explicit and sequential fast Fourier transformations ensure quick and easy computation for arbitrary three-dimensional bathymetries. A method for separating near and far fields ensures computational convergence under the appropriate radiation condition.

\end{abstract}

\section{Introduction}
\label{sec:introduction}

Intriguing distortions can often be seen on the surface of flowing water, caused by unevenness in the bed beneath. 
Lord Kelvin, then Sir William Thomson, was the first to publish 
expressions governing such flow distortions
\citep{kelvin1886stationary}, along with myriad other contributions to hydrodynamics, now cemented in history and standard in popular text books \citep[e.g.,][]{lamb1932hydrodynamics}.  
Parallel to this, Lord Rayleigh derived 
seminal 
theory for similar, closely related problems, such as the standing waves which form atop flowing water when the water is excited by a steady pressure disturbance  \citep{rayleigh1883_standing_waves}.
\arev{
We review the extensive literature below, but note already that no study to date includes the possibility of a velocity of nontrivial depth-dependence, which is indicated by recent studies to affect the stationary waves behind obstacles significantly \citep[e.g.][]{li_2016_ship_wake_finite_depth}.
}

\arev{
Linear theory of the kind adopted by Lords Kelvin and Rayleigh
is particularly tractable, 
making 
it a powerful tool for examining fundamental features of bed--surface wave interactions. 
This was demonstrated by \citet{kennedy_1963_sand_dunes_due_to_waves}
in his study of sand dune formations. 
Coupled with a simple, slow time scale sediment transport equation, \citeauthor{kennedy_1963_sand_dunes_due_to_waves} was able to make good prediction about the wavelength and velocity features of dunes.
Researchers have since resorted to 
more sophisticated techniques in order to enrich their studies with nonlinear dynamics;
conformal mapping 
approaches have
been popular, particularly in predicting flows over sharp protrusions. 
Conformal mapping is not
restrictive in terms of steepness but is confined to irrotational, two-dimensional flows  
and the mapping must be made to match the shape of the particular bathymetry.
A relatively simple such example is the
curvilinear co-ordinate transformations adopted by 
\citet{benjamin1959_shear_wavy_bed} in his study of boundary layer shear friction over sinusoidal bathymetries.
Transformations that allow the problem 
to be described in an integro--differential system have 
featured regularly during
the last few decades,
often following in the steps of \citet{forbes_1982_potential_semicircle} and \citet{vanden-broeck_1997_boundary_integral_sluice_gate}.
}

\arev{
Boundary integral methods have been adopted also for three-dimensional nonlinear  free-surface flows \citep{Forbes_1989_3D_boundary_integral_method}.
\citet{buttle_2018_river_boundary_integral_method}
employed a Jacobian-free Newton-Krylov scheme for this purpose to study the stationary three-dimensional wave patterns forming behind a Gaussian-shaped flow obstacle.
\citet{buttle_2018_river_boundary_integral_method} further points out that this problem is analogous to the ship wake of a moving surface pressure source, 
about which a rich body of literature already exists. 
Fully numerical approaches require for three-dimensional problems significant CPU-time and memory storage (often involving supercomputers) but can provide reliable and accurate solutions.
In contrast, 
linearised solutions, as derived herein, are computed at almost negligible cost and afford 
much wider exploration of 
the physical aspects of our problem.
These solutions are naturally approximate and 
caution is needed with regard to the range of their validity.

Alongside boundary integral representations, perturbative weakly nonlinear approximations
including higher-order interactions among normal modes
are frequently adopted in recent literature. 
Some of these allow the problem to be cast in terms of a forced Korteweg--de Vries (KdV) equation.
The forcing can for example be generated by the bathymetry, a sluice gate or some other internal or surface pressure source 
\arev{
\citep{%
akylas_1984_first_fKdV,%
Cole_1985_waves_forming_around_a_bump,%
mei_1986_solitons_off_of_slender_bodies,%
wu_1987_upstream_advancing_solitons,%
katsis_1987_upstream_advancing_solitons,%
dias_2002_critical_flows_over_obstacle_KdV,%
Binder_2006_KdV_boundary_integral_step,%
binder_2013_bathymetry_inverse_problem,%
binder_vanden-broeck_2007_KdV_boundary_integral_sluice_gate}.
}

Among the bottom topographies commonly considered 
is the half-cylinder \citep{forbes_1982_potential_semicircle,Forbes1988_semi_circle_supercritical, zhang_1996_flow_over_half_cylinder},
the triangular obstacle  \citep{dias_1989_channel_triangle_shape_conformal,Chuang_2000_bethymetry_generalized_schwarz_christoffel,binder_vanden-broeck_2007_KdV_boundary_integral_sluice_gate}
and the step \citep{king_1987_potential_step,Binder_2006_KdV_boundary_integral_step}.
}
\arev{Typical of these problems is the way in which their asymptotic behaviour  depends on the criticality of the flow;
a notable example (which falls outside the scope of the present paper) is the case of an hydraulic fall
\citep{dias_1989_channel_triangle_shape_conformal,Forbes1988_semi_circle_supercritical}.
\citet{binder_2013_bathymetry_inverse_problem} provide an overview of 11 such asymptotic flow types.

}

The present work is founded on linearisation and Fourier's principle, as in Lords Kelvin and Rayleigh's pioneering work. 
This approach is however extended to bathymetries of finite amplitude.
A significant advantage of this 
\arev{linearisation strategy} 
is that it is not restricted to potential theory and so permits rotational flows. 
Furthermore, it does not require conformal transformations, thus 
\arev{permitting three-dimensional arbitrary bathymetries.}
 
\arev{We shall in the present work consider currents which, unperturbed, are represented by a power-law}
\begin{equation}
\arev{
\bU = (U,0,0); \quad
U = z^q
\qquad [q,z\geq0], 
}
\label{eq:U}
\end{equation} 
$z$ being a vertical distance and $q$ a constant. 
\arev{The power-law is well suited for analytical evaluation, as will be shown. 
The earliest examples of its utilization in the present context that we are aware of are}
\citet{lighthill_1953__zover1/7_profile_paper}
and \citet{fenton_1973_zover1/7_profile_paper}
\arev{who adopted a value $ q=1/7$ }
and touched upon many of the results encountered herein. 
\arev{We too adopt the $1/7$ exponent in several of our numerical examples.}
Inspiration has also been drawn from 
\arev{\citet{phillips_1996_family_of_zq_CL2_instability} and \citet{phillips_1996_Langmuir_qoverz_profile} who, allowing $q$ to remain arbitrary, utilized the power-law profile family to neatly represent a multitude of current profiles within the same analysis.
These papers deal with the flow stability to longitudinal vortices, related to the oceanographic phenomenon on Langmuir circulation under strong shear.
}
The uniform current most commonly considered is then recovered by setting $q$ to zero.
A linear current profile is recovered when $q=1$.
This profile has been investigated frequently in the literature since it is---in 2D---the only rotational flow for which potential theory is applicable.
In between, profiles resembling that observed in turbulent 
\arev{bottom boundary layer}
flows 
reside, while flows with a surface shear layer may be modelled with $q > 1$.

\arev{
The power-law \eqref{eq:U} has alongside the log-law traditionally been used to represent statistically averaged boundary and shear layers.
Prandtl first suggested the power-law exponent $q=1/7$ for low Reynolds numbers turbulence over smooth boundaries.
A range of values have since been suggested for turbulent flows,
ranging between $1/3$ to $1/12$ depending on roughness and Reynolds number \citep{Cheng_2007_power_laws,Chen_1991_power_laws,dolcetti_2016_channel_directional_spectrum}.
These values generally increase with the relative roughness of the boundary (shallowness of the flow).
\citet{barenblatt_1993_power_law_theory}
made the since celebrated 
conjecture
that the power-law exponent 
should be
inversely proportional to the logarithm of the Reynolds number.
The log-law can thus be regarded as an asymptotic limit of the power-law as the Reynolds number approaches infinity.
\citeauthor{barenblatt_1993_power_law_theory} 
stressed that
both the log- and power-law are supported by an equally rigorous theoretical foundation, differing only in physical assumptions.
There now seems  to be a general consensus that the
power-law preforms notably better than the log-law over rough boundaries or at low Raynolds numbers where the overlap layer in relatively narrow \citep[][and references therein]{bergstrom_2001_power_law_low_Re,djenidi_1997_advantages_power_vs_log_law,George_1997_extended_power_law,hinze_1975_turbulence}.
}%

\arev{ Explicit analytical solutions exist also for exponential current profiles}
\citep{abdullah_1949_exp_current_profile,lighthill_1957_fundamental_throry_3D_disturbances_on_parallel_shear_flow}.
Predictions for \textit{weak} arbitrary currents may be attainable using perturbation techniques, as was proven for surface waves by \citet{shrira1993_shear_current_to_nth_order}.
\arev{
Approximate methods and methods of numerical integration are available for strong arbitrary currents; 
see brief reviews in 
\citet{ellingsen_2017_new_Kirby_Chen} and 
\citet{li_2019_DIM}, respectively.
}
\\

We here consider the situation where a sheared current with a free surface travels over a bottom of varying depth.
Variations of the bottom topography do not need to be small compared to the water depth  nor is its steepness restricted. 
However, we do assume the perturbations of the water surface have low steepness so as to allow linearisation.  
The reason is not that allowing also higher order surface effects would involve undue difficulty, but rather that, as indicated by \citet{akselsen_ellingsen_2019}, the interaction of shear and higher surface deformation harmonics introduces a further range of phenomena which would increase the scope of this study beyond reason. 
These questions we leave for future investigations.

The paper is structured as follows: 
Our problem is modelled in Section~\ref{sec:model_eq}
and its linearised solution derived in Section~\ref{sec:linear_sol}.
This solution is further analysed in Section~\ref{sec:asymptotic} where its asymptotic behaviour and extension to spectral bathymetry representations are considered.
The linear solution of Section~\ref{sec:linear_sol} is further extended to bathymetries of finite amplitude in Section~\ref{sec:exact_lower_BC}.
Results are presented in Section~\ref{sec:results}, followed by a \arev{discussion and} summary in Section~\arev{\ref{sec:discussion} and \ref{sec:summary}, respectively.}

\section{Model equations }
\label{sec:model_eq}

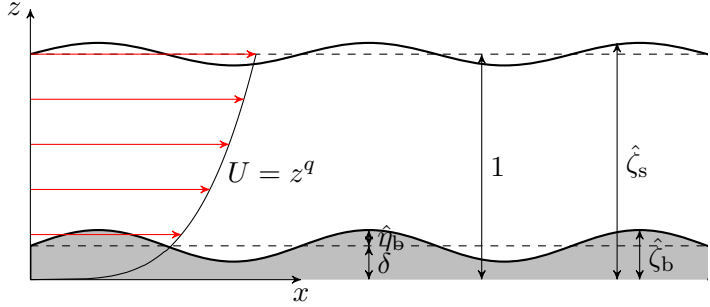
\begin{figure}%
\centering

%
%


\begin{tikzpicture}[scale=3.,font=\large]

\def\q{(1/4)}; 

\begin{scope}
\clip (0,-.1) rectangle (3,1.1);

\def\y{.07}
\def\x{.3}
\def\d{.15}
\draw[thick,name path=ground,fill=lightgray] (0,0)-- (0,\d) sin +(\x,\y) cos +(\x,-\y) sin +(\x,-\y) cos +(\x,\y) 
						sin +(\x,\y) cos +(\x,-\y) sin +(\x,-\y) cos +(\x,\y)
						sin +(\x,\y) cos +(\x,-\y) sin +(\x,-\y) cos +(\x,\y) -- (3,0);

\draw[dashed] (0,\d)-- (3,\d);
\draw[<->,font=\large] (1.5,0) -- (1.5,\d) node[right, midway] {$\delta$};
\draw[<->,font=\large] (1.5,\d) -- (1.5,\d+\y) node[right, midway] {$\petab$};
\draw[<->,font=\large] (2.7,0) -- (2.7,\d+\y) node[right, midway] {$\pzetab$};

\def\y{.05}
\draw[thick]
(0,1)
sin +(\x,\y) cos +(\x,-\y) sin +(\x,-\y) cos +(\x,\y) 
						sin +(\x,\y) cos +(\x,-\y) sin +(\x,-\y) cos +(\x,\y)
						sin +(\x,\y) cos +(\x,-\y) sin +(\x,-\y) cos +(\x,\y);
						
\draw[dashed] (0,1)-- (3,1);
\draw[<->] (2,0) -- (2,1) node[right, midway] {$1$};
\draw[<->] (2.6,0) -- (2.6,1+\y) node[right, midway] {$\pzetas$};

\end{scope}


\draw[name path global=U,-,domain=0:1,samples=250,mylabel=at 0.6 above right with {$~U=z^q$}] plot  ({(\x)^\q},\x);
\foreach \z in {0.2,0.4,...,1.0}{
\draw[red,->] (0,\z)--({(\z)^\q},\z);
};

\draw[->] (0,0)-- (1.2,0) node[below]  {$x$};
\draw[->] (0,0)-- (0,1.2) node[left]  {$z$};

\end{tikzpicture}

\caption{Sketch of the problem setup.}%
\label{fig:schematic}%
\end{figure}

The model considered in this work applies to stationary, incompressible flows where viscosity and surface tension effects are ignored. 
Our problem is readily converted to nondimensional form using the profile depth $h$ and surface current velocity $U\s$, as follows:
\begin{align}
(x,y,z)&\mapsto (x,y,z)\,h,
&
\bk&\mapsto \bk\,h^{-1},
&
\p \bu\tot &\mapsto  \p \bu\tot\,U\s,
&
\p p\tot &\mapsto \p p\tot\, \rho U\s^2,
\end{align}
$\bk=(k_x,k_y)$ is the wavenumber in the surface plane, 
$\p\bu\tot$ the fluid velocity and $\p p$ the pressure.
A solution will be constructed in Fourier space by considering a single mode.
Intuitively, one may in this frame regard the problem configuration
as that of water flowing over a sinusoidal bed, as sketched in figure~\ref{fig:schematic}. 
The shear flow profile is represented by the power-law family
\eqref{eq:U}.
The mean vertical position of the bed itself is placed at some elevation 
$z=\delta>0$ above the shear profile origin. 
Thus, the two parameters $q$ and $\delta$ govern the current profile felt by the bed and the Froude number $\Fr$ its strength. 
The bathymetry undulations redirects the current energy to perturb the velocity field and otherwise flat free surface.
With the prescribed nondimemtionalisation, the elevation of the surface, $\pzetas(\bmr)$,  is oriented about $z=1$, just as the bed elevation, $\pzetab(\bmr)$, is oriented about $z=\delta$.
Symbols $\p\eta$ are used when referring the elevation relative to the unperturbed surface height.

The problem to be solved consists of the stationary incompressible Euler equations, along with an impermeability condition at the bed and dynamic and kinematic boundary conditions at the surface;
\begin{subequations}
\begin{alignat}{2}
\left.
\begin{aligned}
( \p\bu\tot \tcdot \p\nabla)\p \bu\tot + \p\nabla \p p\tot  &= - \Fr^{-2} \bm e_z\\
\p\nabla \tcdot \p \bu\tot &=0
\end{aligned}
\right\};&
\quad& \pzetab &\leq z\leq\pzetas, \label{eq:problem:Euler}\\
\left.
\begin{aligned}
\p \bu\tot \tcdot \p\nabla  \pzetas &= \p w \\
\p p\tot 
&= 0
\end{aligned}\right\};
& & z &= \pzetas, \label{eq:problem:BC_s}\\
\p \bu\tot \tcdot \p\nabla  \pzetab = \p w\; ;
& & z &= \pzetab, \label{eq:problem:BC_b}
\end{alignat}
\label{eq:problem}%
\end{subequations}%
with 
$\Fr=U\s/\sqrt{g h}$ being the Froude number and $\p\nabla=(\pp_x,\pp_y,\pp_z)$.

\section{Linearised solution}
\label{sec:linear_sol}
Begin by separating the perturbed part of the internal flow from the total,
\begin{subequations}
\begin{align}
\p\bu\tot\of{\bmr,z} &= \bU\of z  +  \p\bu\of{\bmr,z},\label{eq:bu}
\\
\p p\tot\of{\bmr,z} &= (1-z)\Fr^{-2} +  \p p\of{\bmr,z},
\label{eq:p}
\\
\pzetab\of \bmr &= \delta + \petab\of \bmr,
\\
\pzetas\of \bmr &= 1 + \petas\of \bmr,
\end{align}%
\label{eq:bu_p}%
\end{subequations}%
and assume that the perturbed part is much smaller in magnitude than that adhering to the free stream.  
(The reference surface pressure has been made to equal zero.)
Now let the bottom topography 
be described in terms of a Fourier spectrum 
\begin{equation}
\etab\of\bk=\F_\bk \petab\of\bmr, \quad \etab\of{\bm 0} = 0.
\end{equation}
$\F_\bk$ denotes a Fourier transform onto a wave vector plane $\bk$.
Real-space flow fields become
\begin{equation}
(\p\bu,\p p,\petas) =  \F\inv_{\bk} (\bu, p,\etas),
\end{equation}
by virtue of the lower boundary condition \eqref{eq:problem:BC_b}.

\subsection{Internal flow}
After linearisation, 
the  Euler equations read
\begin{subequations}
\begin{align}
\rmi k_x U \bu + \bU' w +\nabla p &= 
O[\ldots],
\label{eq:Euler:bu}
\\
\nabla\cdot \bu &= 0 \label{eq:Euler:cont}
\end{align}\label{eq:Euler}%
\end{subequations}%
where
$\nabla=(\rmi k_x,\rmi k_y,\pp_z)$.
The error estimate is $O[(k^3+k^{-1})w^2]$ for the $x$-component and $O[(k^2+1)w^2]$ for the $y$- and $z$-components, assuming $k_x\sim k_y\sim k$.
Eliminating $u$, $v$ and $p$ yields the Rayleigh equation
\begin{equation}
w''-\br{k^2+U''/U}w = 
O[ (k^3+k^{-1})w^2],
\label{eq:Rayleigh}
\end{equation}
$k=(k_x^2+k_y^2)^{\frac12}$. 
Upon substituting $w(z) = \sqrt z W(z)$ \arev{\citep{phillips_1996_Langmuir_qoverz_profile}},
the linear part of \eqref{eq:Rayleigh} reduces to the Bessel equation
\begin{equation}
W'' + \frac{W'}{z}-\br{k^2 + \frac{(1-2q)^2}{4z^2}}W = 0
\end{equation} 
whose homogeneous solutions are well known;
\arev{two such exist which are superposed with the coefficients $c^+$ and $c^-$, to be determined from the boundary conditions. }
The horizontal velocity components are found by integrating \eqref{eq:Euler:bu}.
We write the solution for a \arev{generic} flow variable $\phi$ on the form
\begin{equation}
\phi = \epsilon \sum_\pm \phi^\pm c^\pm 
\label{eq:velocity_field_O1}
\end{equation}
with
\begin{align}
w^\pm &= \rmi\sqrt z I_{\pm(q-\frac12)}\of{k z},
&
p^\pm &=  \frac{k_x}{k}  U(z) \sqrt z I_{\pm(q+\frac12)}\of{k  z},
&
\bu\_h^\pm &= \frac{\rmi \bm e_x U' w^\pm-\bk p^\pm}{k_x U},
\label{eq:velocity_field_O1_pm}
\end{align}
$I_\alpha$ being the modified Bessel function of the first kind of order $\alpha$ 
and $\bu\_h=(u,v)$ is the horizontal velocity vector.
$\epsilon$ is here the intensity of the internal flow perturbations, to be related to the 
\arev{intensity of the flow disturbance from the bathymetry.
This we have separated from the weight coefficients $c^\pm$ which are related to the unperturbed flow alone. 
$\ex = (1,0)$ is the horizontal unit vector while the 
summation over `$\pm$' means a sum is taken over `$+$' and `$-$', being either signs or labels as context requires.
}
(
This form \arev{of \eqref{eq:velocity_field_O1_pm}, written in terms of the modified Bessel function,} becomes linearly dependent at $q=1/2$, but is adopted here for symmetry rather than resorting to the second kind of modified Bessel function. 
\arev{Limit values $q\rightarrow1/2$ are evaluated in place of $q=1/2$.}
)
Note that a stream function 
$\p u = \pp_z \p\psi, \p w = -\pp_x \p\psi$ 
may easily be constructed in the case of two-dimensional flow.
One finds
\begin{equation}
\p\psi\of{x,z} = \frac{z^{q+1}}{q+1}
+ \F\inv_{\bk}  
\br{\rmi w/k_x};
\quad [k_y\equiv 0].
\label{eq:streamfunction}
\end{equation}

\subsection{Boundary condition}
The lower kinematic boundary condition \eqref{eq:problem:BC_b},
linearised about $z=\delta$, reads
\begin{equation}
w(\delta) = \epsilon 
+ O[( k + k^3)\etab^2]
\label{eq:BC_kin_b}
\end{equation}
in Fourier space, 
where
\begin{equation}
\epsilon=\rmi k_{x}U(\delta) \etab.
\label{eq:epsilon}
\end{equation}
Linearised about $z=1$, 
the upper boundary conditions \eqref{eq:problem:BC_s}
yield
\begin{subequations}
\begin{align}
\etas - Fr^{2} p\of 1 &= 
O[\Fr^2 w(1)^2],
\label{eq:BC_dyn_s}
\\
\rmi k_x\etas - w\of1 &= 
O[(k+k^{-1})w(1)^2 ].
\label{eq:BC_kin_s}
\end{align}	
\label{eq:BC_s}%
\end{subequations}%
Equation~\eqref{eq:Euler} has here been used and
$U$ derivatives assumed order unity  to simplify error estimates.
It's tempting to relate these estimates back to amplitudes of the bed, but, as will be seen, these variables may not scale in a linear manner.

Either of the surface conditions \eqref{eq:BC_s} yields $\etas$ directly.
The remaining 
coefficients
$c^+$ and $c^-$ are 
\arev{readily determined by}
the other surface condition and the kinematic condition at the bed, \eqref{eq:BC_kin_b}.
In terms of the functions in \eqref{eq:velocity_field_O1_pm}, 
\begin{equation}
c^\pm = 
\sqbrac{
w^\pm\of\delta
-\frac{w^\pm\of 1-\rmi k_x Fr^2 p^\pm\of 1}{w^\mp\of 1 - \rmi k_x Fr^2  p^\mp\of 1}w^\mp\of\delta
}^{-1}.
\label{eq:coeffs}
\end{equation}
Note that $c^\pm$ approaches finite values both as $\Fr^2\rightarrow 0$ and $\Fr^2\rightarrow \infty$.

\subsection{Critical Froude number}
\label{sec:Frcrit}
\begin{figure}%
\subfigure[$\delta=0.1$]{
\includegraphics[width=.5\columnwidth]{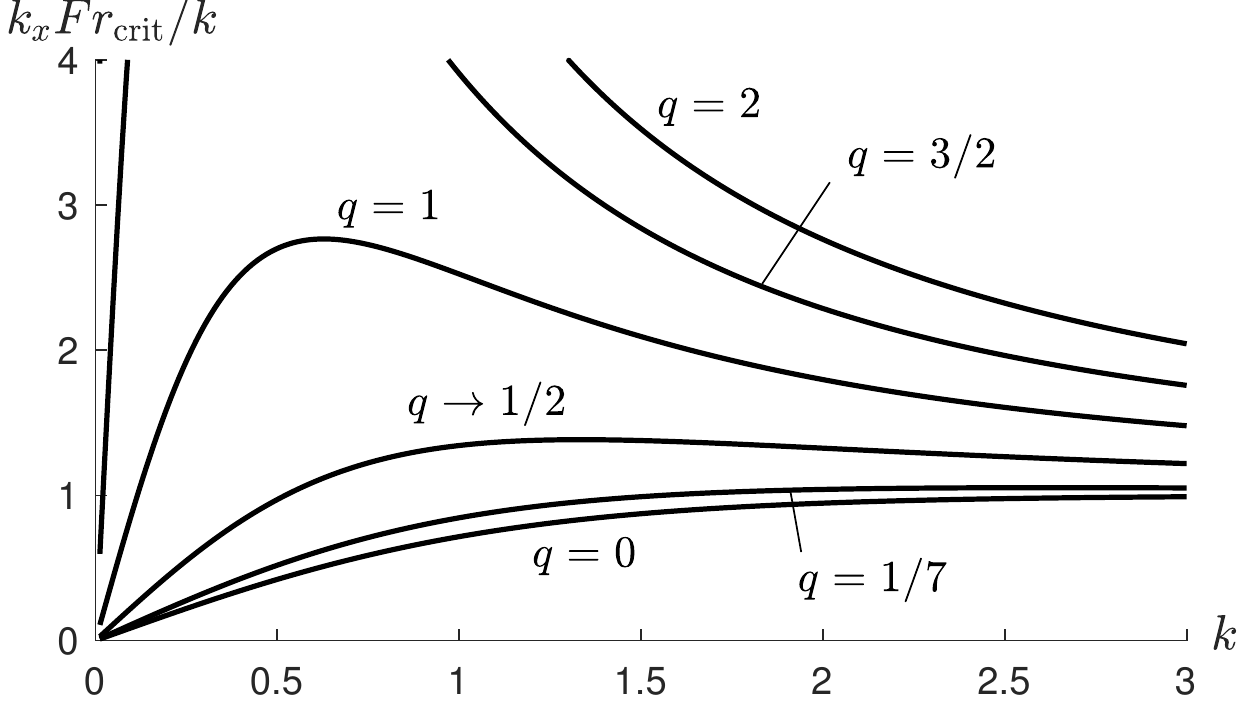}%
\includegraphics[width=.5\columnwidth]{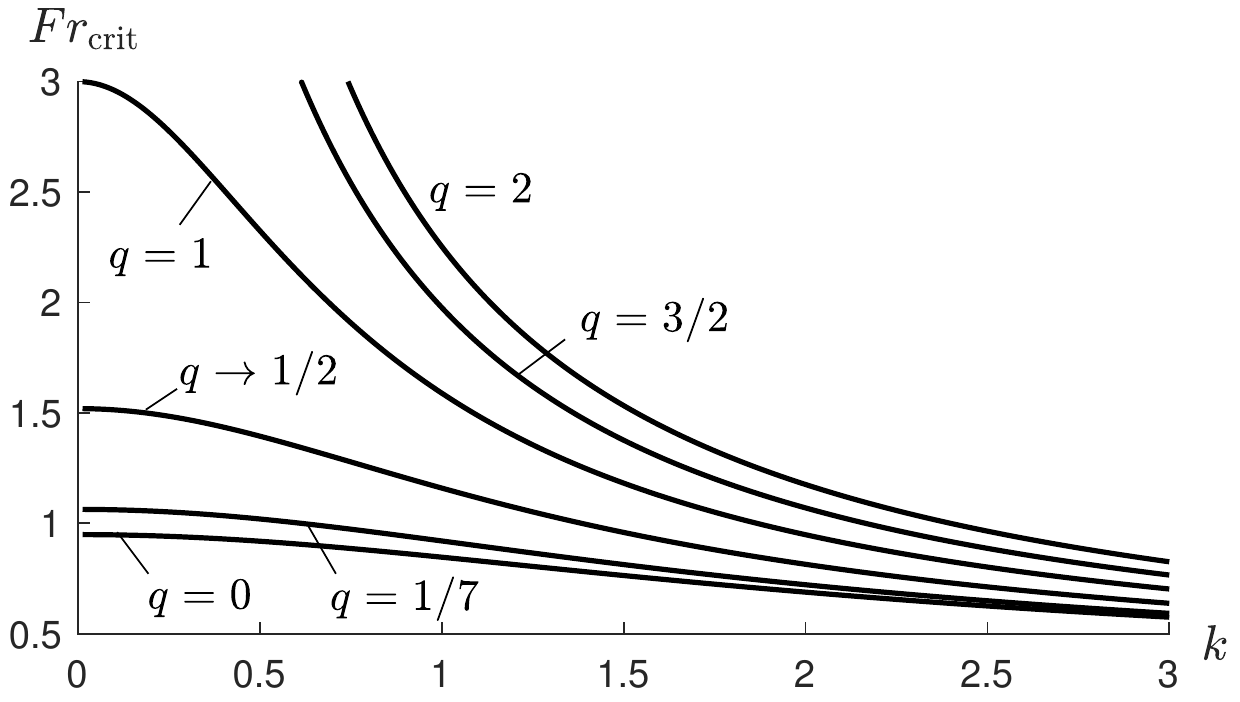}%
}%
\\
\subfigure[$\delta=0.5$]{
\includegraphics[width=.5\columnwidth]{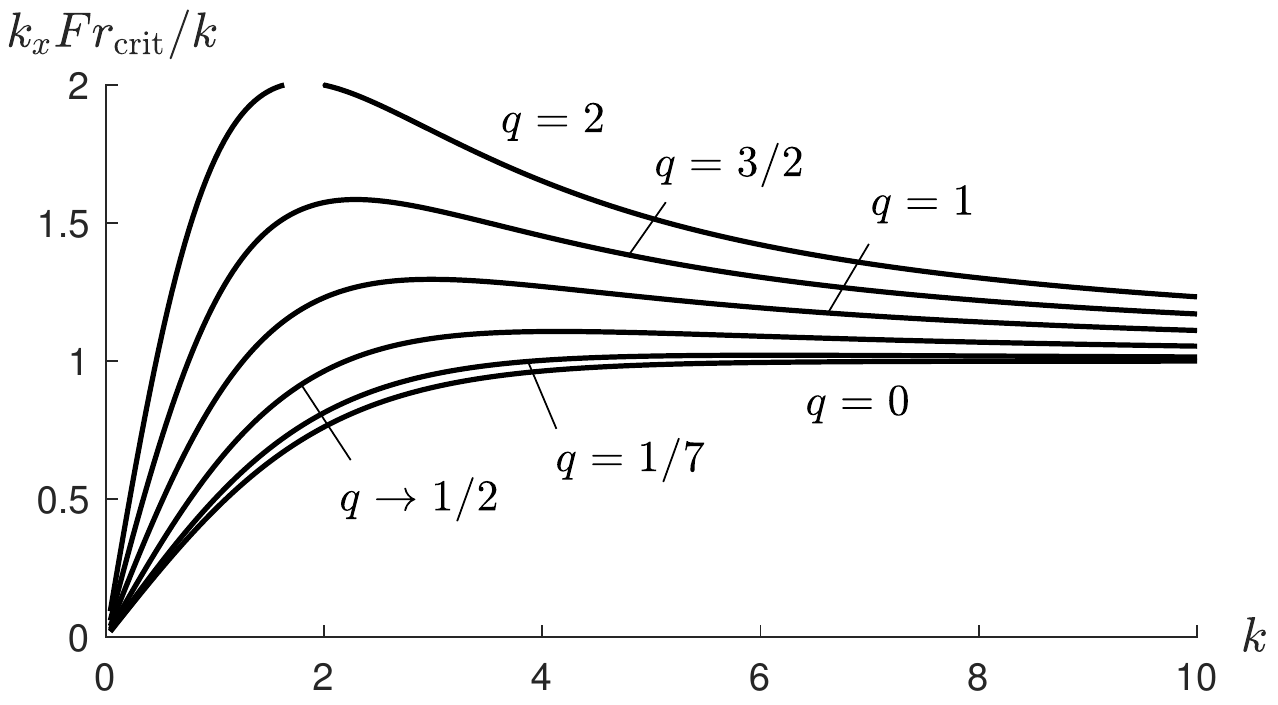}%
\includegraphics[width=.5\columnwidth]{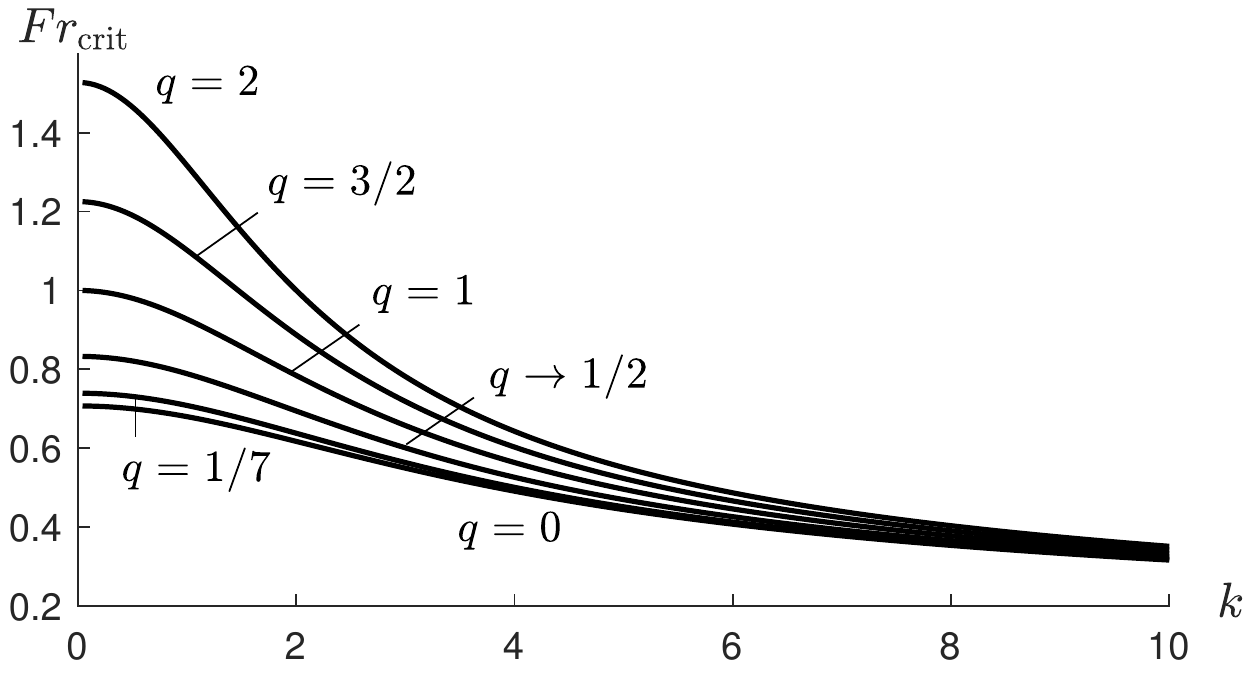}%
}%
\caption{$\frac{k_x^2}{k}\FrCrit^2$ and $\FrCrit(k_y=0)$ for two values of $\delta$.
In all panels the graphs are, top to bottom,
$q=2$, $q=3/2$, $q=1, q\to 1/2, q=1/7$ and $q=0$.
}%
\label{fig:Fr_crit}%
\end{figure}

From \eqref{eq:coeffs} it is seen that there may exist particular parameter values for which both field 
coefficients diverge. 
We shall later examine the impact of such singular cases in a spectral bathymetry representation. 
At present though, a single mode is considered.  
The \textit{critical} Froude number $\FrCrit$, 
at which the coefficient denominator \adl{(or system determinant)} is zero, is
\begin{equation}
\frac{k_x^2}{k}\FrCrit^2 \of \bk
= 
\frac{
I_{q-\frac12}\of{k} I_{-q+\frac12}\of{k\delta}
-I_{-q+\frac12}\of{k} I_{q-\frac12}\of{k\delta}
}{
I_{q+\frac12}\of{k} I_{-q+\frac12}\of{k\delta}
-I_{-q-\frac12}\of{k} I_{q-\frac12}\of{k\delta}
},
\label{eq:Fr_Crit}
\end{equation}
which is found to be always real positive for $q>0$, $0<\delta<1$.
Plots of $\FrCrit^2$ are shown in figure~\ref{fig:Fr_crit}.
For uniform current profiles, $q=0$, one recovers the well know result
\citep{kelvin1886stationary}
$(k_x\FrCrit)^2/k = \tanh[k(1-\delta)]$ and $\{\coth[k(1-\delta)]-k^{-1}\}^{-1}$ for linear ones ($q=1$).
All profiles $0\leq q\leq 1$ have the short wave asymptote $(k_x\FrCrit)^2/k \rightarrow 1$ as $k\rightarrow\infty$.
From figure~\ref{fig:Fr_crit} it is further seen that $\FrCrit$ diminishes monotonically with increasing wavenumber. 
The largest critical Froude number is 
\begin{equation}
\FrCritMax = \lim_{k\rightarrow0}\FrCrit = \frac{k}{|k_x|} \sqrt{\frac{1-\delta^{1-2q}}{1-2q}}.
\label{eq:Fr_Crit_max}
\end{equation}
$\FrCritMax$ marks the distinction between sub- and supercritical flows over bathymetries represented with an infinite Fourier spectrum, as shall be expounded upon later.
Also, $(k_x\FrCrit)^2/k \rightarrow\frac{I_{-(q-\frac12)}(k)}{I_{-(q+\frac12)}(k)}$ in the limit $\delta \rightarrow0$ if $q<1/2$, agreeing with the result of \citet{lighthill_1953__zover1/7_profile_paper} for a standing wave over a flat bed with a $q=1/7$ profile.
The critical Froude number for the $q=1/2$ profile is obtained by considering the limit, yielding
\[
\frac{k_x^2}{k}\FrCrit^2\Big|_{q\rightarrow\frac12} = \frac{
I_{0}\of{k} K_{0}\of{k \delta}
- K_{0}\of{k} I_{0}\of{k \delta}
}{
I_{1}\of{k}K_{0}\of{k \delta} 
+K_{1}\of{k}I_{0}\of{k \delta} 
}.
\]
$K_\alpha$ is the modified Bessel function of the second kind.
In the limit $k_x=k\rightarrow 0$ and $q=0$ one retrieves the expected $\FrCrit^2=1-\delta$, or just $1$ if the Froude number is instead defined based on the actual depth $h-\delta$ in physical units.
(It is also common to define the Froude number based on the mean velocity 
for which $\Fr\_m=\Fr\,(1-\delta^{q+1})/[(1-\delta)(1+q)]$.%
)

\arev{
Singular states are a common artefact of resonances when linearisation has been performed. 
With the allowance of transient dynamics, these singularities are often associated with algebraic wave growth
as current energy is transferred to the perturbed flow field without feedback
\citep{Benney_1961_O2_critical_layer_etc,Craik_1970_Langmuir_myidea,akselsen_ellingsen_2019}.
The asymptotic growth of critical flow over a bump or under a pressure patch goes in linear theory as the cubic root of time \citep{akylas_1984_first_fKdV,Cole_1985_waves_forming_around_a_bump},
but a linear steady state exists if the 
forcing is compact and the flow unrestricted in the spanwise direction \citep{katsis_1987_upstream_advancing_solitons}.
In the setting of a monochromatic bathymetry with a critical Froude number, resonance can be interpreted as the phase velocity being zero relative to the bed, 
denying dispersion of the bathymetry induced disturbances.
This issue can however be resolved and a steady state reached
when considering the full nonlinear problem. 
Classical Stokes wave theory provides a conceptual indication of how;
as the level of nonlinearity increases with growing wave amplitude so does the Stokes wave phase velocity relative to that of a linear wave. 
The shift in phase velocity detunes the Stokes wave such that it cannot remain stationary relative to the bed.
Resonance is thus broken. 
\citet{mei_1969_Fr_crit_steady} was the first to consider the Stokes wave solution near criticality. 
He discovered triple-valued steady state solutions. 
In order to deduce which solutions are likely occur in nature,
\citet{sammarco_1994_mei_1969_Fr_crit_transient} examined the transient problem and its initial stability and long-time evolution.

}

\section{Asymptotic behaviour}
\label{sec:asymptotic}
Consider next the composite solution as a spectrum of modes in a Fourier integral. 
It was shown in Section~\ref{sec:Frcrit} that the linear theory will, in the frame of a monochromatic bathymetry, break down for some particular combinations of parameters as a form of resonance is encountered. 
Excluding these parameter combinations, and their immediate neighbourhoods, the derived solution is valid. 
When, on the other hand, unevenness in bed topography is of limited spatial extent the Fourier integral must cover all values of $\bk$.
If the flow is subcritical (there exists wavenumbers $\bk$ so that $\FrCrit(\bk)>\Fr$)
then singular point will be encountered during the integration.
The problem as it stands is indeterminate until 
a radiation condition is applied.
Following a procedure due to \citet{rayleigh1883_standing_waves} (see also 
\citet[p.~399]{lamb1932hydrodynamics}
) 
the momentum equations are furnished with an artificial frictional force $-\mu (\p \bu-\bm U)$.
The only requirement for this force is that it always dampens perturbations and that it is vanishingly small, i.e., that $\mu$ be infinitesimally small yet positive everywhere. 
This provides a direction of time and 
ensures that causality is obeyed. 

Without loss of generality, let this force vary in $z$ according to $\mu=\muo k_x U(z)$ where $\muo=0^+\sgn(k_x)$.
The resulting modified Euler equations are
\begin{align*}
\rmi k_x U (1-\rmi \muo) \bu + \bU' w +\nabla p &= 0
\\
\nabla\cdot \bu &= 0 
\end{align*}
which, to linear order in $\muo$,
alters the solution \eqref{eq:velocity_field_O1}--%
\eqref{eq:velocity_field_O1_pm}
as follows:
\begin{align}
w^\pm &= \rmi z^{\frac12+\frac\rmi2 q \muo} I_{\pm(\tq-\frac12)}\of{k z},
&
p^\pm &=  \frac{k_x}{k} (1-\rmi \muo) z^{\frac12+\tq} I_{\pm(\tq+\frac12)}\of{k  z},  \label{eq:wp_mu1}
\\
\bu\_h &= \frac{\rmi \bm e_x U' w-\bk p}{(1-\rmi \muo)k_x U},
\end{align}
where $\tq = q(1+\rmi \muo/2)$.
Now rewrite the $c^\pm$-coefficients by inserting \eqref{eq:wp_mu1} into \eqref{eq:coeffs}:
\begin{equation}
c^\pm =\frac{ \fgpm(\bk)}{\fh(\bk)}
\end{equation}
with 
\begin{align}
\fgpm(\bk)&=\pm
\sqbrac{I_{\mp(\tq-\frac12)}(k)-\Fr^2\frac{k_x^2}{k}(1-\rmi\muo)I_{\mp(\tq+\frac12)}(k)},
\label{eq:g}
\\
\fh(\bk)&= \rmi\delta^{\frac12+\frac\rmi2 q \muo}\sum_\pm I_{\pm(\tq-\frac12)}(k\delta) \fgpm(\bk).
\label{eq:h}
\end{align}
The denominator $\fh$ is shared by all flow variables 
and is a dispersion relation for surface waves atop the flowing water.
Its roots 
may be seen as representing waves propagating upstream with phase velocity exactly equal to surface velocity, so that the flow is stationary.

Consider now a two-dimensional flow ($k_y=0$).
A localised unevenness in the bathymetry generates, for any flow variable, a perturbation of the surface of the form 
\begin{equation}
\int_{-\infty}^\infty \!\frac{\dd k_x}{2\pi} \frac{ \fg(k_x)}{\fh(k_x)} \rme^{\rmi k_x x}.
\label{eq:1D_integral}
\end{equation}
For example, if computing the surface elevation $\petas$ from \eqref{eq:BC_dyn_s} then
\begin{equation*}
\fg = -\epsilon\,\Fr^2\frac{k_x}{k^2}(1-\rmi\muo)\frac{2}{\pi} \cos(\pi \tq)
\end{equation*}
---note that Lord Kelvin's solution for a uniform current 
\citep[p.~520]{kelvin1886stationary},
\citep[p.~409]{lamb1932hydrodynamics},
\begin{equation}
\etas = \etab\wigbrac{\cosh[k(1-\delta)]-\frac{k}{\Fr^2 k_x^2}\sinh[k(1-\delta)]}^{-1},
\label{eq:etas_q0}
\end{equation}
is recovered when putting $q=0$.
The integral \eqref{eq:1D_integral} may be evaluated using well known techniques from complex analysis. 
The denominator $\fh(k_x)$ has conjugate roots along the imaginary axis 
whose contributions make up the near field.
Two roots $k_x=\pm  \alpha + \rmi 0^+$ also appear 
if the flow is subcritical.
These
are both shifted slightly away from the real axis into the upper imaginary plane by the artificial friction force.
Subcritical flow here means that there exists a wavenumber such that $\Fr<\FrCrit(\bk)$.
As seen before, this is equivalent to checking whether $\Fr<\FrCritMax$, \eqref{eq:Fr_Crit_max}.
Above this Froude number current transport is too great to permit
disruptions of the surface to propagate upstream or remain stationary.

\begin{figure}%
\centering
\includegraphics[width=.65\columnwidth]{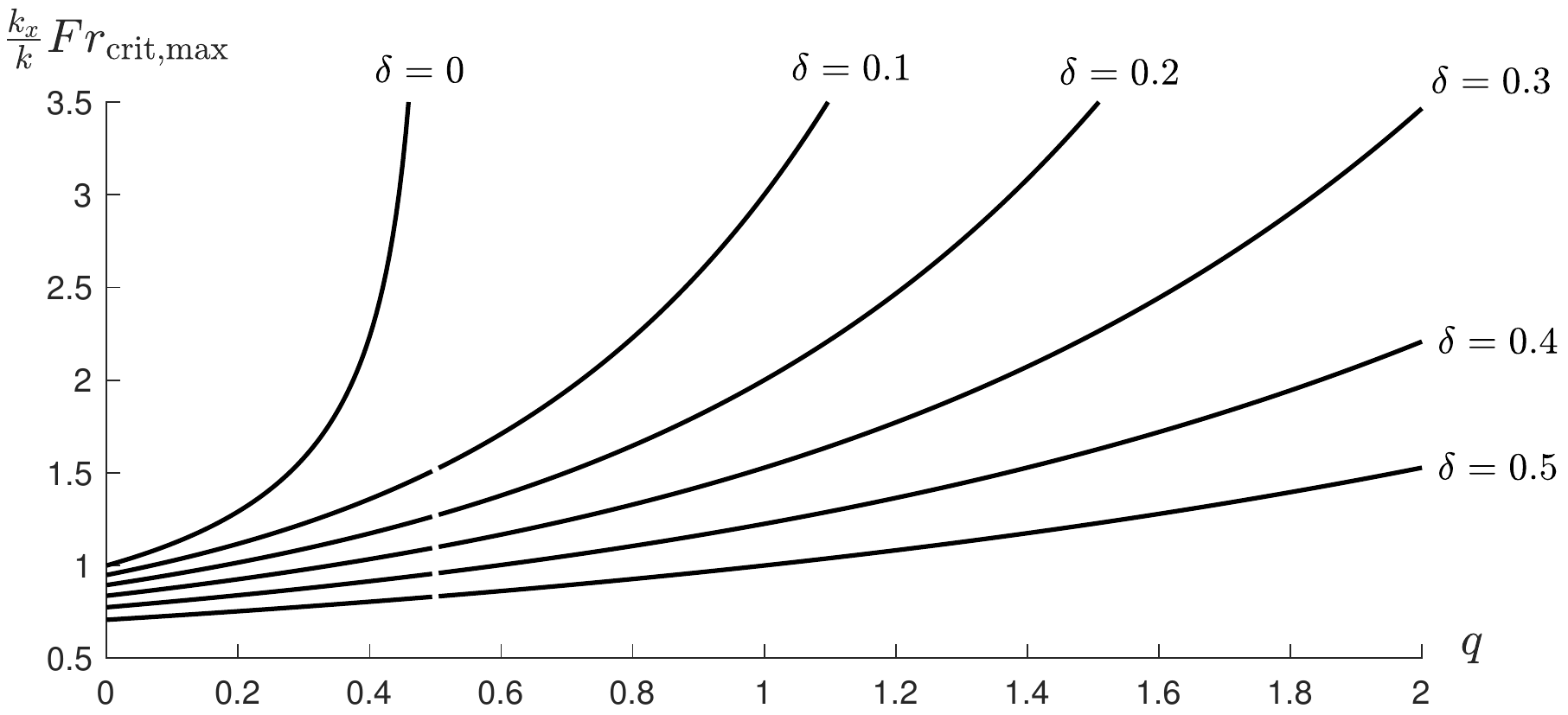}%
\caption{Critical Froude number $\FrCritMax$, \eqref{eq:Fr_Crit_max}, for the existence of a far field wave train.
}%
\label{fig:Fr_crit_kzero}%
\end{figure}

Although it is possible to express the integral \eqref{eq:1D_integral} considering these singular points only \citep[p. 410]{lamb1932hydrodynamics}, a form more suitable for computation is obtained by adding and subtracting
the pole terms (leading term of the integrand's Laurent series expansion) inside the integral 
whenever the flow is subcritical
and
using Cauchy's integral theorem to 
write \eqref{eq:1D_integral} in the form
\begin{align}
\int_{-\infty}^\infty 
\!\frac{\dd k_x}{2\pi}
&\sqbrac{
\frac{ \fg(k_x) }{\fh(k_x)}
- \sum_\pm\frac{ \fg(\pm\alpha)}{\fh'(\pm\alpha)(k_x\mp\alpha)} 
}\rme^{\rmi k_k x}
+\rmi\sum_\pm\frac{ \fg(\pm\alpha)}{\fh'(\pm\alpha)}\rme^{\pm\rmi\alpha x} \Theta(x)
\label{eq:1D_integral_evaluated}
\end{align}
when $\FrCritMax>\Fr$. 
A prime denotes differentiation with respect to $k_x$ and
$\Theta$ is the Heaviside function.
The poles that hinder numerical convergence are thus eliminated from the integrand and their effect, a wave train emitted downstream,
appears explicitly in the solution.
If the bathymetry obstruction is localised and centred near $x=0$, we may regard the integral term of \eqref{eq:1D_integral} as a `near field' and the oscillating term as a `far field'.
No alteration to  \eqref{eq:1D_integral} is needed when $\FrCritMax<\Fr$ (supercritical) since this flow consists of a near field only, the real axis being free of poles.

\begin{figure}%
\subfigure[$\Fr=1.0$ fixed]{\includegraphics[width=.48\columnwidth]{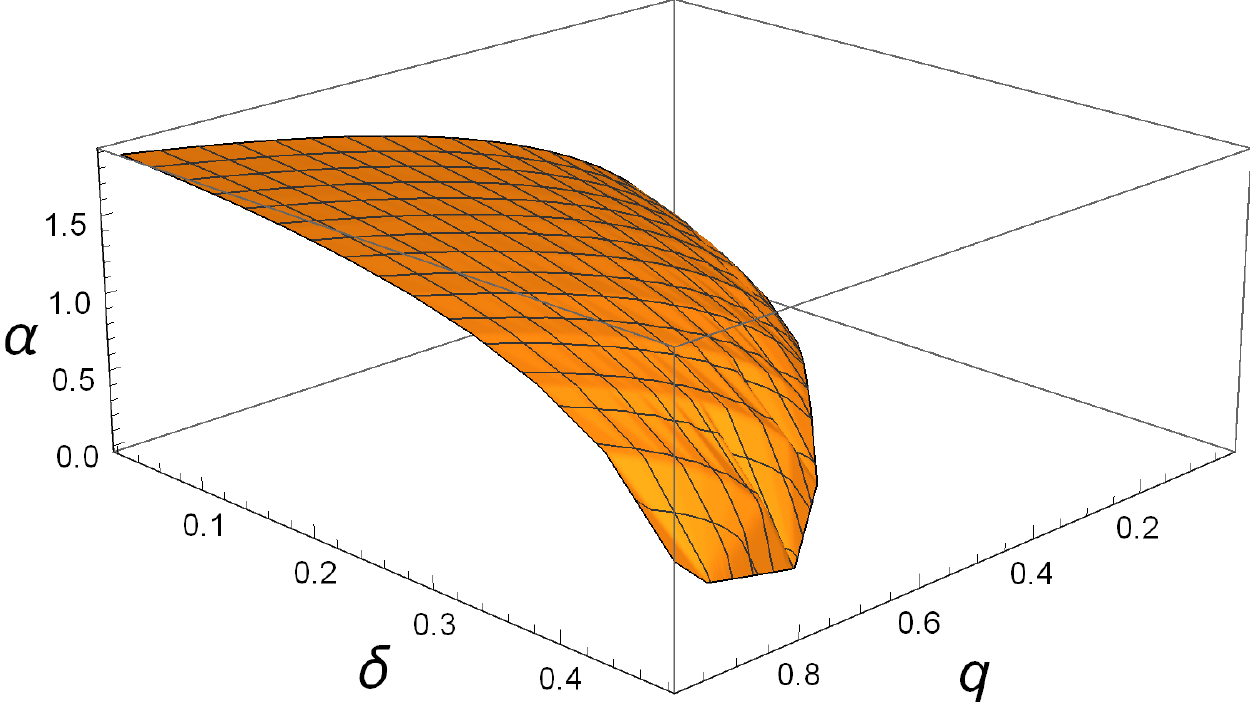}}%
\hfill
\subfigure[$\delta=0.1$ fixed]{\includegraphics[width=.48\columnwidth]{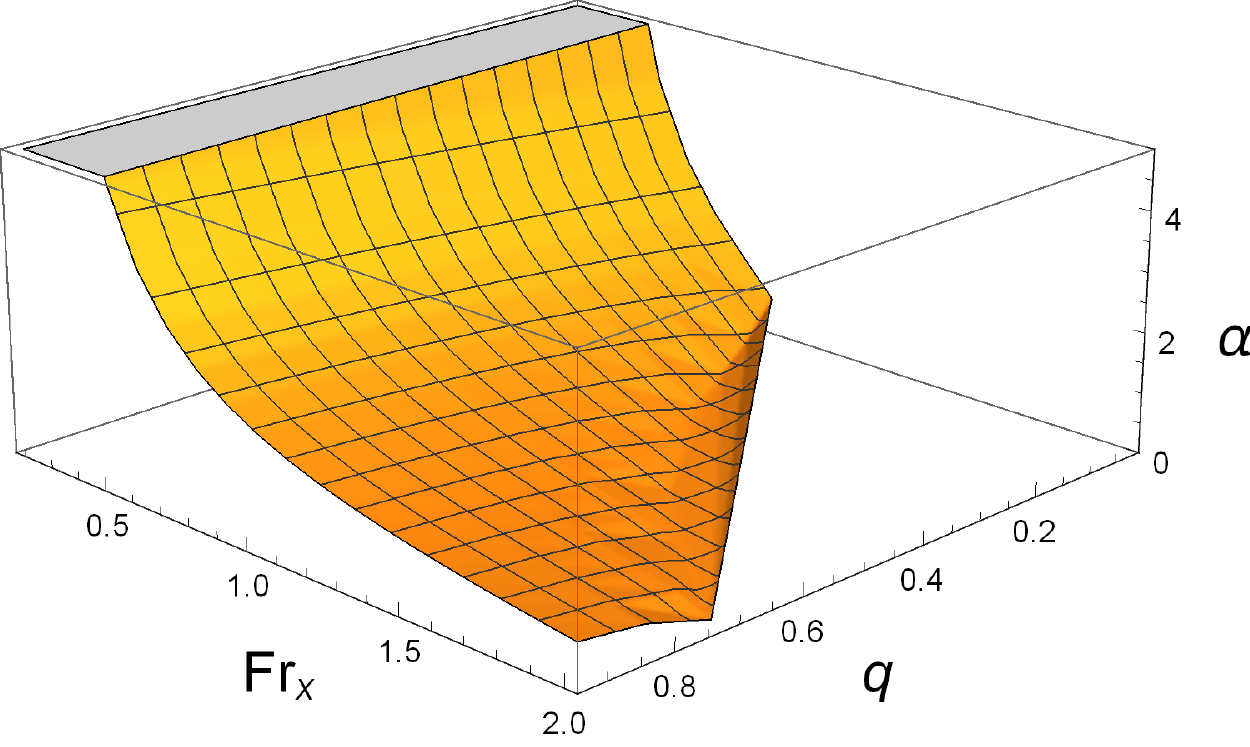}}%
\caption{Wavenumber $\alpha$ of far field wave train as function of $q$, $\delta$ and $\Fr$.}%
\label{fig:k_crit}%
\end{figure}

It's worth noting
that when the flow is subcritical
the shape of the bed deformation is of consequence only for the amplitude of the emitted wave train while the current profile alone dictates the  wavelength.
In a dynamic sense the downstream far-field wave will have exactly the wavelength of an upstream wave propagating towards the left at the surface velocity so as to be stationary in the `lab' frame. 
The dispersion relation of such a wave is a functional of 
the shear profile $U(z)$ and is, for arbitrary $U(z)$,
readily calculated numerically; c.f., e.g. \citet{li_2019_DIM}. 
Figure~\ref{fig:k_crit}, which is essentially an inversion of figure~\ref{fig:Fr_crit}, shows the wavenumber of the  wave train for given parameters $q$, $\delta$ and $\Fr$. 
Supercritical flow states, for which wave trains cease to appear, are excluded from the figure. 

\begin{figure}%
\includegraphics[width=1\columnwidth]{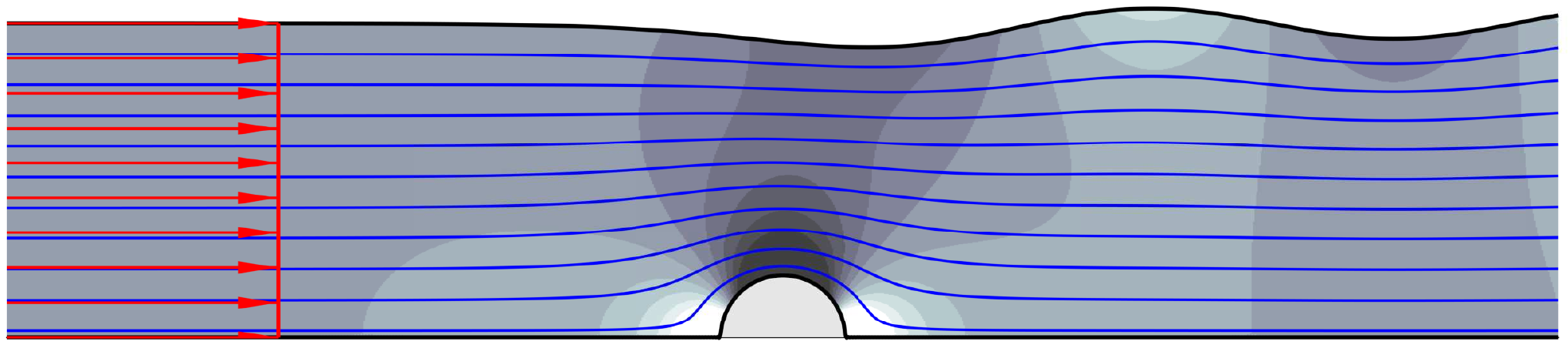}
\caption{
Uniform flow over a cylinder of radius $0.2$. 
The surface, bed, streamlines and dynamic pressure field $\p p + \Fr^2 z$ are shown. 
$\Fr=0.5$.
The extended lower boundary conditions of Section~\ref{sec:exact_lower_BC} low are applied, which does not affect the features of the wave train.
}%
\label{fig:cylinder}%
\end{figure}

In two-dimensional problems, equation \eqref{eq:1D_integral_evaluated} is readily computed numerically with
the derivative  of $\fh\of{k_x}$ evaluated with discrete differentiation.
An example
is shown in figure~\ref{fig:cylinder}. 
The extended method for obtaining accuracy in adherence to conditions at
the lower boundary, presented in Section~\ref{sec:exact_lower_BC}, is here employed.

In three dimensions the field separation procedure which yielded \eqref{eq:1D_integral_evaluated} becomes more involved as roots of $\fg\of\bk$ form a continuous path in $\bk$-space along which an integration must be performed.  
Instead, we opt in three dimensional problems for enforcing the  radiation condition by allocating a small but finite  value of the artificial friction coefficient $\muo$ such that the original Fourier integral converges.
This is a tried and trusted method which is equivalent to introducing a low level of artificial viscosity, see, e.g., 
\citet{moisy_rabadu_2014_finite_imag_perturbation}.
The artificial friction thus introduced must be made strong enough for the far field to die out before re-entering the periodically bounded numerical domain, yet weak enough not to mask any salient features pertaining to the inviscid flow
or noticeably affect the far-field wavelength, which obtains a correction of order $\mu_1^2$.
The strategy can 
add to the computational load
since a large 
computational domain  may be required. 
Many more sophisticated methods are available  such as non-reflecting boundary conditions, yet given the modest overall computational effort in our examples the simplest approach is adequate.
\\

A final observation is warranted. 
The limit $\delta\rightarrow0$ of the denominator $\fh$ in \eqref{eq:h} is finite 
if $q<1/2$ and infinite if $q>1/2$.
This means that infinitesimally small bathymetry perturbations centred around the current stagnation point $z=0$ are felt within the flow only if  
$\dd U/\dd z>1/\sqrt z$ at $z=0$.
Profiles developing slower than this
are not felt by a bed of infinitesimal amplitude.  
The feature can be seen in \eqref{eq:Fr_Crit_max} and figure~\ref{fig:Fr_crit_kzero} where critical Froude numbers are infinite for $q>1/2$ as $\delta\rightarrow0$.

\section{A lower boundary condition of $n$-th order accuracy}
\label{sec:exact_lower_BC}
Recall that the bathymetry $\petab$ is 
prescribed
and notice that the lower boundary condition \eqref{eq:problem:BC_b} therefore contains no products of unknowns. 
This is convenient for describing the lower boundary condition precisely, without requiring that the bathymetry perturbations be very small.

Many classical works have employed coordinate transformation techniques to two-dimensional problems \citep{stokes1880,benjamin1959_shear_wavy_bed,forbes_1982_potential_semicircle}, but these are not so readily extendible to three-dimensional non-potential flow.
A perturbation approach is instead viable, and the evaluation of \eqref{eq:problem:BC_b} at  $z=\pzetab$ is achieved by Taylor expansion about $z=\delta$. 
This generates products between variable derivatives and powers of $\petab$,
and so a perturbation solution 
\begin{equation}
\p\phi = \sum_{n=1}^\infty \p\phi_n,
\qquad \text{where}\qquad
\p\phi_n = O(\petab^n);
\quad 
\phi\in\{\bu,p,\etas\}
\label{eq:series_expansion}
\end{equation}
is postulated.
Inserting this series into the bottom boundary condition \eqref{eq:problem:BC_b} and Taylor expanding it about $z=\delta$, one obtains
\begin{equation}
\sum_{m=0}^\infty \frac{\petab^{m}}{m!}\sqbrac{
U^{(m)}\pp_x \petab  + \sum_{n=1}^\infty
\br{  \p\bu_n^{(m)}\tcdot\p\nabla\petab - \p w_n^{(m)}
}}_{z=\delta} = 0, 
\label{eq:BC_expanded}
\end{equation}
where
parenthesized superscripts indicate order of derivative in $z$.
Sorting into consecutive orders of $\petab$ and imposing that equality holds for each order individually, one finds the lower boundary condition
\begin{equation}
\p w_n(\delta)= \p\epsilon_n 
\end{equation}
where
\begin{equation}
\p\epsilon_n =
\petab^n\sqbrac{
\frac{U^{(n-1)}}{(n-1)!} 
\frac{
\pp_x \petab
}{\petab}
+ \sum_{m=1}^{n-1}\frac{1}{m!}\br{ m  \frac{\p\bu^{(m-1)}_{{\mr h},n-m}}{\petab^{n-m}}  \tcdot \frac{\p\nabla \petab}{\petab}  - \frac{\p w^{(m)}_{n-m}}{\petab^{n-m}}} 
}_{z = \delta}.
\label{eq:pdn}
\end{equation}
For bathymetries consisting of many Fourier modes (typically an obstruction localized in space) the most computationally efficient procedure is to go back and forth between physical and Fourier space using fast Fourier transforms. 
That is, having computed $\bu_{n-1}$ in Fourier space, 
the physical field is readily obtained via transformation $\p\bu_{n-1}^{(m)}=\F\bu_{n-1}^{(m)}$.
After evaluating \eqref{eq:pdn}, the result is transformed back to Fourier space: $\epsilon_{n} = \F\inv\p\epsilon_{n}$.
The only required adjustment to the linear solution presented in Section~\ref{sec:linear_sol} is to replace in \eqref{eq:velocity_field_O1} 
$\epsilon$ with increasing orders of $\epsilon_{n}$.
The horizontal derivatives of the velocity field are readily available via the Fourier transformation, or simply using finite differences.
$z$-derivatives are computed in Fourier space---we have constructed a building block function 
$
f_{a,b,m} \equiv \pp^m_z \big[z^{b-\frac12} I_{a}\of{k z}\big]
$
evaluated using binomial statements, 
\begin{equation}
f_{a,b,m}\of z = \sum_{j=0}^m 
\br{\frac{k}{2}}^{j} 
{m\choose m-j}\frac{\Gamma(b+\frac12)z^{b+j-m-\frac12}}{\Gamma(b+j-m+\frac12)}   \sum_{i=0}^{j}{j\choose i} I_{a+j-2i}\of{k z},
\label{eq:f_abm}
\end{equation}
whence
\begin{subequations}%
\begin{align}
w_n^{(m)}\of z &= \rmi \dn \sum_{\pm}f_{\pm(q-\frac12),1,m}\of z c^{\pm},
\\
\bu_{{\rm h},n}^{(m)}\of z &= - \dn \sum_{\pm} \sqbrac{ 
\frac{q \bm e_x}{k_x} f_{\pm(q-\frac12),0,m} \of z
+ \frac{\bk}{k} f_{\pm(q+\frac12),1,m} \of z
}c^{\pm},
\\
p_{n}^{(m)}\of z &=  \label{eq:bu_mn:p}
\dn \frac{k_x}{k}\sum_{\pm}f_{\pm(q+\frac12),q+1,m}\of z c^{\pm},
\\
U^{(m)}\of z &= z^{q-m}\prod_{j=0}^{m-1} (q-j)= z^{q-m}\frac{\Gamma(q+1)}{\Gamma(q-m+1)},
\end{align}
\label{eq:bu_mn}
\end{subequations}%
$m=0,1,2,\ldots$

The convergence the Taylor expansion of the above expressions will by Taylor's theorem be limited to the disc within which the represented function is analytical. 
There is a singularity at $z=0$ 
if the current profile is curved ($q$ different from $0$ or $1$)
which imposes the restriction $\eta_b\leq\delta$.
Even so, 
since the first few terms of the Taylor series typically decrease in magnitude before growing again, 
good approximations are often obtained  by truncating the Taylor series 
one term before the smallest in the manner recommended by 
\citet{bender_orszag_1991_approximate_methods}, section 3.5.

We can still compute cases where fully curved current profiles encounter large--amplitude obstructions by instead adopting an iterative approach.
One then evaluates an error to the boundary condition \eqref{eq:problem:BC_b} in real space after having computed $\p\bu\of\pzetab = \F\inv\{\bu\of\pzetab\}$ at all points $(x,y)$.
This error is then Fourier transformed and fed back into the amplitudes as an augmentation of $\epsilon$; $\epsilon\_{new} = \epsilon\_{old} + \text{error}$.
Convergence properties are usually good, presumably because $\epsilon=w(\delta)\sim \rmi k_x U(\delta)$ dominates over $\bu\_h(\delta) * \rmi \bk \etab$ in the cases tested herein.
\\

The products in \eqref{eq:pdn} constitute convolutions in Fourier space. 
Going back-and-forth between physical and Fourier space circumvents computing  the multi-dimensional mode interactions which appear within the convolution.
If instead individual mode interactions are considered, the solution takes the form
\begin{equation}
\p\bu_n =  \F\inv_{\bk_1}\F\inv_{\bk_2}\ldots\F\inv_{\bk_n}\bu_n
\end{equation}
where $\bu_n$ is a function of the interaction wave vector 
\begin{equation}
\bm \kappa_n=\sum_{m=1}^n k_n
\label{eq:kappa}
\end{equation}
(replacing $\bk$ in \eqref{eq:f_abm}--\eqref{eq:bu_mn}) except for in $\dn$
where
\begin{equation}
\dn =
\sqbrac{
\rmi k_{nx}\frac{U^{(n-1)}}{(n-1)!}
\prod_{j=1}^n\etab(\bk_j)
+ \sum_{m=1}^{n-1}\frac{1}{m!} \br{ m\, \rmi \bk_n \tcdot 
\bu^{(m-1)}_{{\mr h},n-m} - w^{(m)}_{n-m} }
\!\!\prod_{j=n-m+1}^n\!\!\etab(\bk_j)
}_{z = \delta}.
\label{eq:dn}
\end{equation}
\arev{The lower order velocity components $\bu_{m-n}(z;\bm \kappa_{m-n})$ are in turn functions of $\epsilon_{m-n}$ and so on.}
Higher harmonics are therefore present even when the bathymetry is a sinusoidal shape provided its amplitude is finite.
We will later demonstrate the appearance of higher harmonics in a flow over a sinusoidal bed of finite amplitude and the possibility of higher-order resonance. 
\\

Although the nonlinear lower boundary condition greatly extends the range of well approximated flow scenarios, the solution will still be restricted to linear order within the flow field and at the surface. 
\arev{
However, there are two cases, $q=0$ and 2D flow with $q=1$,  where the undulated velocity field $\p\bu$ is inherently irrotational.
Potential theory is permissible in these two cases which are consequently the two scenarios where the most theoretical headway has been made. 
Irrotationality
will in the Rayleigh equation \eqref{eq:Rayleigh} manifest 
by its inhomogeneous right-hand term, assumed small in linear theory, becoming identically zero so that the Rayleigh equation  reflects the Laplace equation of potential theory \citep[see e.g.][]{akselsen_ellingsen_2019}.
In the first case, when also the
unperturbed current is uniform and irrotational ($q=0$),
flow irrotationality 
is intuitive since inviscid perturbations are incapable of generating new vorticity.
The second case is 
when the unperturbed current is linear and perturbations are everywhere aligned with the current (two-dimensional flows with $q=1$). 
This latter case 
becomes rotational, however, once the flow has nontrivial spanwise dependence, in particular 
if the bathymetry is three-dimensional or not parallel to the current
because
current vortex lines are made to twist if they pass through an undulated velocity field \citep{Ellingsen_vorticity_paradox}.
Except for in these special cases, 
the linearised internal flow puts a limit on the steepness  $k \etabnil$ of the bathymetry, meaning that 
perturbations should be shallow ($k\ll1$)
when current profiles are curved.
}
Surface boundary conditions are also approximate to linear order so that 
perturbations in the region $z=1$ ought to be made small.

Perturbation strategies for resolving the weakly nonlinear dynamics in the entirety of the flows are straightforward in principle (see e.g. \citet{akselsen_ellingsen_2019} for an example), but they  quickly become unwieldy.

\section{Results}
\label{sec:results}

First,
figure~\ref{fig:wave_train_benchmark} has been generated as a benchmark against the conformal mapping result in figure~3 of 
\citeauthor{forbes_1982_potential_semicircle}' (\citeyear{forbes_1982_potential_semicircle}) paper.
It shows the surface elevation profile $\pzetas$ as a uniform current flows over a half-cylinder of radius $0.2$---the example used for illustration in figure~\ref{fig:cylinder}.
A qualitative comparison with \citet{forbes_1982_potential_semicircle} is straightforward;
amplitudes and wavelengths closely resemble those of \citeauthor{forbes_1982_potential_semicircle}' `linearised solution'---that is, linearised within the framework of a conformal mapping. 
In comparison, a fully linear perturbation solution in Euclidean space ($N=1$) is far too low in amplitude (shown as a dotted line).
The Stokes wave shape that \citeauthor{forbes_1982_potential_semicircle} obtained in their fully nonlinear solution (with increased amplitudes, sharper peaks and flatter troughs) is not captured as this a phenomenon related to surface nonlinearity, disregarded in the present model.

\subsection{Flow over sinusoidally corrugated beds}
Infinite, cosinusoidal bathymetries are considered in the next part of this section.
A sinusoidal bathymetry
has the spectrum $\etab\of\bk=
\arev{\tfrac12(2\pi)^2} 
[\etabnil\tilde\delta(\bk-\bknil)+\etabnil^*\tilde\delta(\bk+\bknil)]$.
Asterisk represents the complex conjugate and $\tilde\delta$ is here the Dirac delta function.
$\etabnil$ is the bottom perturbation amplitude.
\\

\begin{figure}%
\centering

\includegraphics[width=.8\columnwidth]{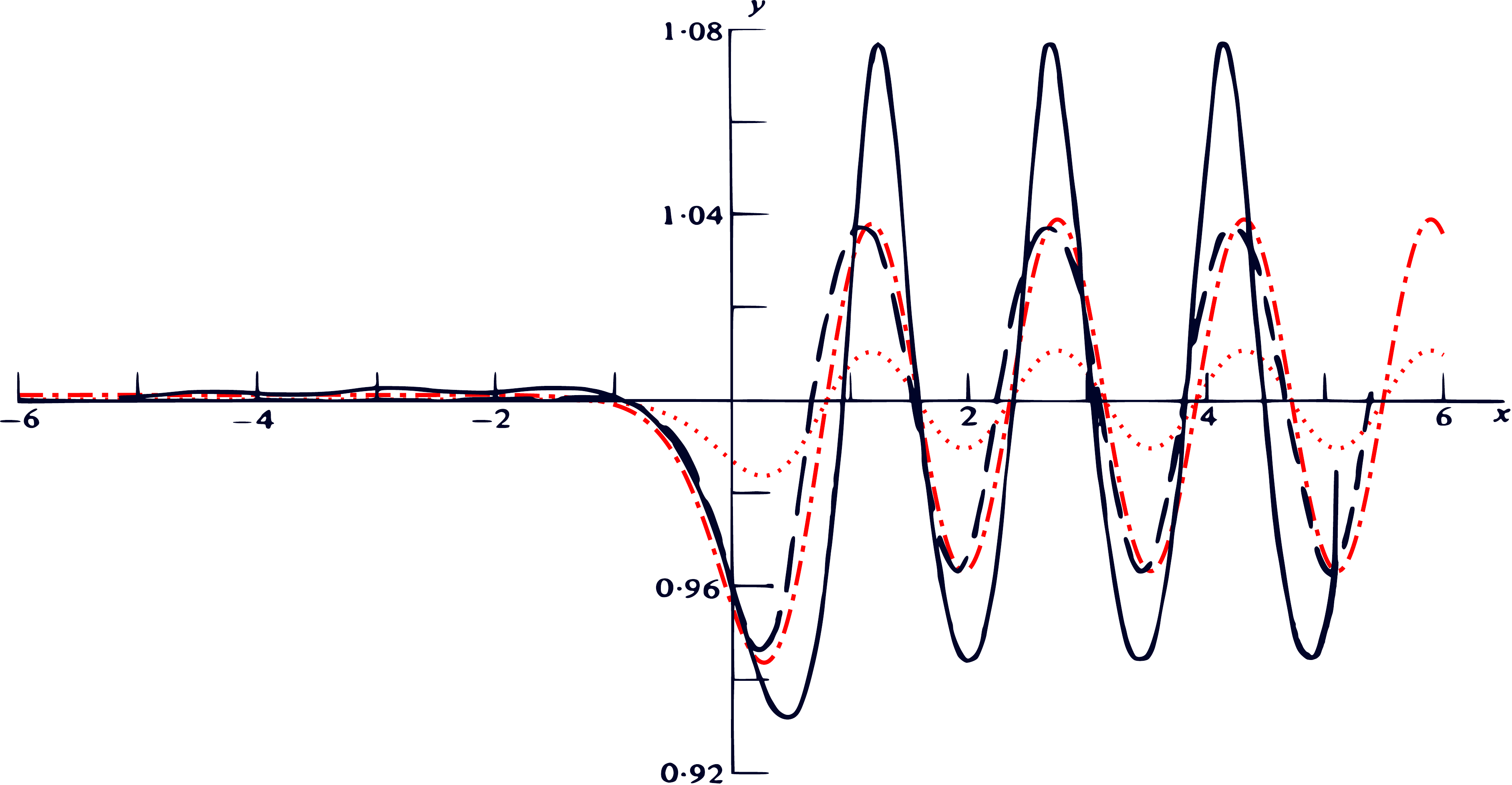}
\caption{Surface elevation profile $\pzetas$ generated by a uniform current flowing over a half-cylinder of radius $0.2$ at $z=0$. $\Fr=0.5$, as in the example of figure~\ref{fig:cylinder}.
Black solid and dashed lines are taken from figure 3 in \citet{forbes_1982_potential_semicircle} where the
solid line is the fully nonlinear solution computed with a numerical method and the dashed lines correspond to their `linearised solution'
where the free-surface boundary condition is linearised while the bottom condition is satisfied exactly by virtue of the conformal mapping.
Red dot-dashed and dotted lines are the present model.
Dotted line is with the linear lower boundary condition and dashed line is with the extended boundary procedure of Section~\ref{sec:exact_lower_BC}.
}%
\label{fig:wave_train_benchmark}%
\end{figure}

\begin{figure}%
\begin{tabular}{l cc}
&Subcritical---$\Fr\_m=0.25$ & Supercritical---$\Fr\_m=4.00$\\
\rotatebox{90}{\hspace{10mm}$\knil=0.1$}
&
%
\includegraphics[width=.45\columnwidth]{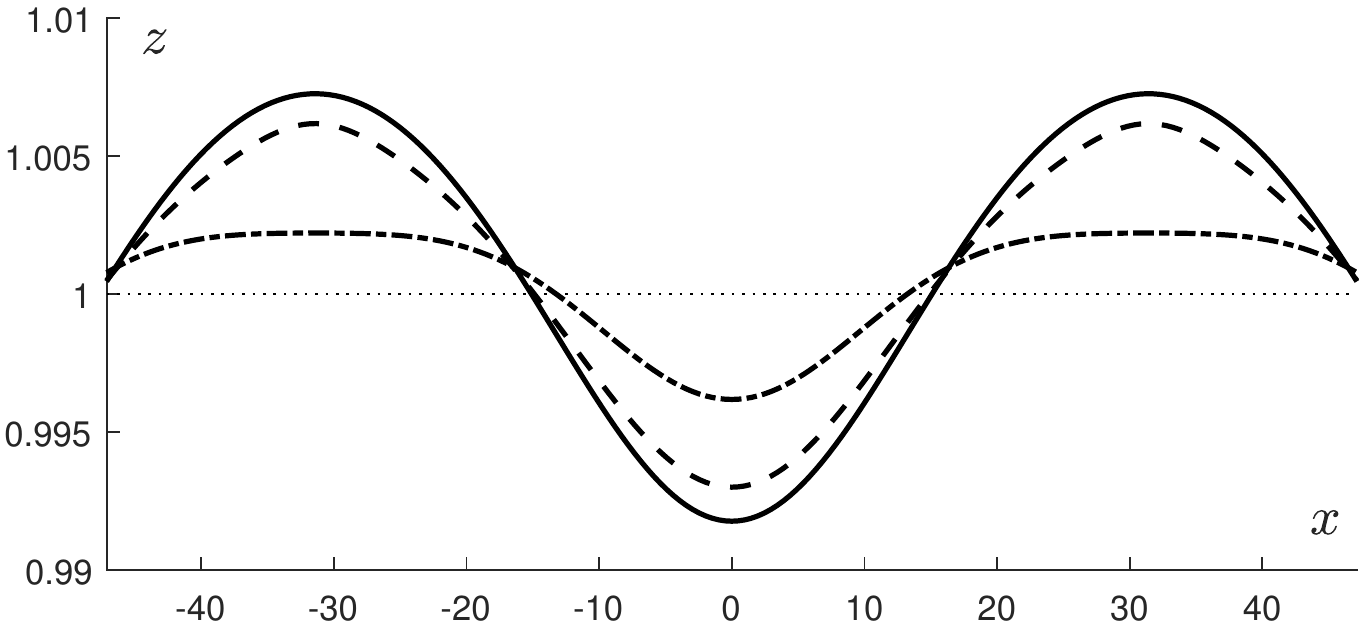}
&
\includegraphics[width=.45\columnwidth]{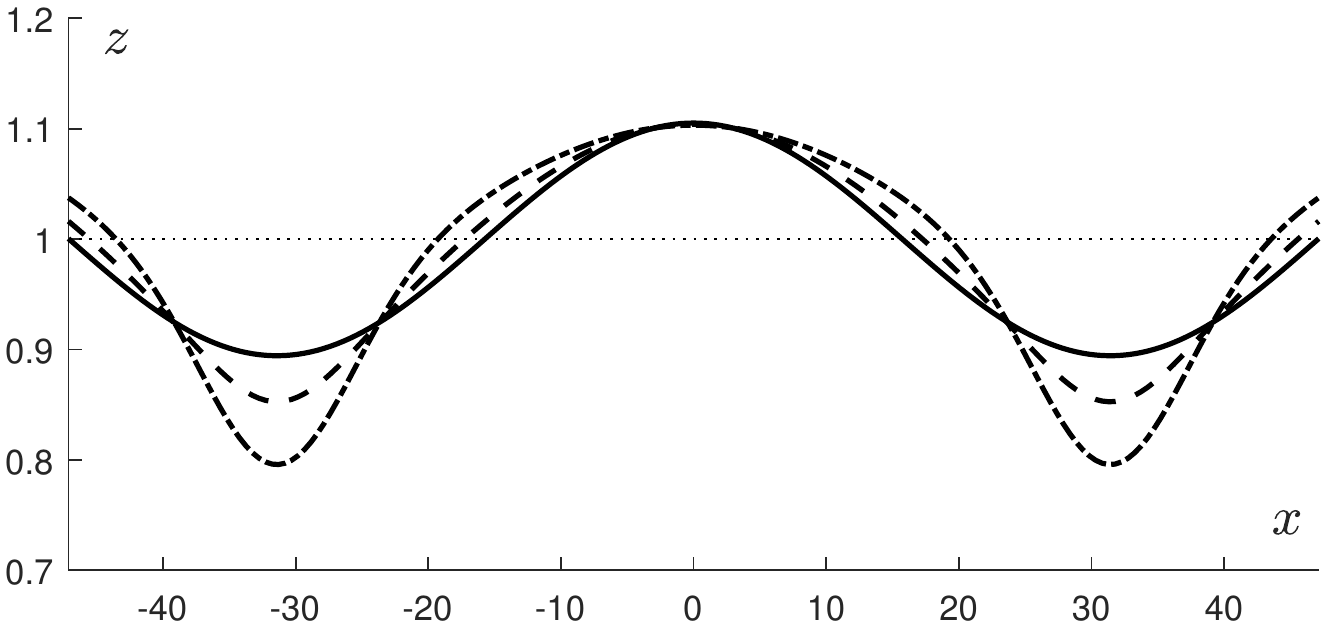}
\\
\rotatebox{90}{\hspace{10mm}$\knil=1.0$}
&
\includegraphics[width=.45\columnwidth]{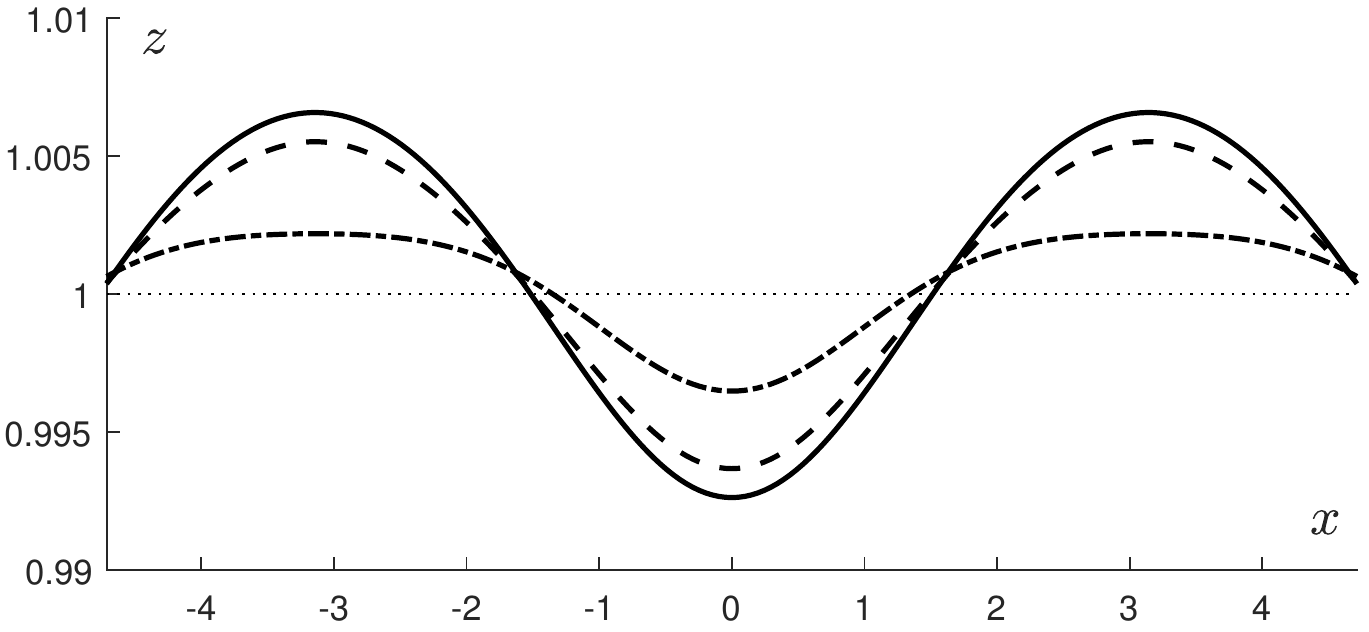}
&
\includegraphics[width=.45\columnwidth]{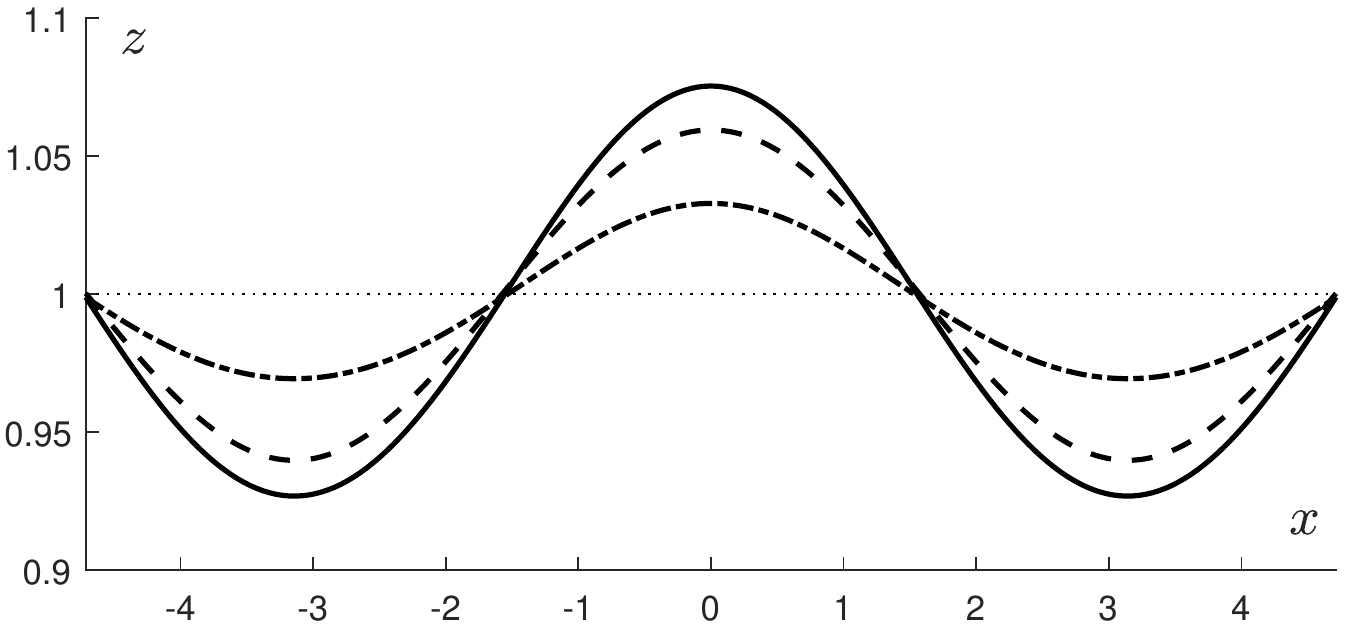}
\\
\rotatebox{90}{\hspace{10mm}$\knil=\pi$}
&
\includegraphics[width=.45\columnwidth]{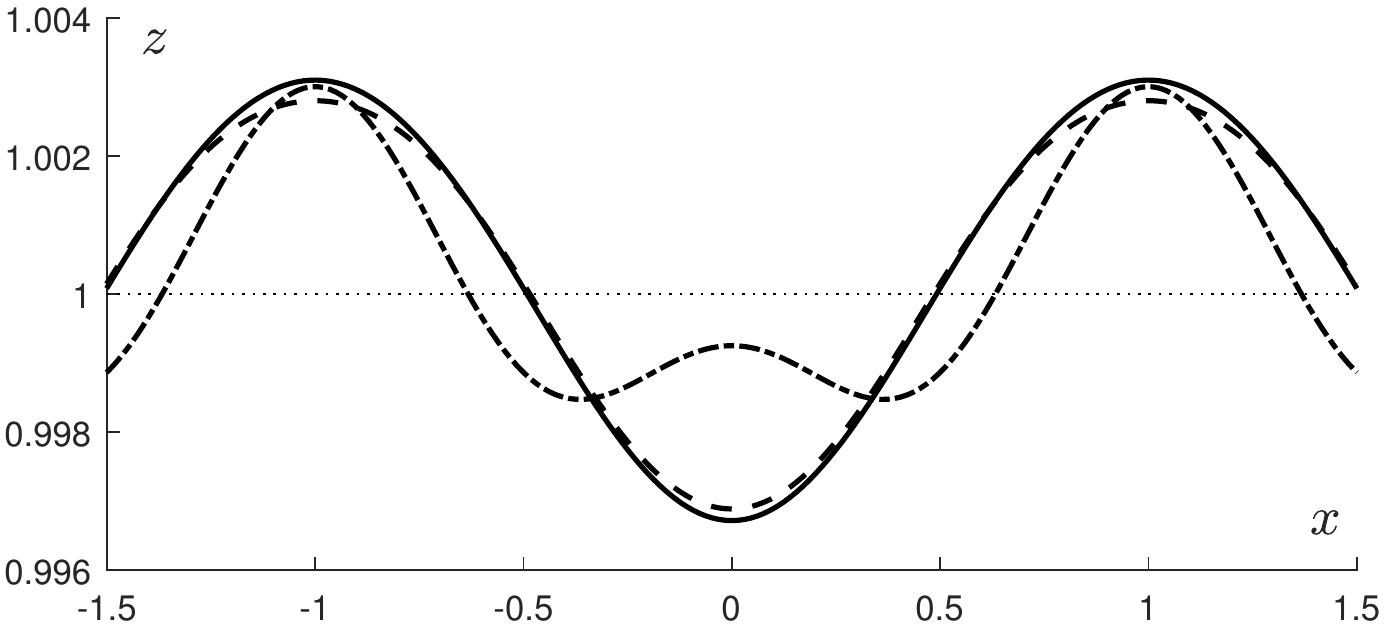}
&
\includegraphics[width=.45\columnwidth]{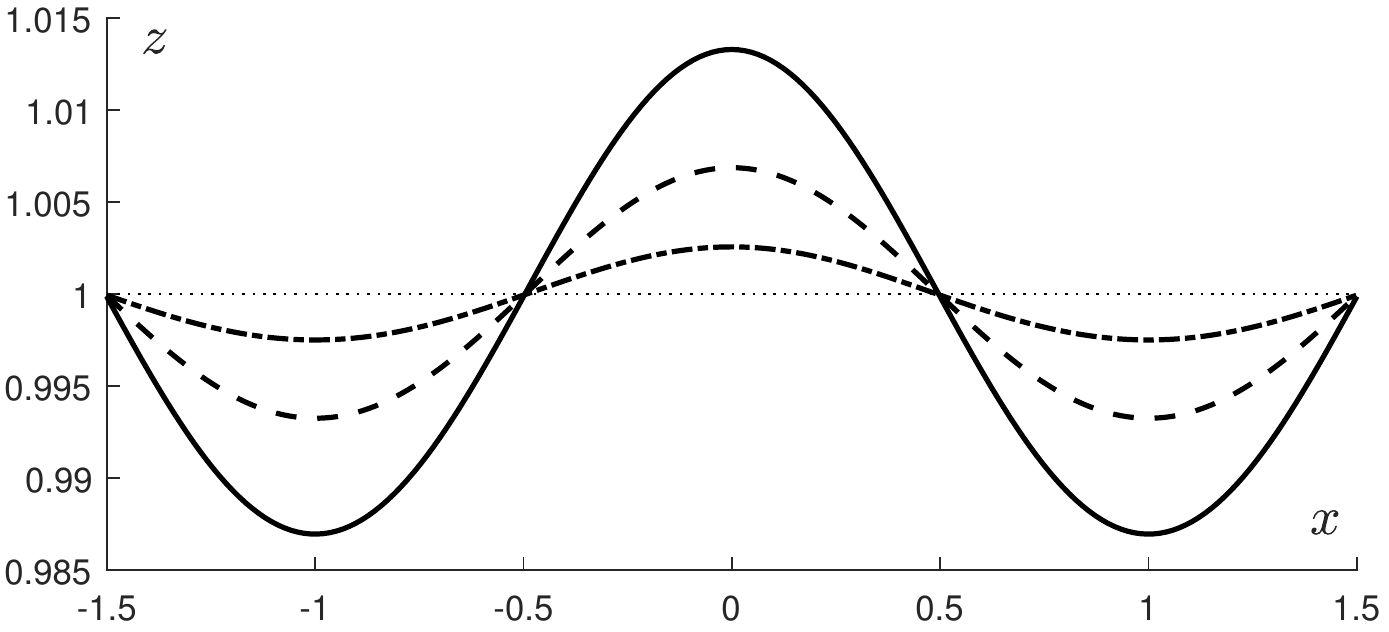}
\end{tabular}
\caption{
Surface undulations over a sinusoidal bed;
$\petab = \etabnil \cos (\knil x)$.
Three current profiles are shown---%
solid: $q=0$; dashed: $q=1/2$; dash-dotted: $q=1$.
$\delta=0.125$ and $\etabnil = 0.10$.}%
\label{fig:eta_s}%
\end{figure}

Figure~\ref{fig:eta_s} shows surface undulations over a sinusoidal bed $\petab = \etabnil \cos x$. 
Three current profiles, uniform, linear and intermediate, are displayed in sub- and supercritical flow regimes. 
Mean Froude number
$\Fr\_m=\Fr\,(1-\delta^{q+1})/[(1-\delta)(1+q)]$ (based on the mean current velocity) 
has been chosen as the flow intensity parameter for better scaling of the amplitudes.
(One of the main characteristics governing undulation magnitude is the state's proximity to a resonant state $\Fr\approx\FrCrit$.)
In regard to figure~\ref{fig:eta_s}, we make some general observations by varying the parameters 
$\Fr$, $q$ and $k$. 
\begin{itemize}
\item The undulations of the surface are in phase with those of the bed if the flow is supercritical and in antiphase if the flow is subcritical. 
This has long since been established for uniform flows 
\citep[e.g.][p.~409]{lamb1932hydrodynamics}.
The same is true also of higher order harmonics, 
although this together with the sign of the correctional amplitude $\dn$ from \eqref{eq:dn} will determine whether the higher order harmonic is in phase or anti-phase with the bed and thus whether maxima or minima align with the  peaks and troughs of the principal harmonic. 
\item Shallower flows (longer wavelength relative to depth) usually translates into higher amplitudes (relative to depth), but the degree to which this is notable or not depends on the Froude number. 
Exceptions will arise in the vicinity of resonant states.  
\item 
The surface shape in subcritical flows depends on the flow state relative to the resonant harmonics; 
in figure~\ref{fig:eta_s} it is seen that the nonlinear effect is to widen peaks and sharpen troughs when $\knil = 0.1$ and $1.0$, yet the opposite effect dominates for $q=1$ when $\knil = \pi$. 
Supercritical harmonics will eventually be encountered as the active wavenumber \eqref{eq:kappa} grows higher with increasing perturbation order, even if low-order modes are subcritical.
(The most common feature in subcritical flow seems to be a widening of the peaks and sharpening of the troughs, 
which was the observation reported by   
\citet{mizumura_1995_potential_wavy_bed} for uniform currents
with weakly nonlinear boundaries.)
\item
Conversely, \textit{all} higher harmonics are of a supercritical nature when the principal harmonic is supercritical, and so no critical (resonant) states can then be encountered by the higher harmonics.
Still, the influence of higher order harmonics will depend on the sign of the correctional amplitudes $\dn$ which in turn is observed to be strongly wavenumber dependent; supercritical profiles with $\knil=1.0$ are in figure~\ref{fig:eta_s} very slightly sharpened at the peaks (too little to be clearly visible) while the opposite effect is evident for $\knil=0.1$. 
Note therefore that the  effect reported by \citet{mizumura_1995_potential_wavy_bed} for supercritical flows based on numerical and experimental observations (a sharpening of the peaks and widening of the troughs in resemblance to a Stokes wave)
is here observed only intermittently. 
This incongruence was already pointed out in the benchmark test of figure~\ref{fig:wave_train_benchmark} and  is not surprising%
. The sharpening of crests and widening of troughs is a hallmark of Stokes waves, an effect of nonlinear orders of the free-surface steepness.  We cannot capture this phenomenon, however, having linearised the water surface 
to limit the scope of the present work as discussed in out introduction.  
\\
\end{itemize}

\begin{figure}%
\subfigure[$\Fr=0.9$]{\includegraphics[width=.5\columnwidth]{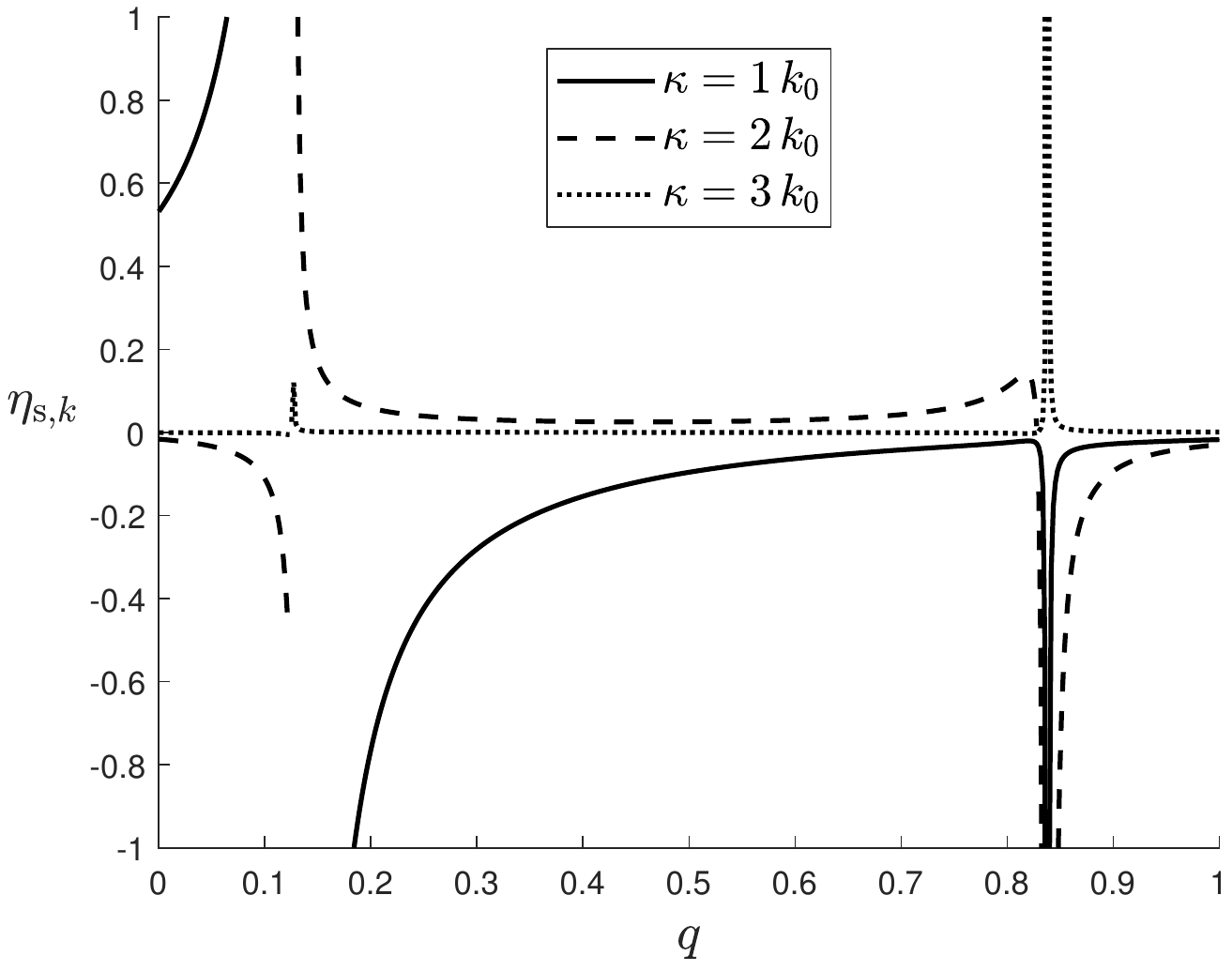}
\label{fig:etas_graph:q}}
\subfigure[$q=1/7$]{\includegraphics[width=.5\columnwidth]{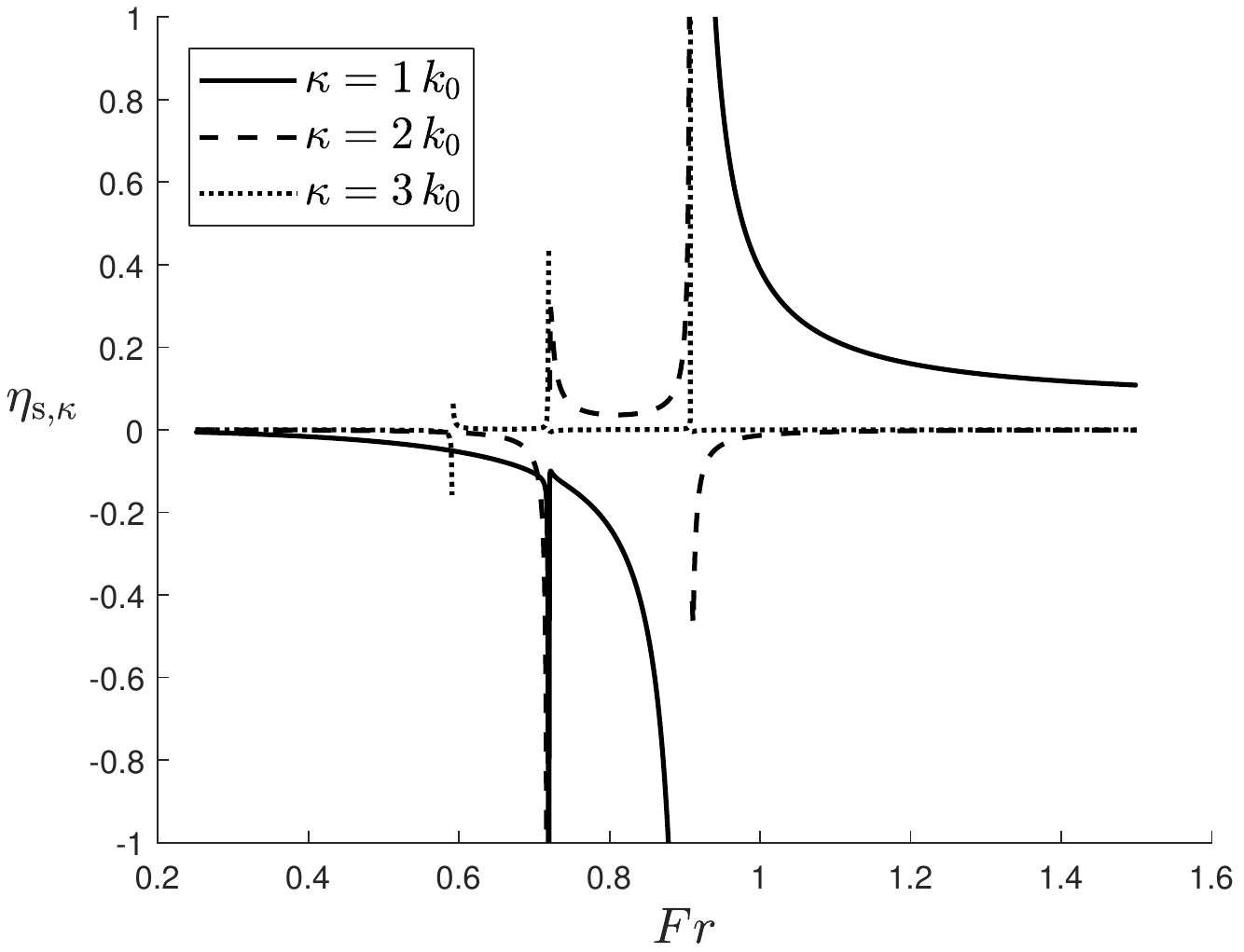}
\label{fig:etas_graph:Fr}}
\caption{
The amplitudes of the first, second and third harmonics $\etask$ of the surface modulation, displayed as function of $q$ and $\Fr$.
$\delta=0.125$, $\etab=0.1$, $\knil=1$.
}%
\label{fig:etas_graph}%
\end{figure}

Consider next the possible resonance of higher harmonics, in the sense of approaching a critical Froude number as described in Section~\ref{sec:Frcrit}.
Figure~\ref{fig:etas_graph} illustrates this in terms of the harmonic amplitudes $\etask$; $\kappa\in\{k,2k,3k\}$.
These are net harmonic amplitudes from which the surface undulation is composed; 
$\petas={\sum}_{j=1}^\infty \eta_{{\rm s},j k} \cos(j k x)$.
Each term 
of the perturbation series \eqref{eq:series_expansion} brings with it a set of harmonics up to and including the order $n$ of that term.
$\etask$ is computed by the summation of these contributions, respective of each harmonic.
Figure~\ref{fig:etas_graph} also reveals that \textit{all} harmonics become singular at the same critical states, not just the resonant one (e.g., the one satisfying \eqref{eq:Fr_Crit}), although the resonant harmonic dominates in the neighbourhood of the critical state. 
The reason for this is that the harmonics are coupled at the lower boundary in terms of the in $\dn$ coefficients, \eqref{eq:dn}.
Thus, $\dn$ becomes large when one of the lower order harmonics is close to resonance.

Now, say, for example, that the flow rate over a long sinusoidal bed of finite amplitude is gradually increased. If the current profile were to obey a fixed power-law, then a number of distinct states of resonance, respective to each harmonic, would be seen at the surface (see figure~\ref{fig:etas_graph:Fr} and figure~\ref{fig:monochrom_fith_order_resonance} below).

\arev{
More generally, 
\citet{sammarco_1994_mei_1969_Fr_crit_transient} showed that,
just as finite bathymetry amplitudes can cause higher order resonance at the surface, so can finite surface amplitudes generate higher order resonance at the bed. 
Consequently, wavelength ratios $\ldots, \tfrac13, \tfrac12, 1, 2, 3,\ldots$ between surface and bed may be resonant when both are of finite amplitude.
}
\\

\begin{figure}%
\centering
\includegraphics[width=.75\columnwidth]{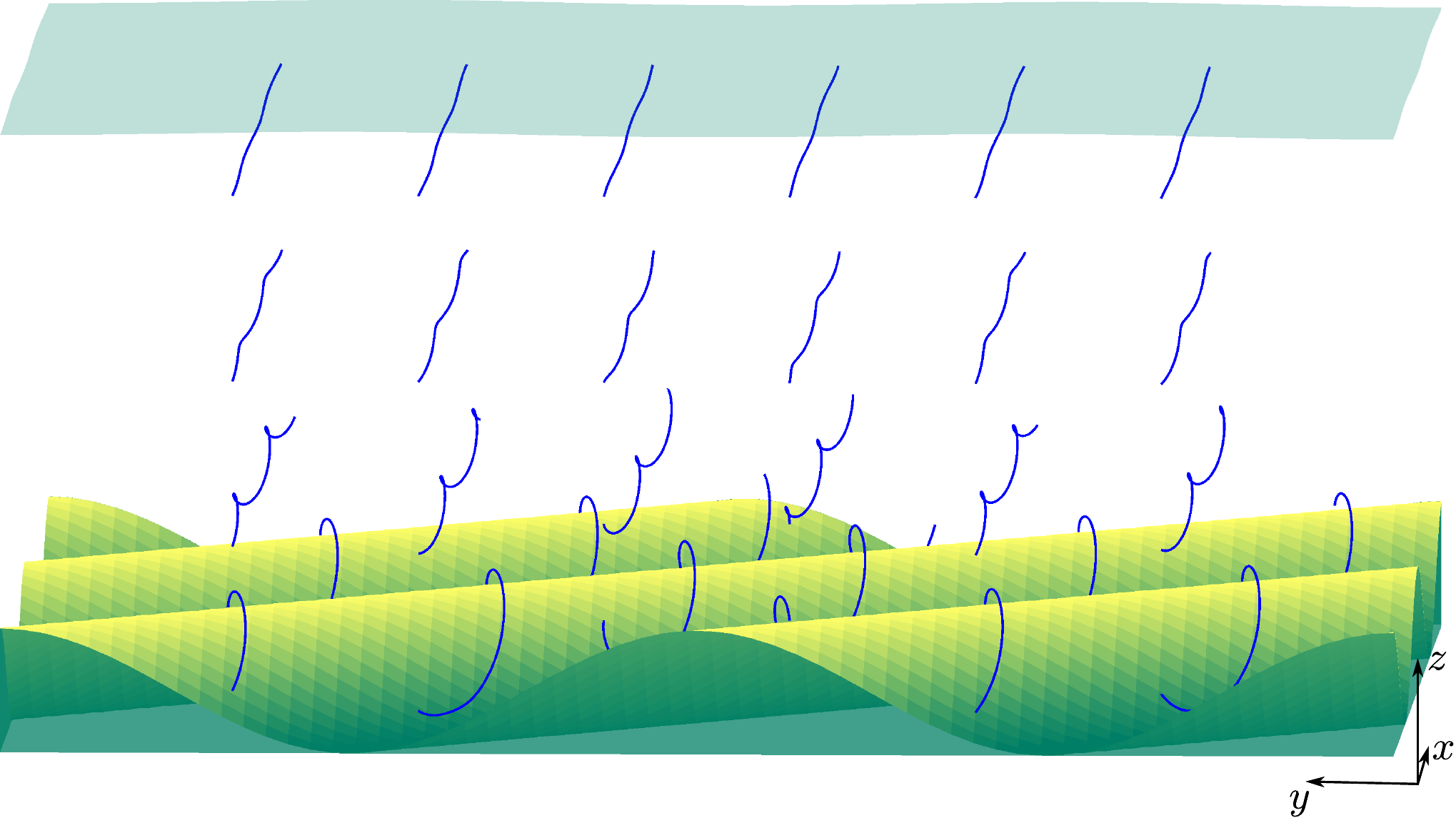}%
\caption{
\arev{
Streamlines (blue) from a current flowing obliquely over a sinusoidal bed (yellow--green surface). 
Free surface shown in transparent blue.
The sinusoidal bed pattern is rotated 45 degrees relative to the direction of the current.
The viewpoint is such that the current is directed straight ahead (`into the paper'). }
$\bknil = (\pi,\pi)$, $\delta = 0.1$,  $\etabnil = 0.1$, $\Fr=0.5$, $q=1/7$. 
}%
\label{fig:45deg_streamlinetwisting}%
\end{figure}

\begin{figure}%
\centering
\includegraphics[width=.75\columnwidth]{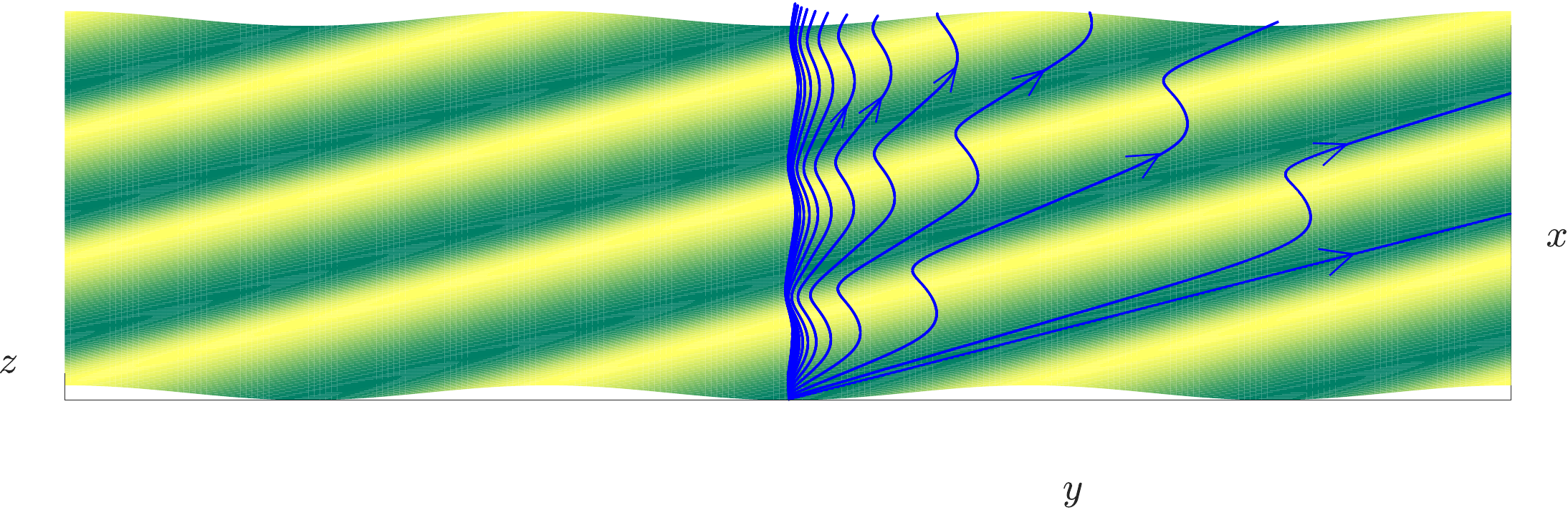}%
\caption{
\arev{
Streamlines (blue) from a current flowing obliquely over a sinusoidal bed (yellow--green surface). 
The sinusoidal bed pattern is rotated 45 degrees relative to the direction of the current.
}
The shown streamlines originate at (from rightmost to leftmost) $z=0$, $.025$, $.05$, $\ldots$.
$\bknil = (1,1)$, $\delta = 0.1$,  $\etabnil = 0.1$, $\Fr=0.5$, $q=1$.
}%
\label{fig:45deg_streamlinetwisting_drift}%
\end{figure}

Next to be considered are currents obliquely incident on a sinusoidal bed.
Figure~\ref{fig:45deg_streamlinetwisting} shows such a bed at $45$ degrees relative to the current. 
As is known from various earlier works in the mathematically analogous case of surface waves \citep[e.g.][]{Ellingsen_vorticity_paradox}, the oblique interaction between linear waves and shear causes stream and vortex lines to twist, as is clearly evident from the image. 
This effect disappears only if the current is shear free, i.e.\ uniform 
($q=0$).
Another interesting phenomenon in an obliquely directed sinusoidal pattern is the streamline migration effect, made clearly visible in figure~\ref{fig:45deg_streamlinetwisting_drift}.
The starting points of our streamlines
are placed along a vertical line with equal spacing throughout the water column, 
the deepest of which sitting right on the bed trough itself
and the others extending upwards to $z=0.35$ with increments of $0.025$.
Streamlines originating from deep within a trough are seen to drift along it, 
whereas streamlines originating from higher up in the flow field are increasingly aligned with the current. 
The lowest points on the bed are at $z = \delta - \etabnil$, which is zero in this case.
At $z=0$, where the unperturbed current is stagnant, 
the streamline remains in the trough indefinitely.
(This will not be the case if 
the trough remains above the current stagnation depth $z=0$.)
The phenomenon is most clearly visible in linear currents ($q=1$), which has been applied here.
\\
 
\begin{figure}%
\begin{tabular}{c|c}
$N=1$ & $N=1$
\\
\includegraphics[width=.5\columnwidth]{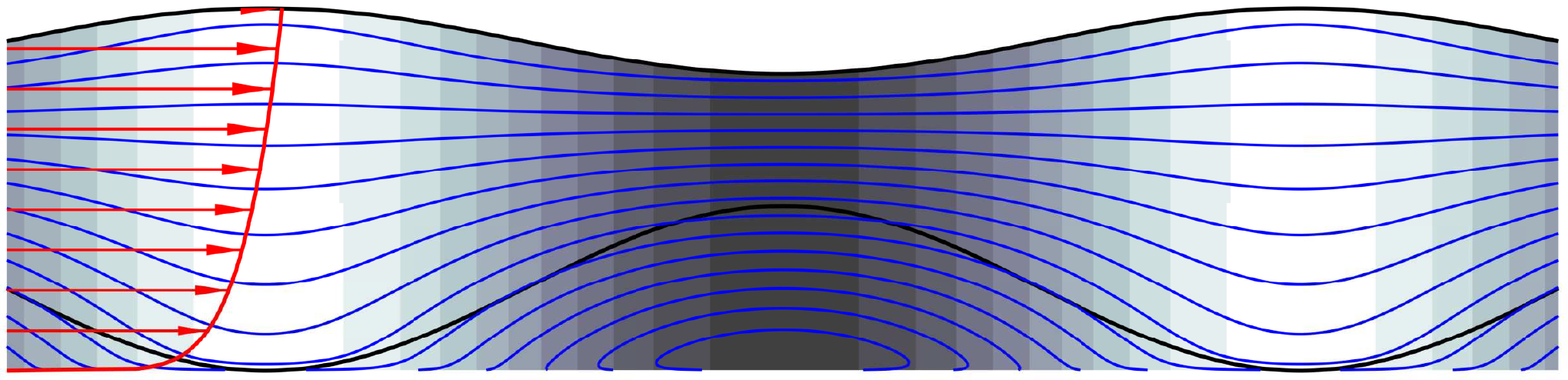}
&
\includegraphics[width=.5\columnwidth]{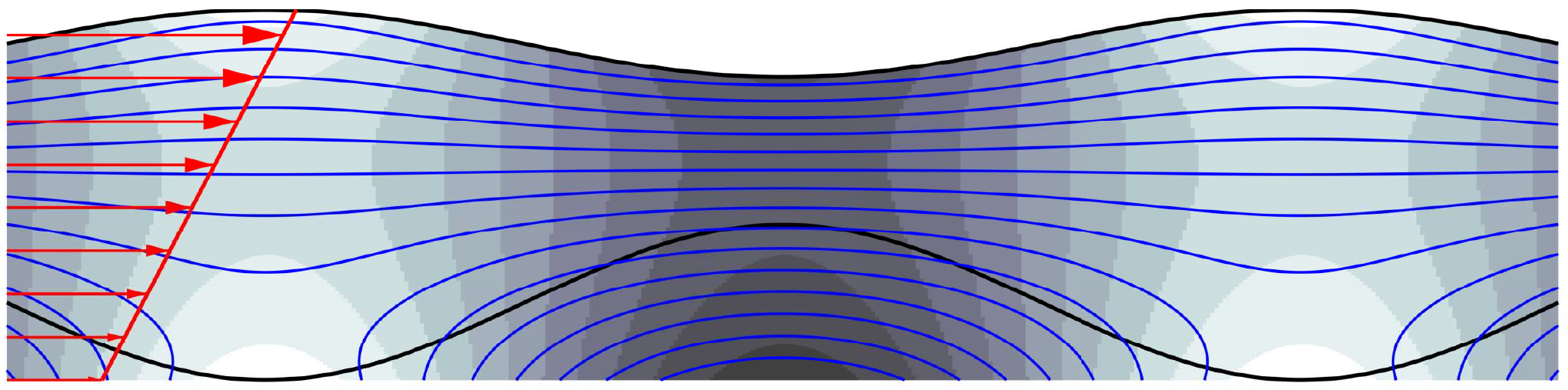}
\\[1.5ex]
$N=2$ & $N=2$
\\
\includegraphics[width=.5\columnwidth]{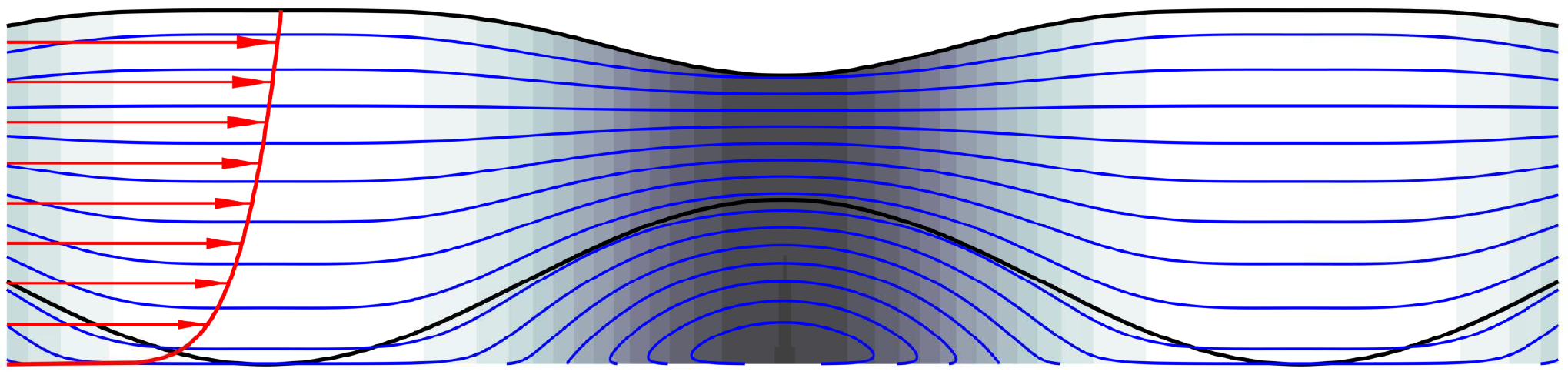}
&
\includegraphics[width=.5\columnwidth]{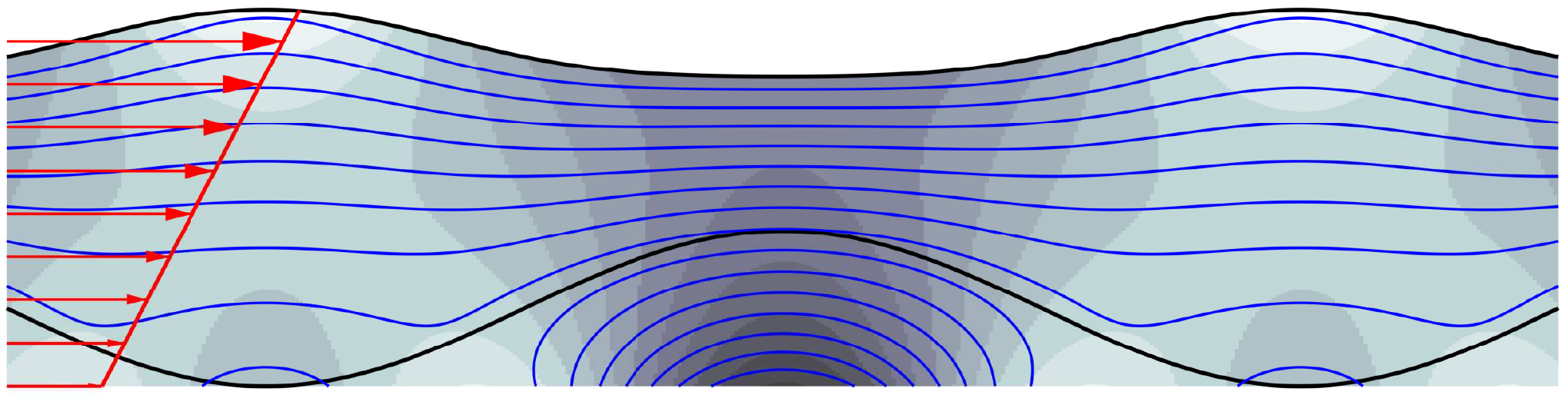}
\\[1.5ex]
$N=3$ & $N=3$
\\
\includegraphics[width=.5\columnwidth]{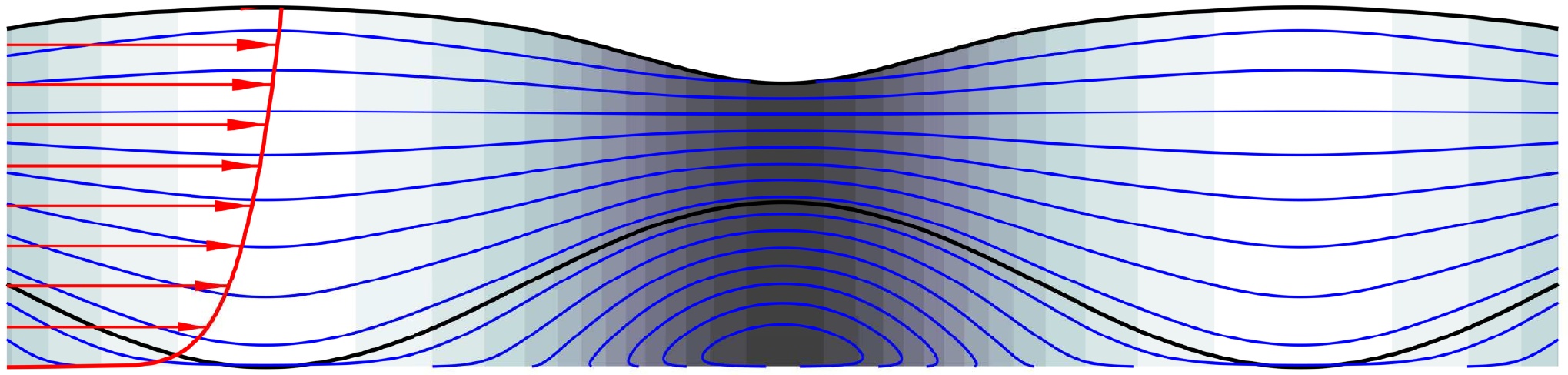}
&
\includegraphics[width=.5\columnwidth]{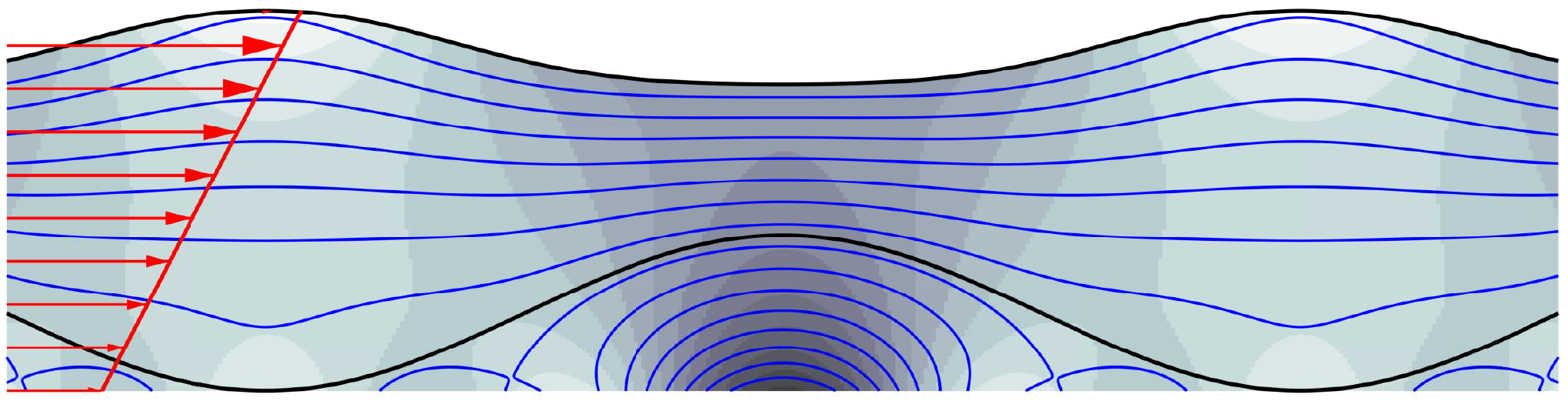}
\\[1.5ex]
$N=6$ & $N=6$
\\
\includegraphics[width=.5\columnwidth]{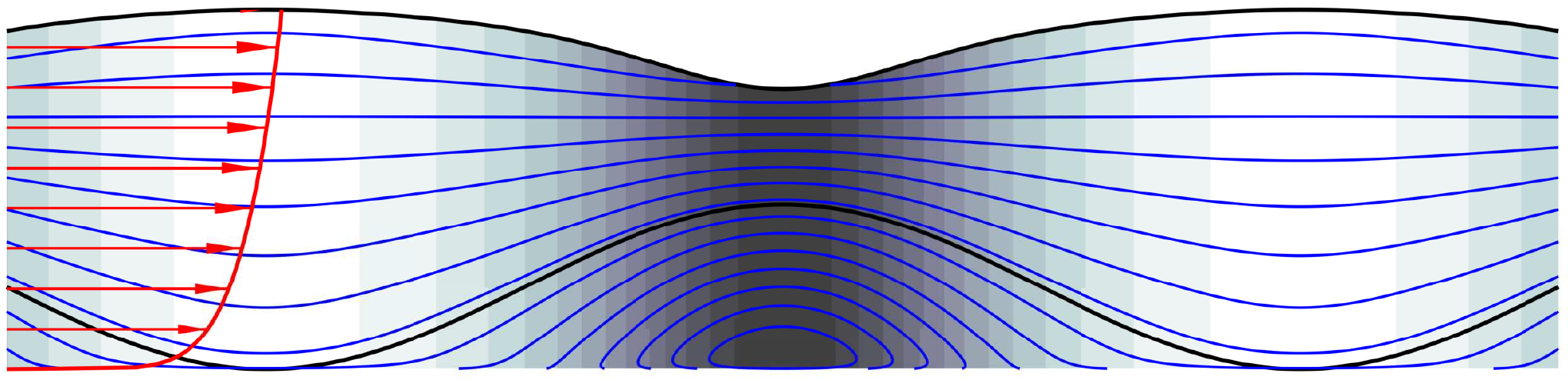}
&
\includegraphics[width=.5\columnwidth]{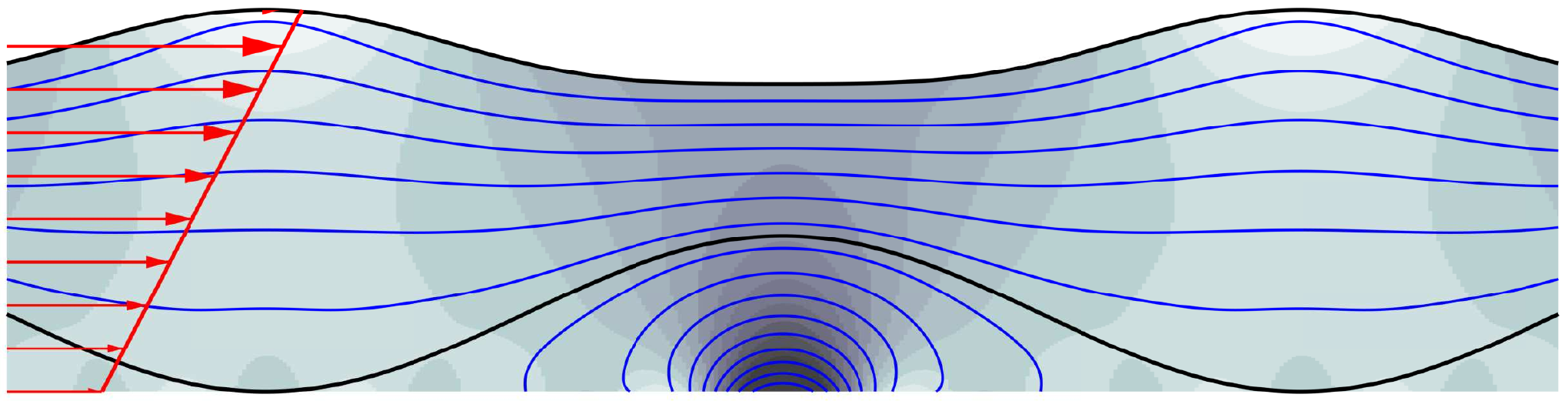}
\\
\subfigure[$q=1/7$, $\knil = 0.1$, $\delta = 0.25$, $\etabnil = 0.25$.
Aspect ratio: 20:1.]{\label{fig:monochrom_Ns:q=1/7} \hspace{.5\columnwidth}}
&
\subfigure[$q=1$, $\knil = \pi$, $\delta = 0.5$, $\etabnil = 0.15$.
Aspect ratio: 1:1.]{\label{fig:monochrom_Ns:q=1} \hspace{.5\columnwidth} }
\end{tabular}
\caption{
Illustration of nonlinear bed solution of Section~\ref{sec:exact_lower_BC} for series truncated at $N=1$, $2$ and $6$, depicted as 
flow field plots showing surface, bed, dynamic pressure field and contours of the stream function \eqref{eq:streamfunction}.  $\Fr=0.5$.
}%
\label{fig:monochrom_Ns}%
\end{figure}

Next, some illustrative examples of accuracy as a function of the truncation order $N$ of the series \eqref{eq:series_expansion} of the lower boundary condition are presented. 
Figure~\ref{fig:monochrom_Ns} shows two examples of subcritical flow over a large-amplitude sinusoidal bed. 
Streamlines are typically flatter in subcritical flows compared to supercritical flows since the surface undulations are in anti-phase to those of the bed.
Higher order harmonics are therefore more visible in subcritical flows.
Cases chosen for figure~\ref{fig:monochrom_Ns} are challanging in the sense that surface amplitudes are large and series convergence slow. 
A $q=1/7$-shear profile 
in a shallow-water stream
is presented in figure~\ref{fig:monochrom_Ns:q=1/7} and a linear shear over an intermediate wavelength bed in figure~\ref{fig:monochrom_Ns:q=1}.
Solutions of the internal flow are in these cases 
accurate since the perturbation steepness is small in the first case and the vorticity uniform in the latter, as pointed out in Section~\ref{sec:exact_lower_BC}. 
The surface elevation, which is similar in the linear solution $N=1$ of the two cases, evolves in opposite ways with increasing truncation orders; 
a widening of the peaks and narrowing of the troughs is observed in figure~\ref{fig:monochrom_Ns:q=1/7}, while the trough flattens as the peaks become sharper in figure~\ref{fig:monochrom_Ns:q=1}.

As discussed 
in relation to figure~\ref{fig:etas_graph},
resonance with higher order harmonics can be encountered in flows where the linear harmonic is subcritical
since the critical Froude number $\FrCrit$ is a monotonically decreasing function of the ever-increasing active wavenumber \eqref{eq:kappa}, see figure~\ref{fig:Fr_crit}.
Figure~\ref{fig:monochrom_fith_order_resonance} exemplifies this, showing a third case where a resonance effect in the fifth-order harmonic is distinct. 
This can be anticipated because $\FrCrit\of{5\bknil}\approx 0.49965$ with the chosen parameters, close to the actual Froude number $\Fr=5.0$.
The proximity to a critical Froude number manifests in the appearance of a 
strong undulation with wavelength one fifth that 
of the bed.
These undulations become apparent only the computation includes fifth-order terms and above ($N\geq5$). 

In figure~\ref{fig:monochrom} we show 
profiles generated by $q>1$ currents for long and intermediate wavelengths.
This may for example be representative of a flow driven by surface stresses
such as the Ekman layer near wind-swept water surfaces \citep[e.g.][]{abdullah_1949_exp_current_profile}.
Recirculation occurs near the topography troughs for the chosen parameters. 
This phenomenon is related to the current vorticity and is not observed with uniform currents.
Recirculation becomes increasingly likely for higher values of $q$.
Linearisation errors in the Euler equations are in the intermediate wavelength range $\knil=\pi$ are not really negligible, but a case (figure~\ref{fig:monochrom:narrow}) has been included as indication of intermediate wavelength behaviour. 
\arev{
A 3D generalisation of figure~\ref{fig:monochrom:wide} is presented in figure~\ref{fig:monochrom:wide_3D}, showing streamline plots near the bed. 
A bulging bathymetry pattern is here considered as a sinusoidal surface of equal amplitude and wavelength is orthogonally superposed on the streamwise one.
Recirculation now occurs in the troughs of the bed, here in the region $z<\delta$, with streamline orbits tilted about planes of symmetry.
}

\begin{figure}%
\centering
\includegraphics[width=.6\columnwidth]{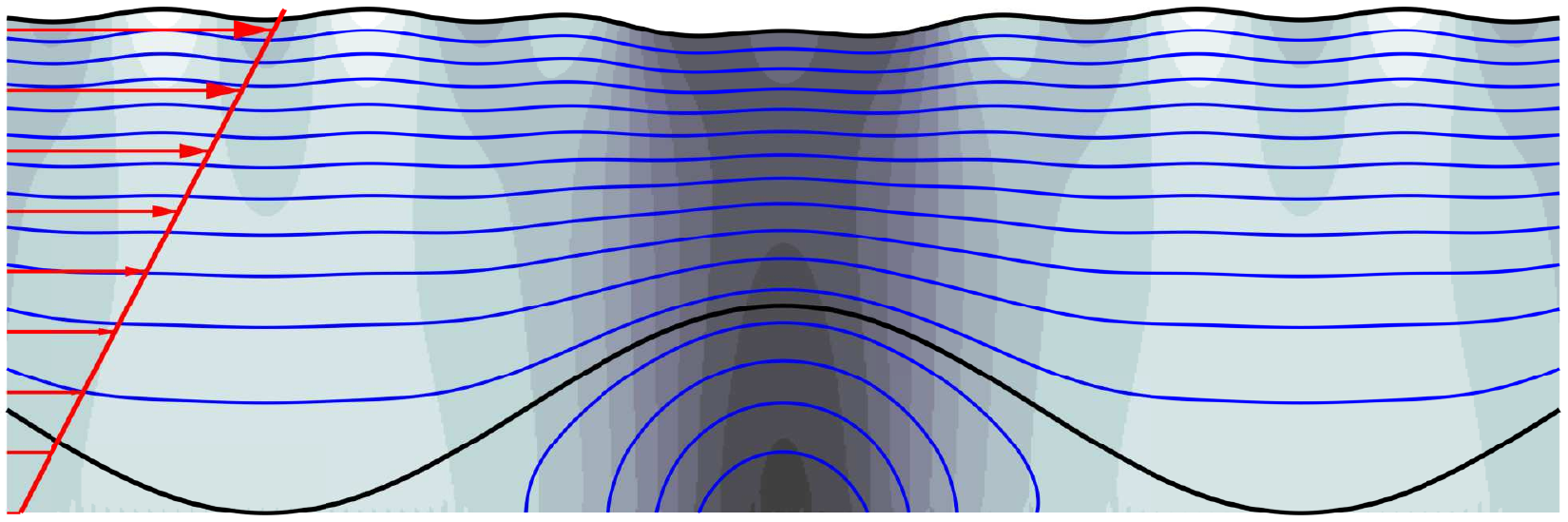}
\caption{
High order flow field plot
showing surface, bed, dynamic pressure field and contours of the stream function \eqref{eq:streamfunction}.
$\delta=0.25$, $\etabnil = 0.20$, $\Fr=0.5$, $\knil=1.0$. Aspect ratio: $\pi$:1.
For these parameters, $\FrCrit\of{5\bknil}\approx 0.49965$. Consequently, a perturbation of one-fifth the wavelength of the bed is clearly discernible in solutions truncated at $N\geq5$.
}%
\label{fig:monochrom_fith_order_resonance}%
\end{figure}

\begin{figure}%
\centering

\subfigure[$q=10$, $\knil=1$, $\delta=0.8$, $\etabnil=0.1$ and $\Fr=0.4$.
Aspect ratio: 5:1.]{
\includegraphics[width=.75\columnwidth]{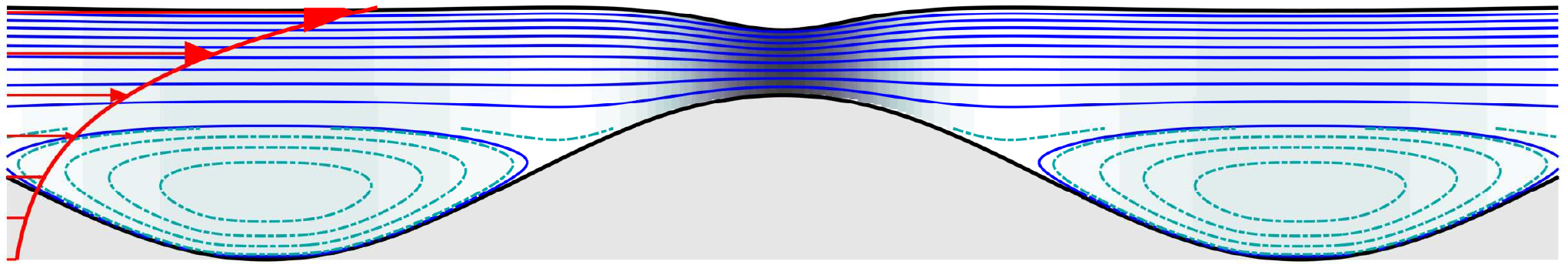}
\label{fig:monochrom:wide}
}%

\subfigure[$q=5$,  $\knil=\pi$, $\delta=0.7$, $\etabnil=0.1$ and $\Fr=0.75$.
Aspect ratio: 1:1.]{
\includegraphics[width=.75\columnwidth]{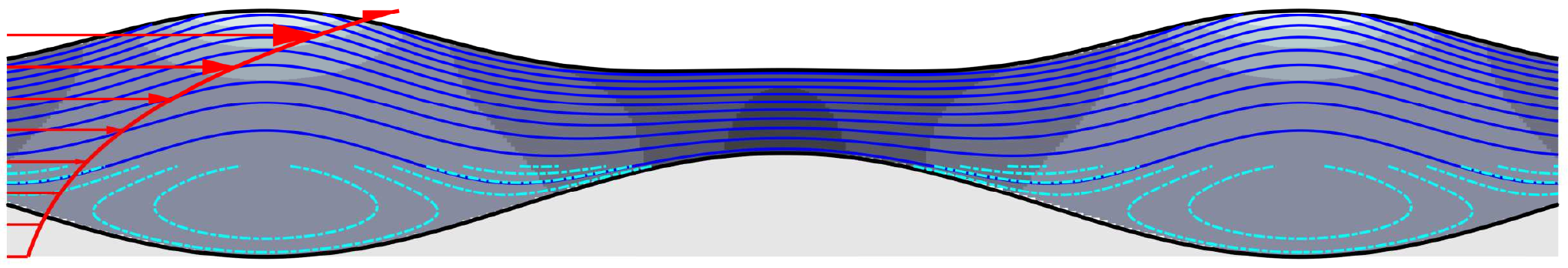}%
\label{fig:monochrom:narrow}%
}
\caption{
Near-exponential shear current over monochromatic bathymetry.
Weak recirculation is observed in the trough.
For visibility, 
additional, more closely spaced streamlines 
with volume flux between streamlines a factor 4 smaller 
are computed in the region $z<\delta$ to increase the contour resolution locally and display the recirculation happening here.
}
\label{fig:monochrom}
\end{figure}

\begin{figure}%
\centering
\includegraphics[width=.7\columnwidth]{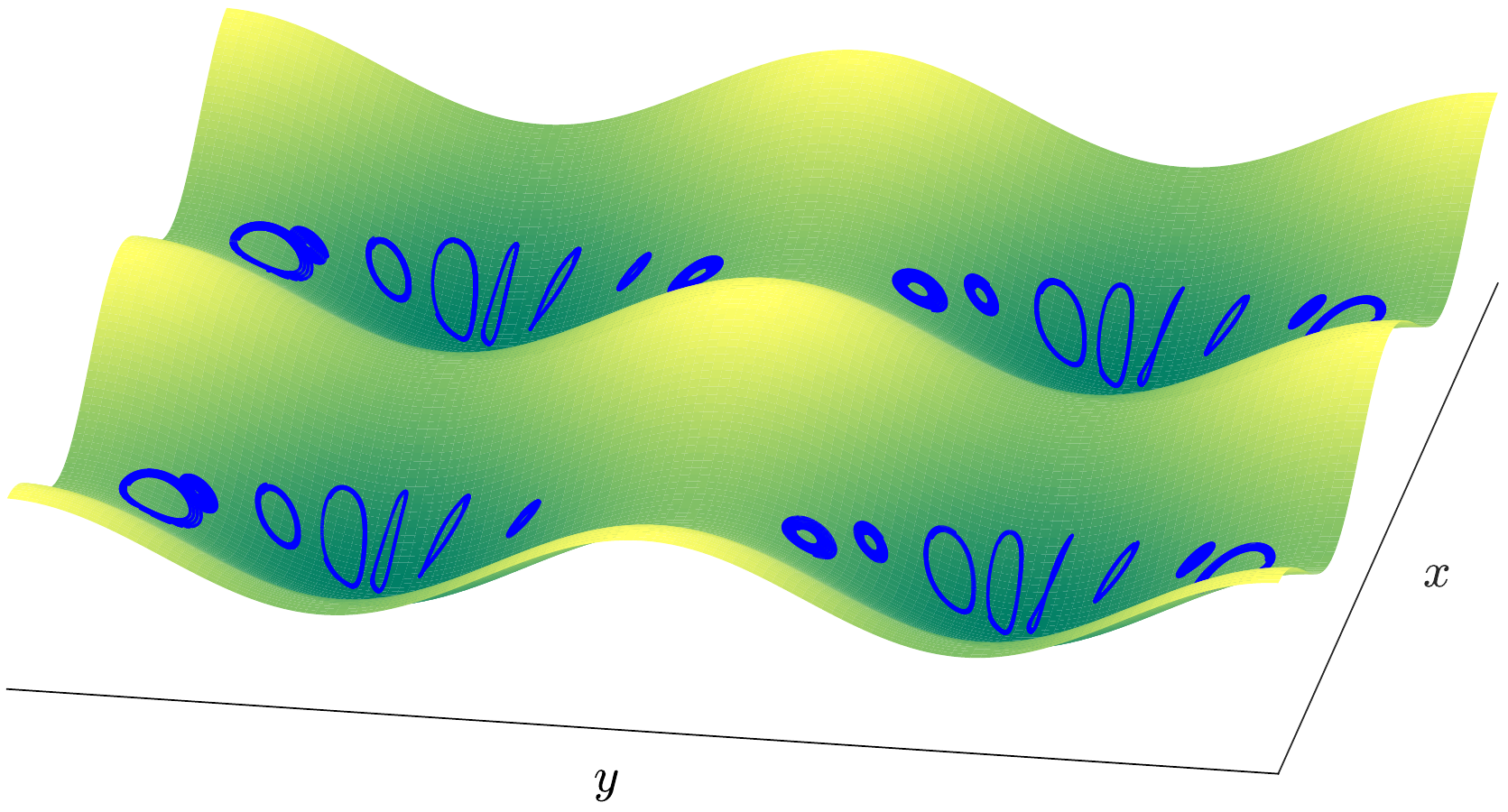}%
\caption{
\arev{
A 3D generalisation of the case presented in figure~\ref{fig:monochrom:wide} showing the near-bed recirculating streamlines (blue).
The bathymetry consists of two superposed  surfaces varying sinusoidally in the streamwise and spanwise direction, respectively, both of amplitude $\tfrac12\etabnil$ and wavenumber $\knil = 1$.
}
}%
\label{fig:monochrom:wide_3D}%
\end{figure}

\subsection{Flow over localised disturbances}

\begin{figure}%
\centering
\subfigure[Gauss. $b\_s=0$.]{
\includegraphics[width=\columnwidth]{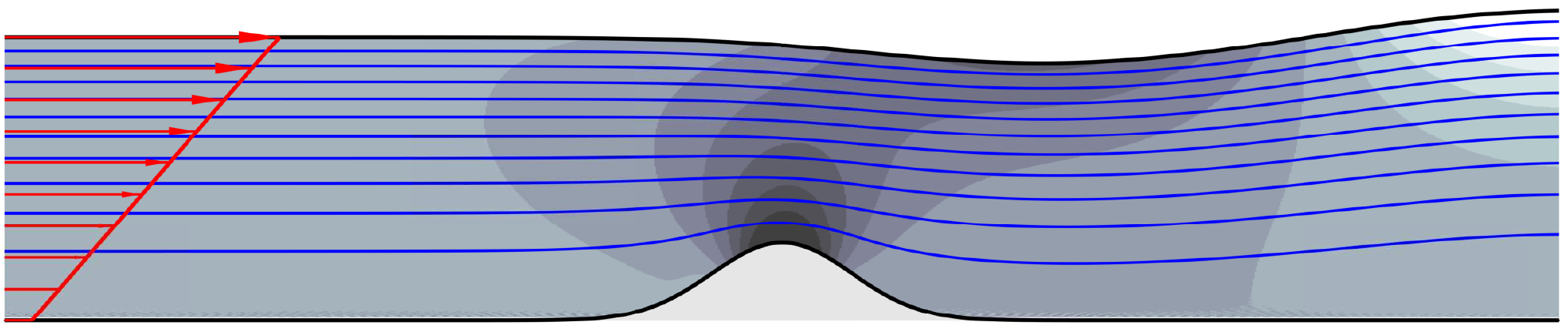}
}
\subfigure[Wedge. $b\_s=0.1\,b$.]{
\includegraphics[width=\columnwidth]{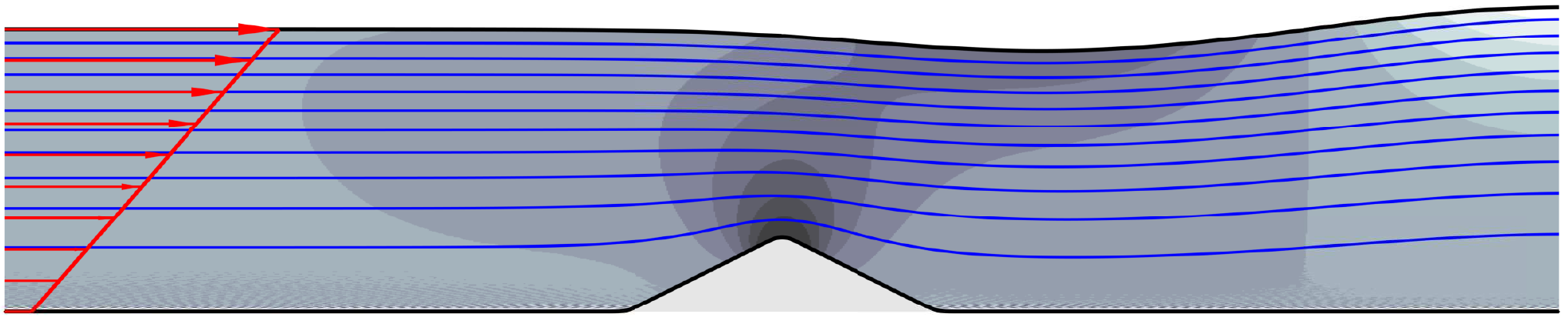}
}
\subfigure[Ellipse. $b\_s=0.1\,b$.]{
\includegraphics[width=\columnwidth]{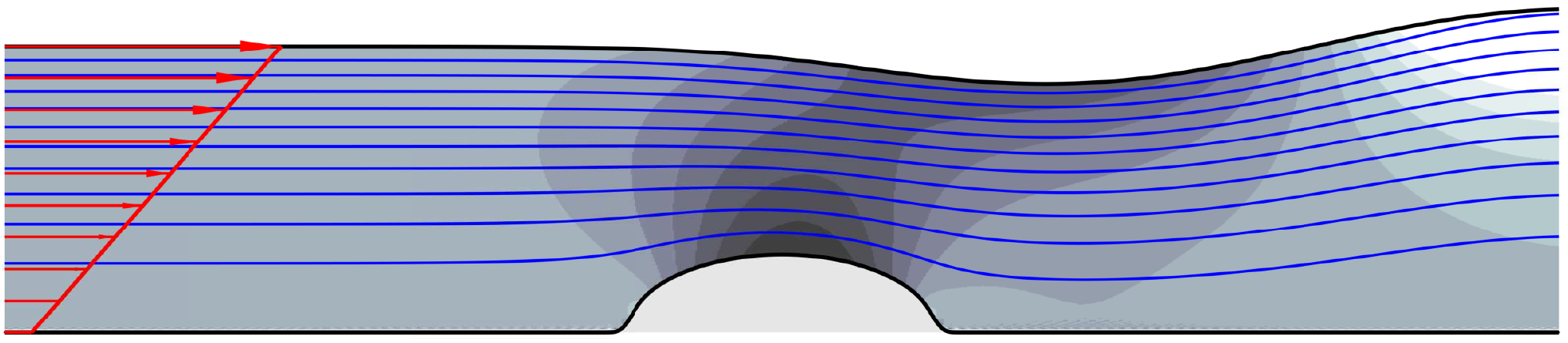}
}
\subfigure[Rectangle. $b\_s=0.25\,b$.]{
\includegraphics[width=\columnwidth]{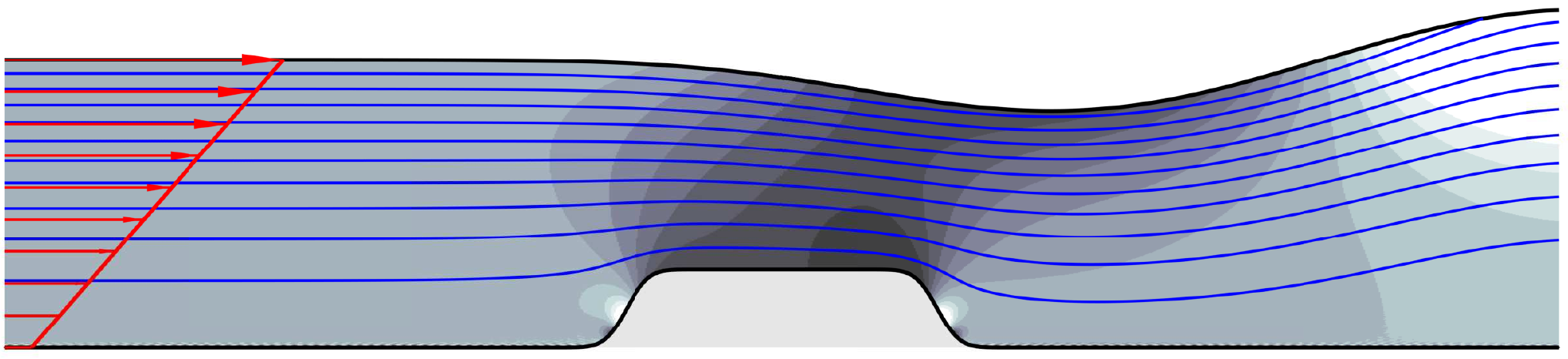}
\label{fig:obstacle_flow:narrow:rectangle}%
}
\caption{Flow over narrow obstacles (see Table~\ref{tab:eta_b}),
showing  dynamic pressure field and contours of the stream function \eqref{eq:streamfunction}.
Aspect ratio: 1:1. Obstacle half-width $b = 0.5$, $\delta = 0.1$, $\etabnil = 0.25$, $q=1$, $\Fr = 1.0$
}%
\label{fig:obstacle_flow:narrow}%
\end{figure}

\begin{figure}%
\centering
\subfigure[Gauss. $b\_s=0$.]{
\includegraphics[width=\columnwidth]{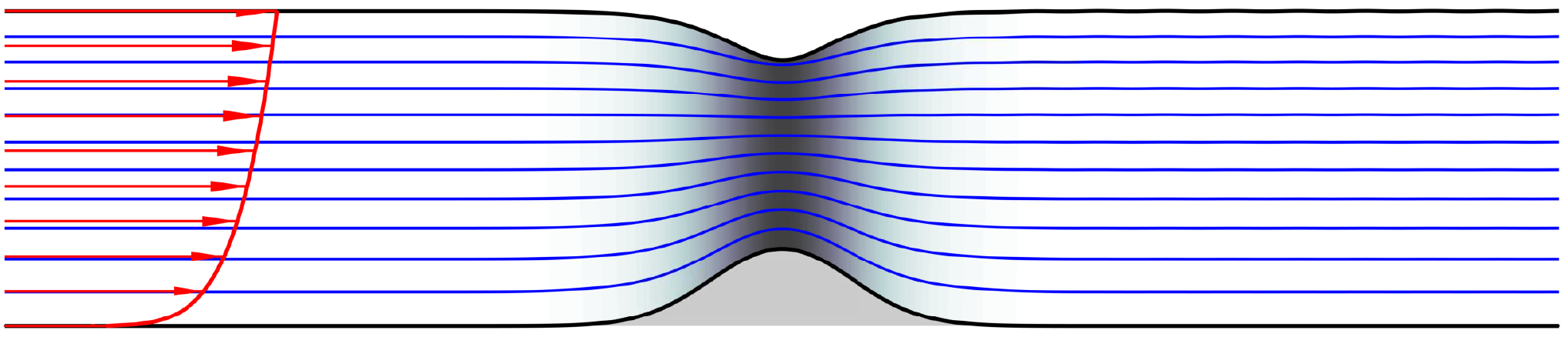}
}
\subfigure[Wedge. $b\_s=0$.]{
\includegraphics[width=\columnwidth]{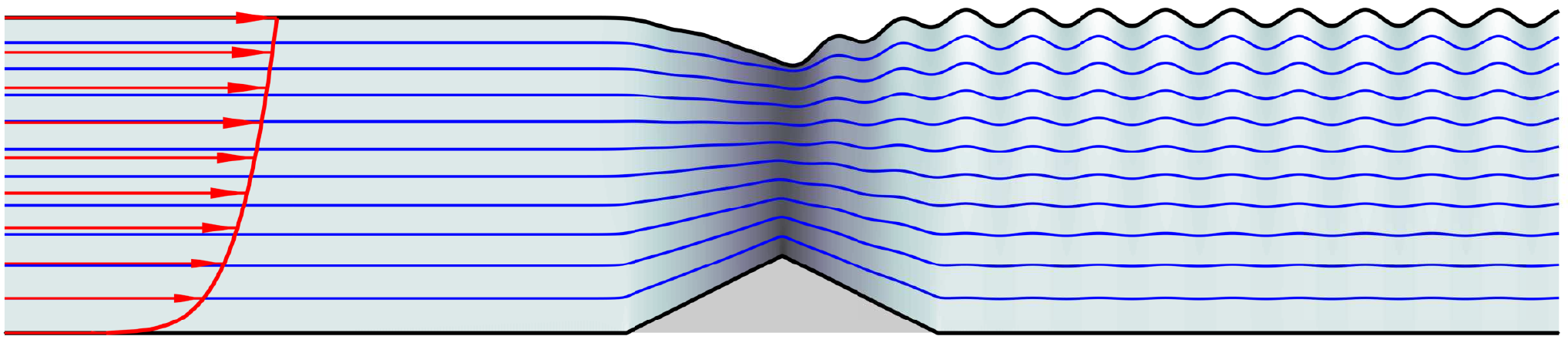}
}
\subfigure[Ellipse. $b\_s=0.05\,b$.]{
\includegraphics[width=\columnwidth]{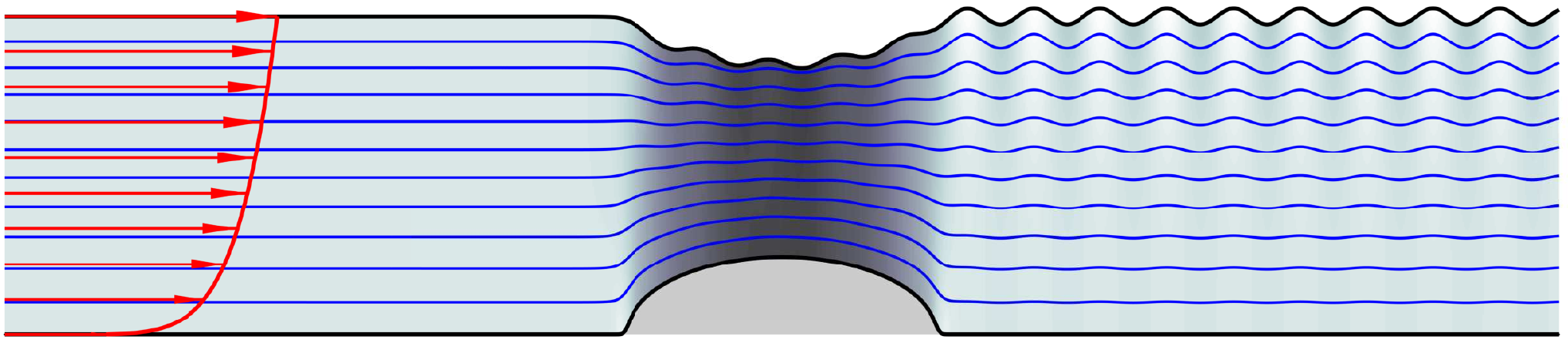}
}
\subfigure[Rectangle. $b\_s=0.1\,b$.]{
\includegraphics[width=\columnwidth]{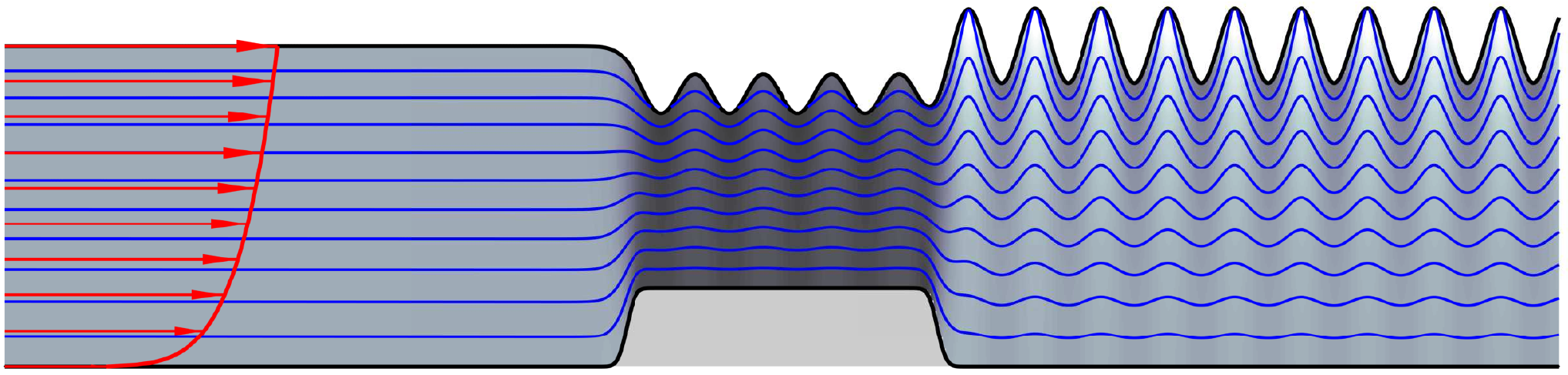}
\label{fig:obstacle_flow:wide:rectangle}%
}
\caption{
Flow over wide obstacles (see Table~\ref{tab:eta_b}),
showing  dynamic pressure field and contours of the stream function \eqref{eq:streamfunction}.
Aspect ratio: 10:1. Obstacle half-width $b = 5.0$, $\etabnil = 0.25$, $q=1/7$, $\Fr = 0.6$, $\delta = 0.001$}%
\label{fig:obstacle_flow:wide}%
\end{figure}

We now consider a current which is perturbed as it flows over a localised bed defect.
Table~\ref{tab:eta_b} gives the functional representation of some two-dimensional obstruction shapes in physical and Fourier space.
An inherent assumption in our model (particularly in the vertical Taylor expansion) is that the length-to-amplitude ratio of undulations is small. 
In practice this places a limit on how steep and unsmooth our obstructions can be. 
(See e.g.\
\citet{Holliday_1977_mode_coupling_compare_Taylor} and \citet{West_1981_lecture_notes_weak_interaction} for similar discussions related to surface waves.)
It can also ultimately limit the convergence of the procedure presented in Section~\ref{sec:exact_lower_BC} as each perturbation order introduces increasingly smaller wavelengths.
We therefore soften some of the sharper shapes by convoluting them with a Gaussian profile of half-width $b\_s$, written in the bottom entry of Table~\ref{tab:eta_b}.
The amount of softening, if any, is given in the subfigure captions.
The flow fields around these are shown in figure~\ref{fig:obstacle_flow:narrow} and \ref{fig:obstacle_flow:wide} for the cases of narrow obstacles within a linear shear current and of wide obstacles within a $q=1/7$ current profile, respectively. 
We have
$\delta\ll\etabnil$ in the latter case which will cause the Taylor series extension of the lower boundary condition to diverge; the alternative iterative approach discussed in Section~\ref{sec:exact_lower_BC} is then instead adopted.
Radiation conditions are imposed through the method of separation into a `far field' and a `near filed', as described in Section~\ref{sec:asymptotic}.

Wavelengths of the train generated behind the obstructions are 
much longer than the obstructions themselves for the flow in figure~\ref{fig:obstacle_flow:narrow}. 
Conversely, in figure~\ref{fig:obstacle_flow:wide} they are significantly shorter than the obstruction. 
The Gaussian shape is `sufficiently smooth' to suppress the relatively shorter wave train in the latter ($\etab\of{k_x}$ decays rapidly) while the same is not true of the other shapes which contain sharp edges
and whose Fourier series representations thus have finite-amplitude contributions of zero wavelength. 
The rectangular, having also vertical side-walls,  shape is the most severe of the set and requires more smoothing compared to the others. 
Anomalies are seen at the face of the narrow rectangle in figure~\ref{fig:obstacle_flow:narrow:rectangle} where the smoothing is insufficient, although the effect of these anomalies doesn't penetrate far out into the flow. 

\begin{figure}%
\centering
\includegraphics[width=.75\columnwidth]{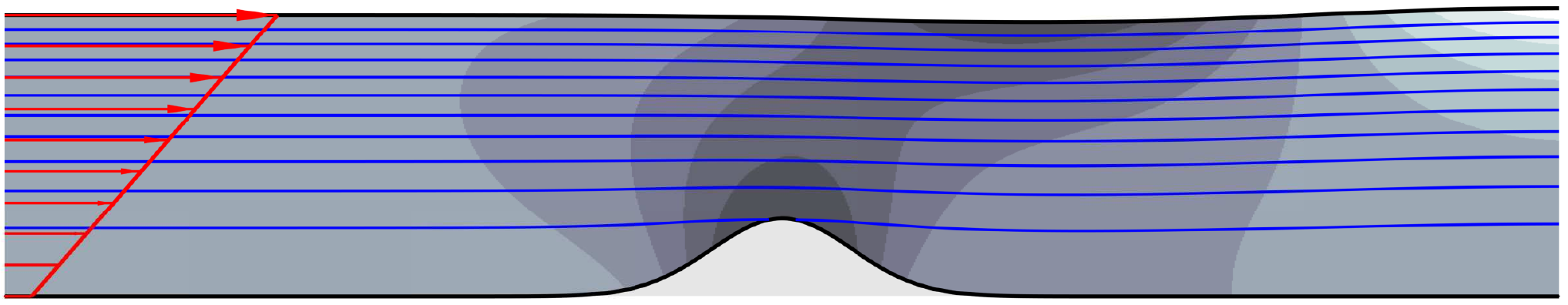}
\caption{The first example of figure~\ref{fig:obstacle_flow:narrow} when all boundaries are fully linearised.}
\label{fig:obstacle_flow:O1_Gauss}%
\end{figure}

Comparing with results from fully linearised boundary conditions $N=1$ (figure~\ref{fig:obstacle_flow:O1_Gauss}), one sees that the extension of the lower boundary is necessary for obtaining acceptable results in these cases as must indeed be expected. 
In terms of computation time, computing the flow fields shown in figures \ref{fig:monochrom_Ns} -- \ref{fig:obstacle_flow:O1_Gauss} calculated to order, say, $N=20$ takes only a few seconds on a  standard laptop computer and is of little relevance here.
\\

\begin{table}%
\setlength{\tabcolsep}{5pt}
\renewcommand{\arraystretch}{1.5}
\centering
\begin{tabular}{lcc}
& $\petab/\etabnil$ & $\etab/\etabnil$
\\\hline
Gaussian 
& $ \rme^{-\pi^2 x^2/(2 b)^2}$ 
& $ ({2b}/{\sqrt{\pi}})\rme^{- b^2 k_x^2/\pi^2}$
\\
wedge
& $ (1-|x|/b)\Heavi{b-|x|}$
& $2 a  [1-\cos(b k_x)]/k_x^2$
\\
ellipse
& $  \sqrt{1-(x/b)^2}\,\Heavi{b-|x|}$
& $ \pi J_1(b k_x)/k_x$
\\
rectangle
& $\Heavi{b-|x|}$
& $2 \sin (b k_x)/k_x$
\\
convolution shape 
& $ (\sqrt{\pi}/{2b\_s}) \rme^{-\pi^2 x^2/(2 b\_s)^2} $
& $\rme^{- b\_s^2 k_x^2/\pi^2}$
\end{tabular}
\caption{Functions for various bathymetry obstacle shapes in real and Fourier space.
$\Theta$ and $J_\nu$ are the Heavisied and Bessel functions, respectively.  }
\label{tab:eta_b}
\end{table}

\begin{figure}%
\subfigure[Gauss. $b\_s=0$]{
\includegraphics[width=.5\columnwidth]{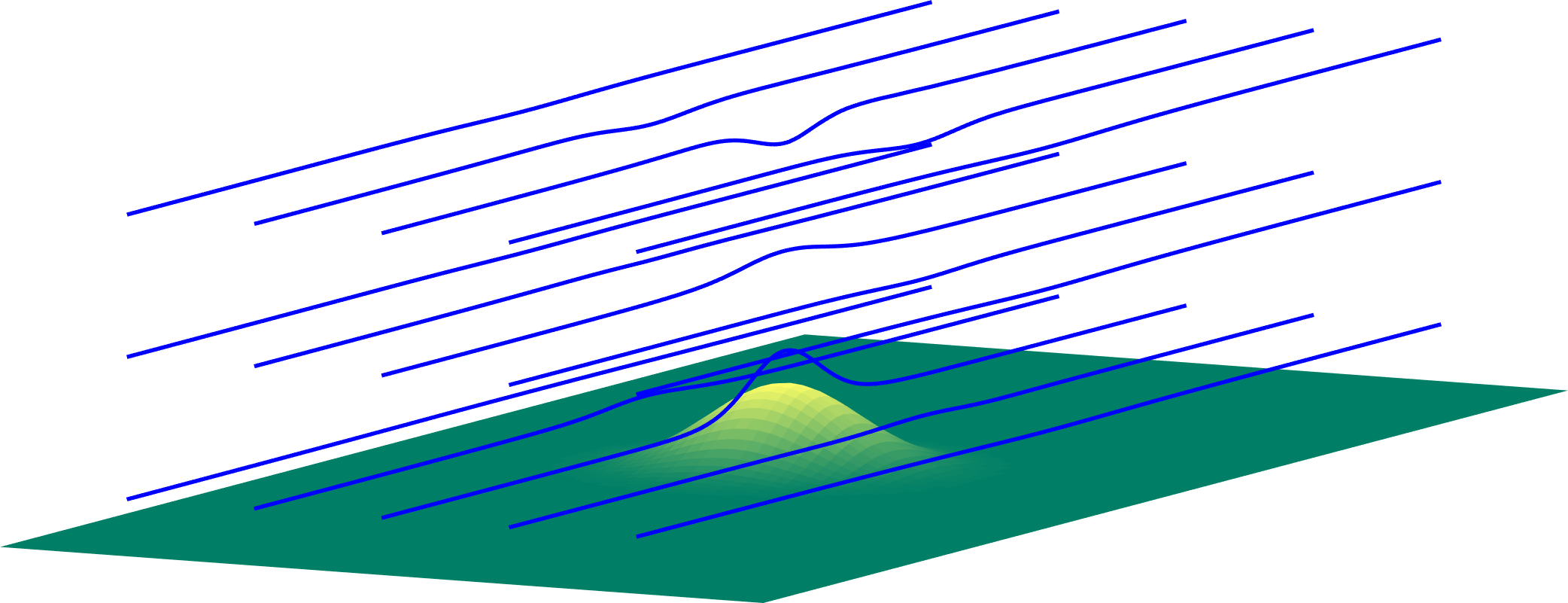}
\includegraphics[width=.5\columnwidth]{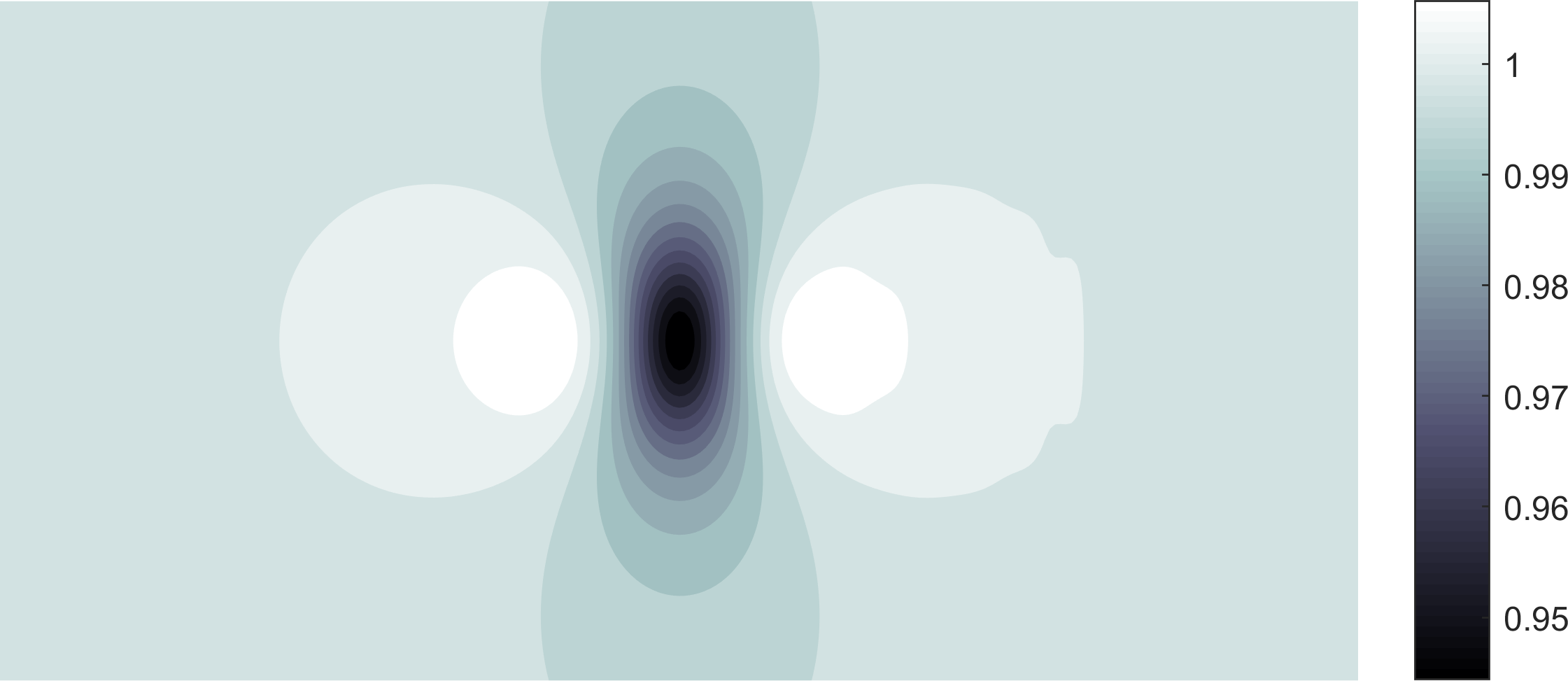}
}
\\
\subfigure[Cone. $b\_s=0$]{
\includegraphics[width=.5\columnwidth]{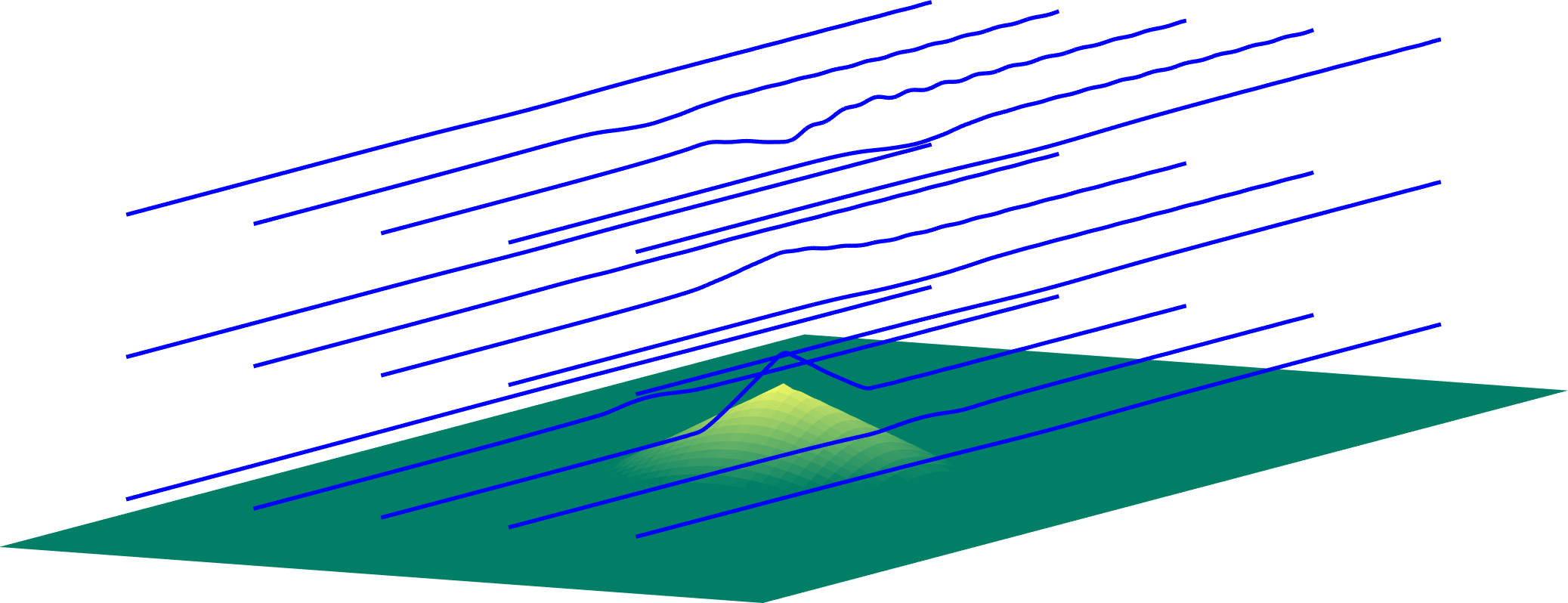}
\includegraphics[width=.5\columnwidth]{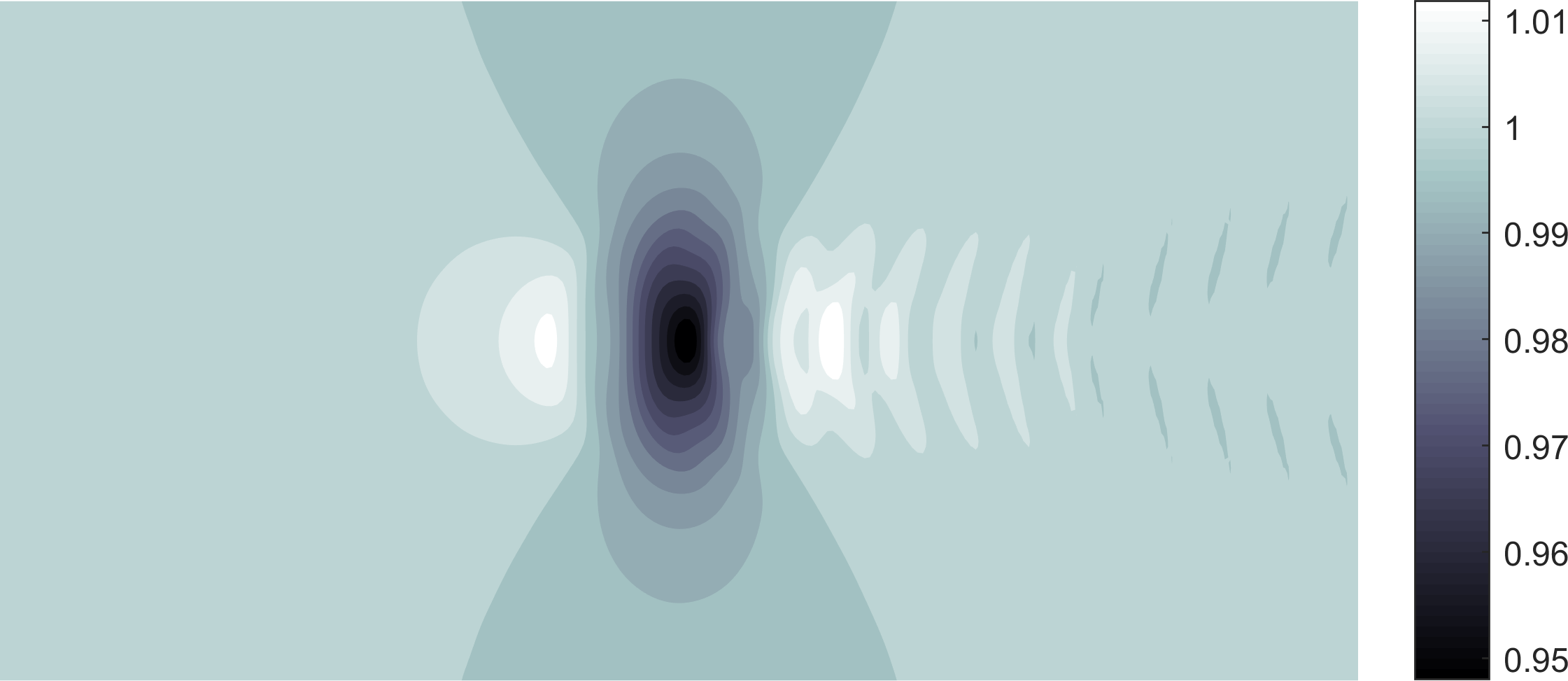}
}
\\
\subfigure[Ellipsoid. $b\_s=0.05$]{
\includegraphics[width=.5\columnwidth]{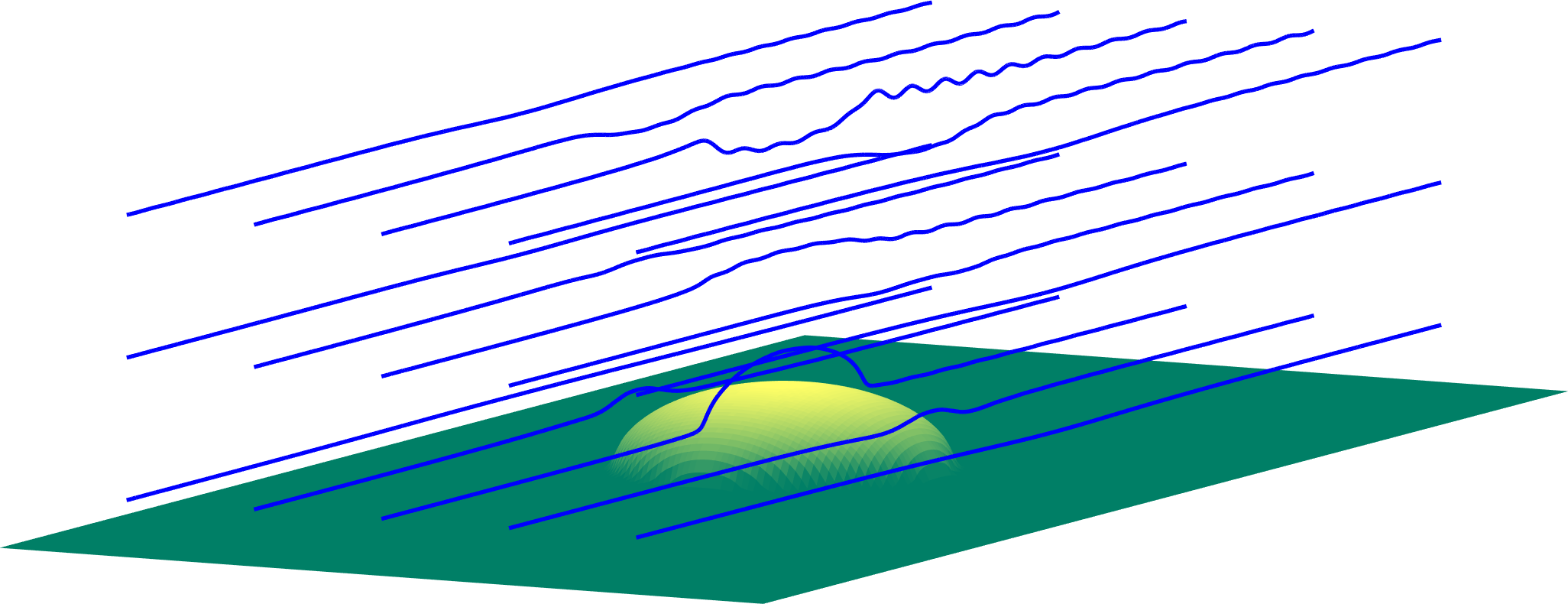}
\includegraphics[width=.5\columnwidth]{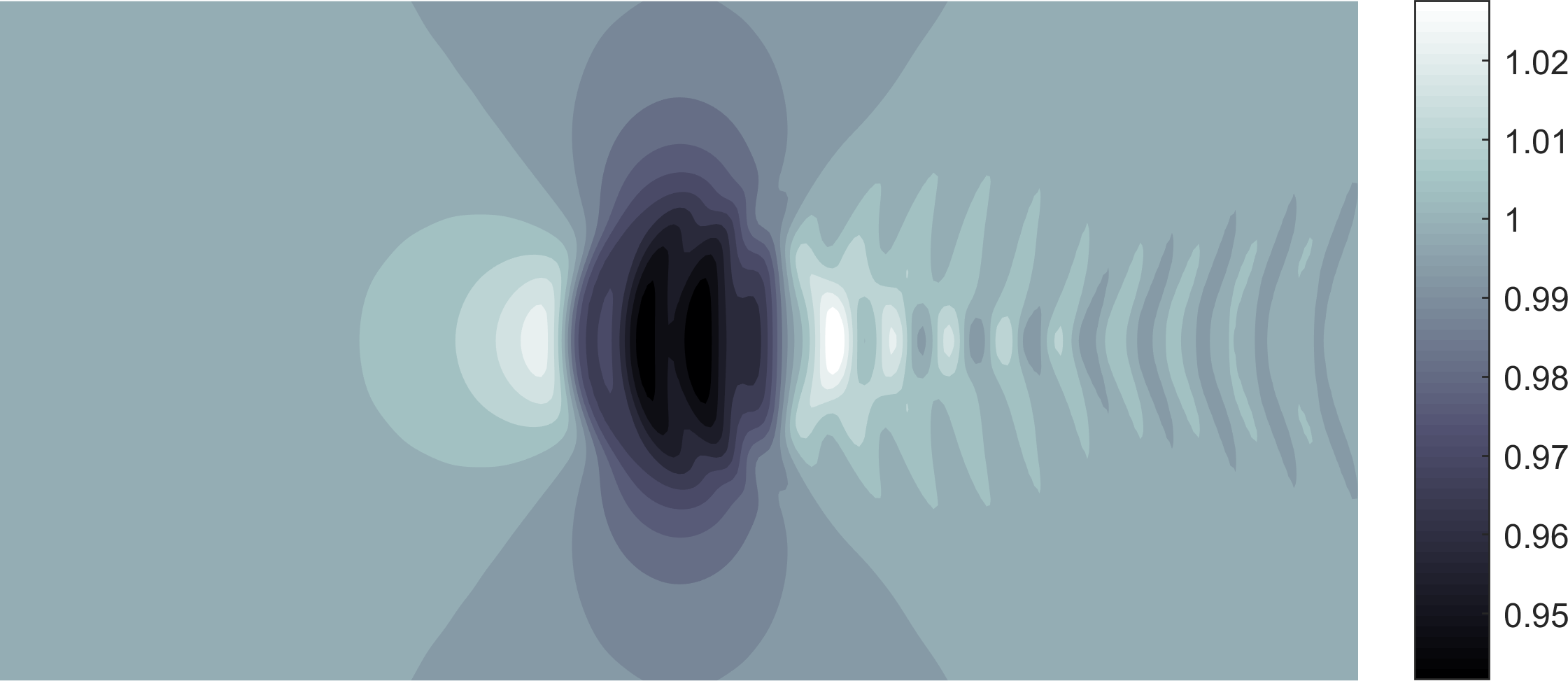}
}
\\
\subfigure[Box. $b\_s=0.1$]{
\includegraphics[width=.5\columnwidth]{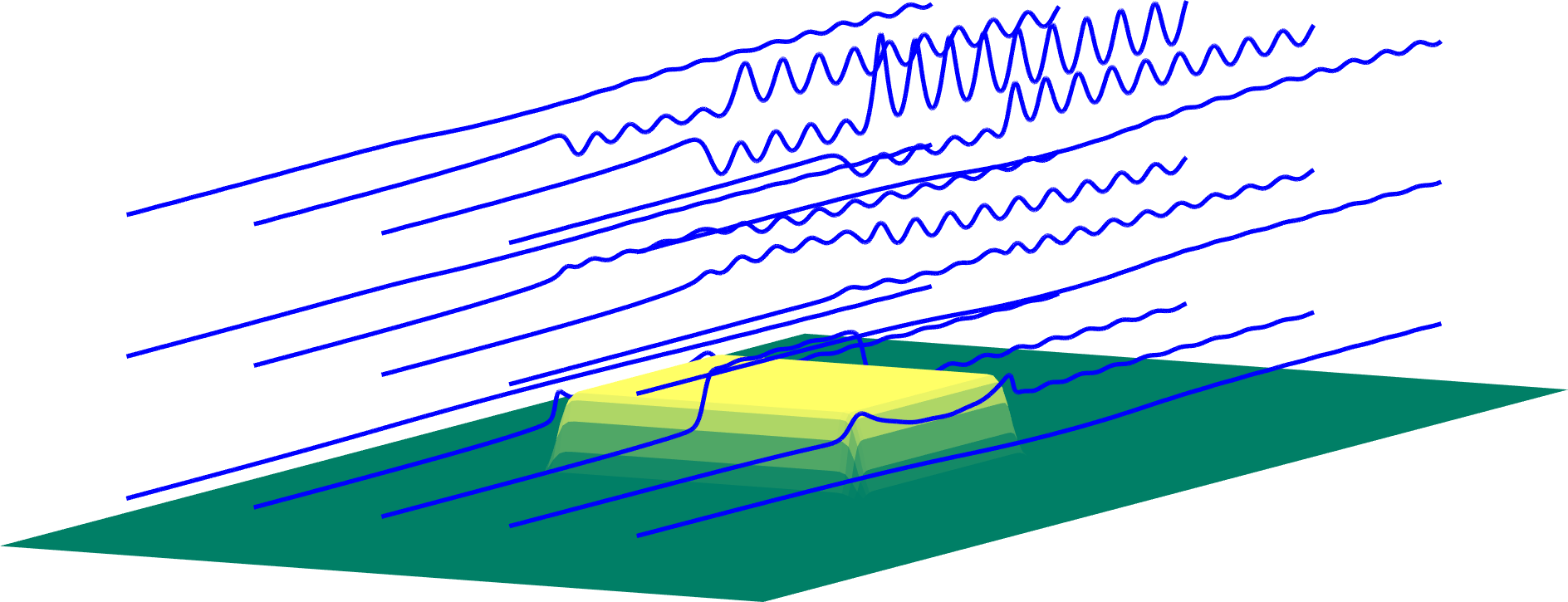}
\includegraphics[width=.5\columnwidth]{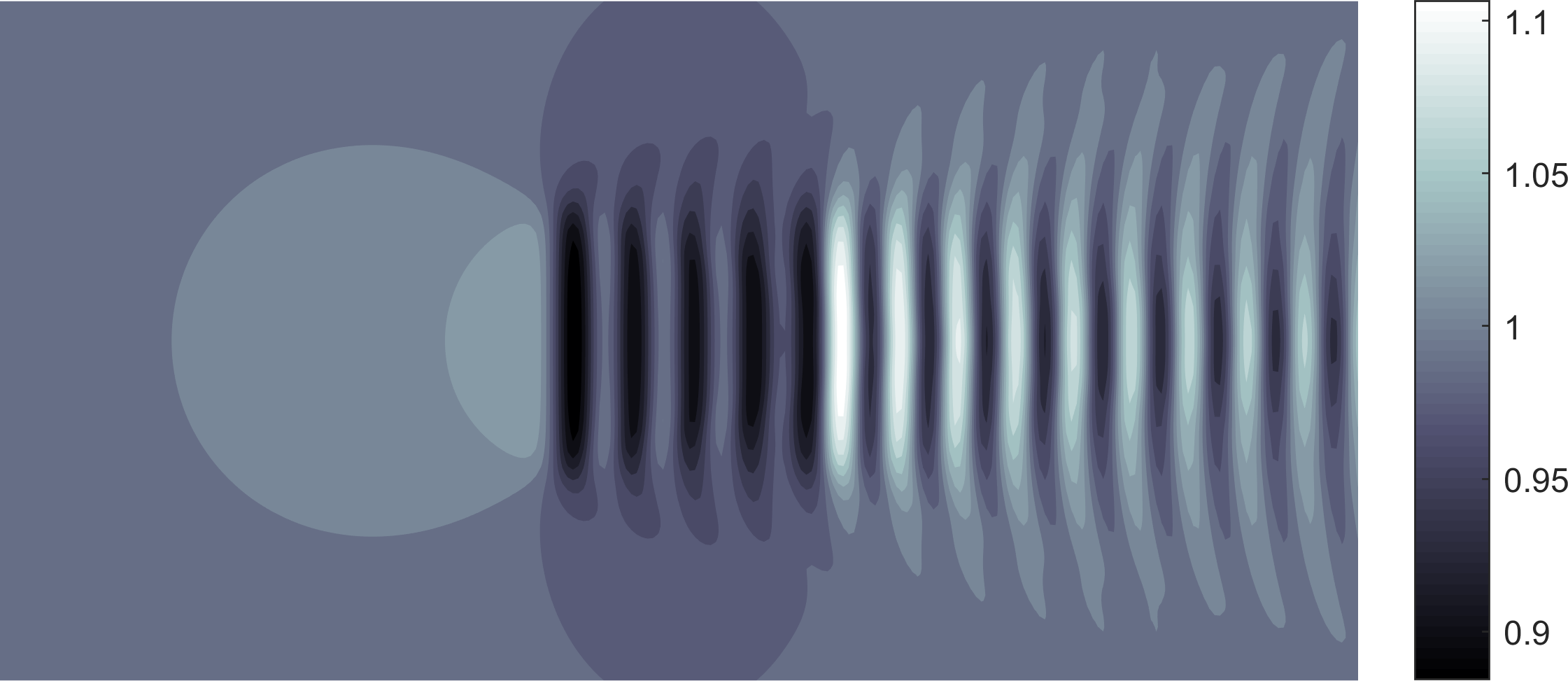}
}
\caption{
Three-dimensional equivalent of figure~\ref{fig:obstacle_flow:wide} 
\arev{($b = 5.0$, $\etabnil = 0.25$, $q=1/7$, $\Fr = 0.6$, $\delta = 0.001$)}
using an artificial friction coefficient $\muo = 0.01$.
Left: Streamline plots \arev{(blue lines) with bathymetry (yellow--green surface)} generated numerically from the velocity field $\p\bu(x,y,z)$. 
Right: Contour plot of surface elevation $\pzetas(x,y)$.
Aspect ratio $(x,y,z)$: 10:10:1.}%
\label{fig:obstacle_flow_3D:wide}%
\end{figure}

Finally, three-dimensional examples similar to those just presented, are shown in figure~\ref{fig:obstacle_flow_3D:wide} in the form of streamline plots of the flow field and contour plots of the surface elevation.
The triangular wedge is here a cone, the ellipse an ellipsoid and the rectangle shapes are boxes (cuboids).
A field separation method is more involved in three dimensions wherefore a finite value $\muo=0.01$ of the artificial friction coefficient is instead adopted.
This force suppresses the wave train before re-entering through the periodic boundaries.
(Only part of the computation domain is shown.)
\arev{The images focus on the near field where the effect of the obstacle shape is most evident;
in the far-field behind the obstacle the well-known Kelvin wake pattern becomes visible, possibly significantly affected by the nontrivial shear profile as shown theoretically \citep[e.g.][]{li_2017_transient_ship} and experimentally \citep{smeltzer_2019_ring_wave}. The far-field is governed primarily by the dispersion relation and its structure is relatively insensitive to the detailed obstacle shape.
Transverse waves dominate in the wake pattern since the Froude number based on the obstacle length is low.}
Again one sees that the Gaussian shape is smooth enough to suppress the wave train and make its amplitude invisible on the plotted scale while the tip of the similar cone-shape generates a ripple which diverges outwards. 

\begin{figure}%
\subfigure[\arev{$\Fr = 2.5$, $q = 0$, $\delta=0.001$.}]{
\includegraphics[width=.5\columnwidth]{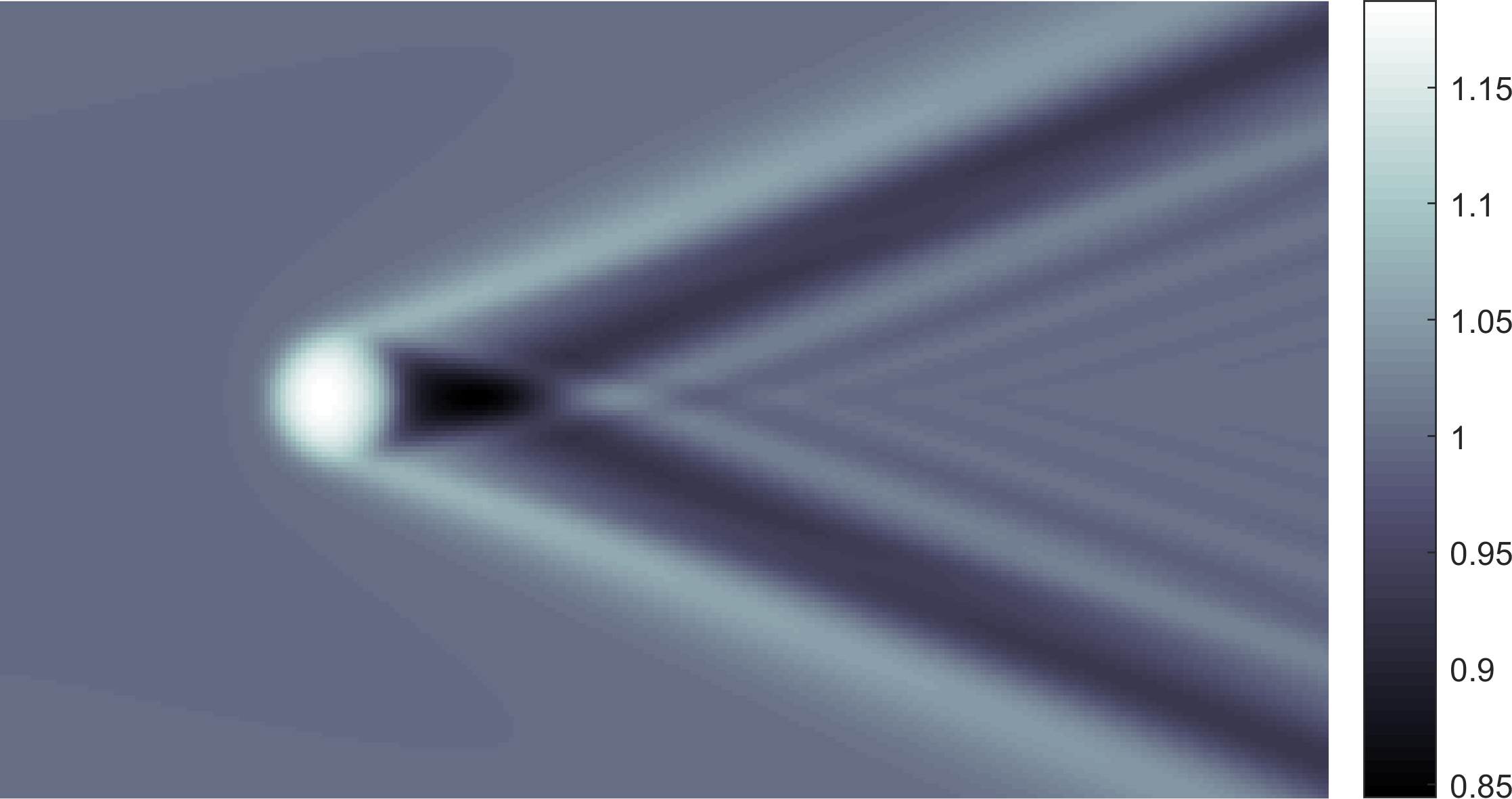}
\includegraphics[width=.5\columnwidth]{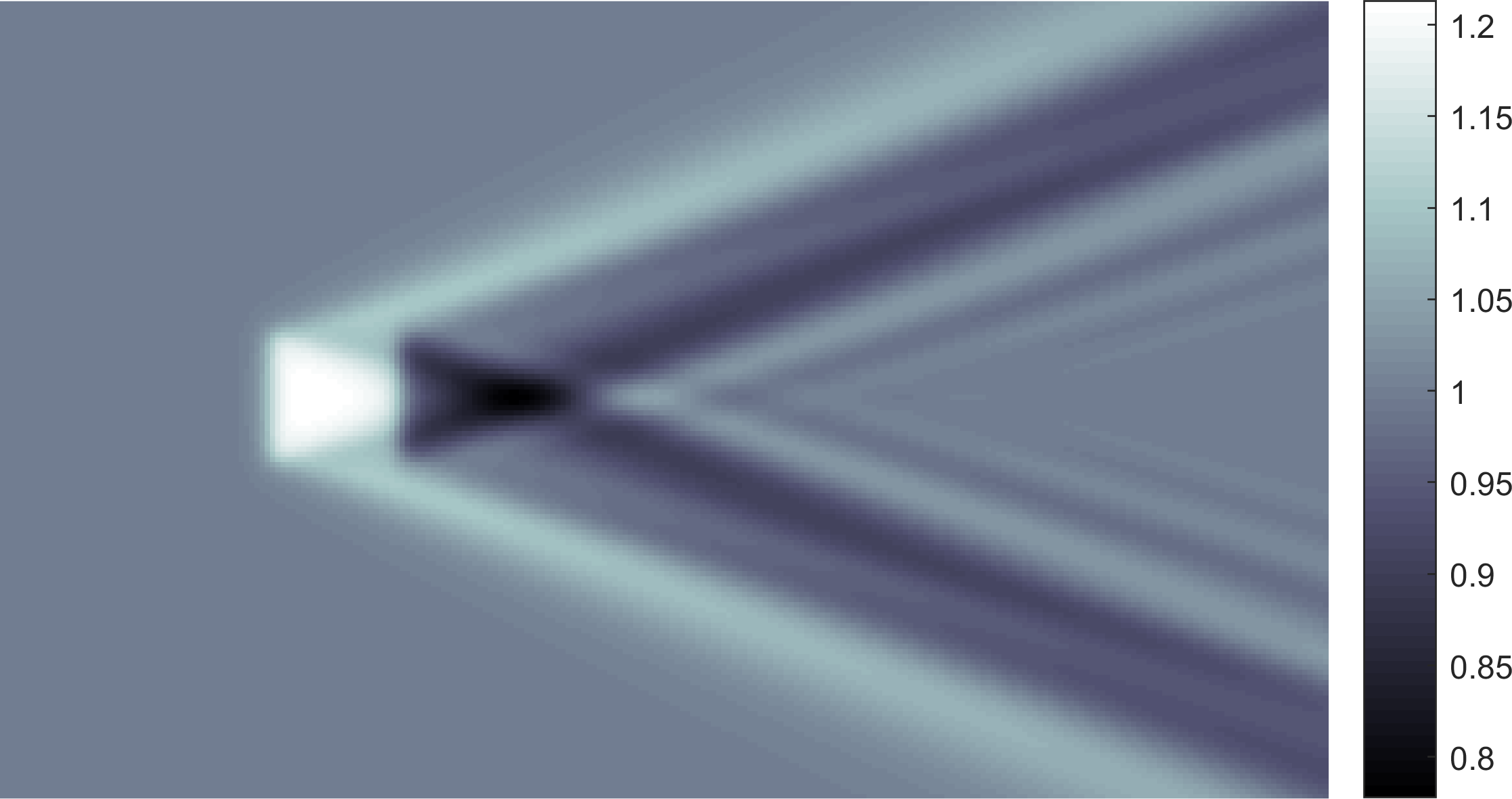}%
\label{fig:obstacle_flow_3D:supercritical:q0}
}
\subfigure[\arev{$\Fr = 4.0$, $q = 1$, $\delta=0.25$.}]{
\includegraphics[width=.5\columnwidth]{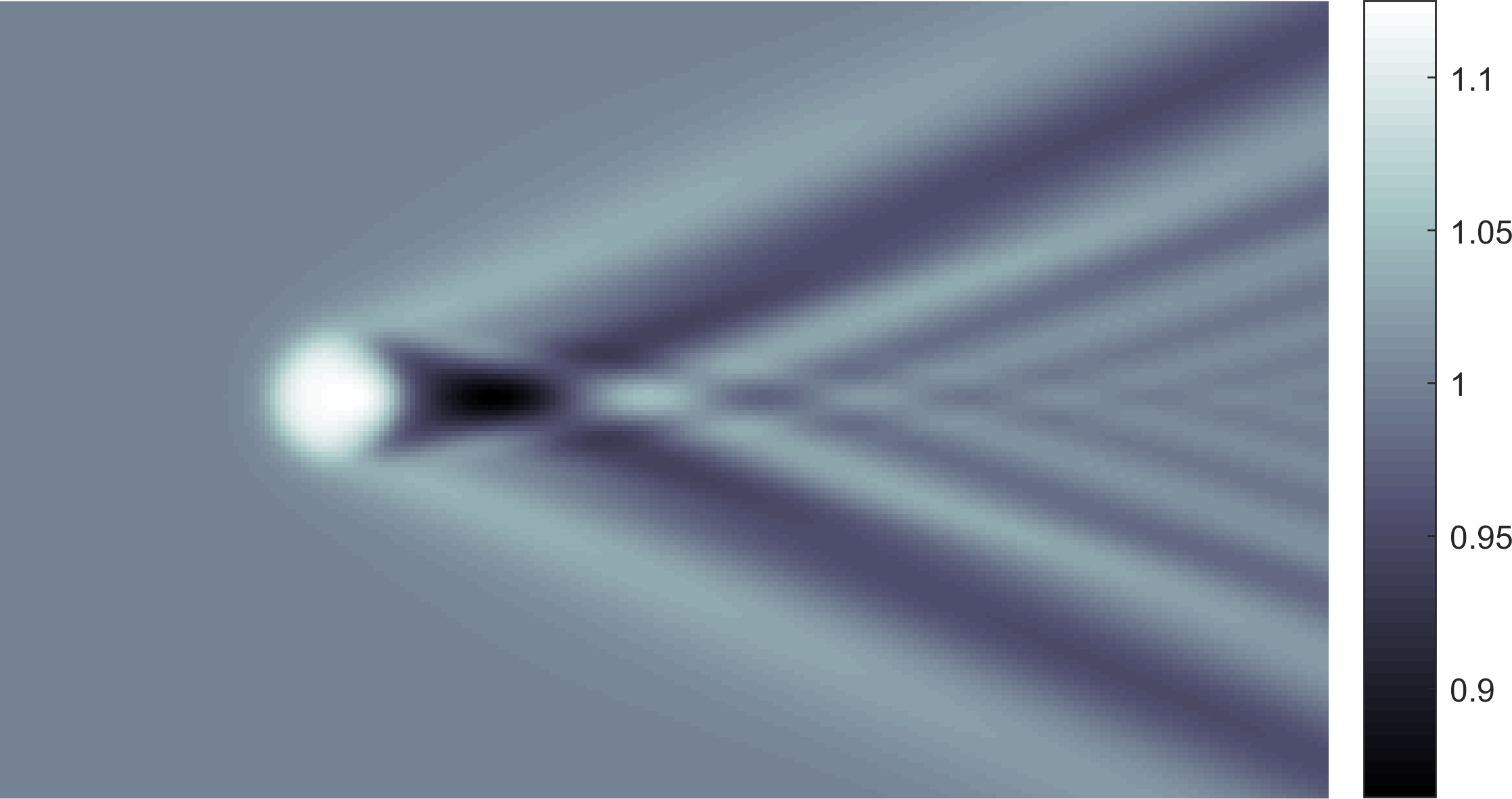}
\includegraphics[width=.5\columnwidth]{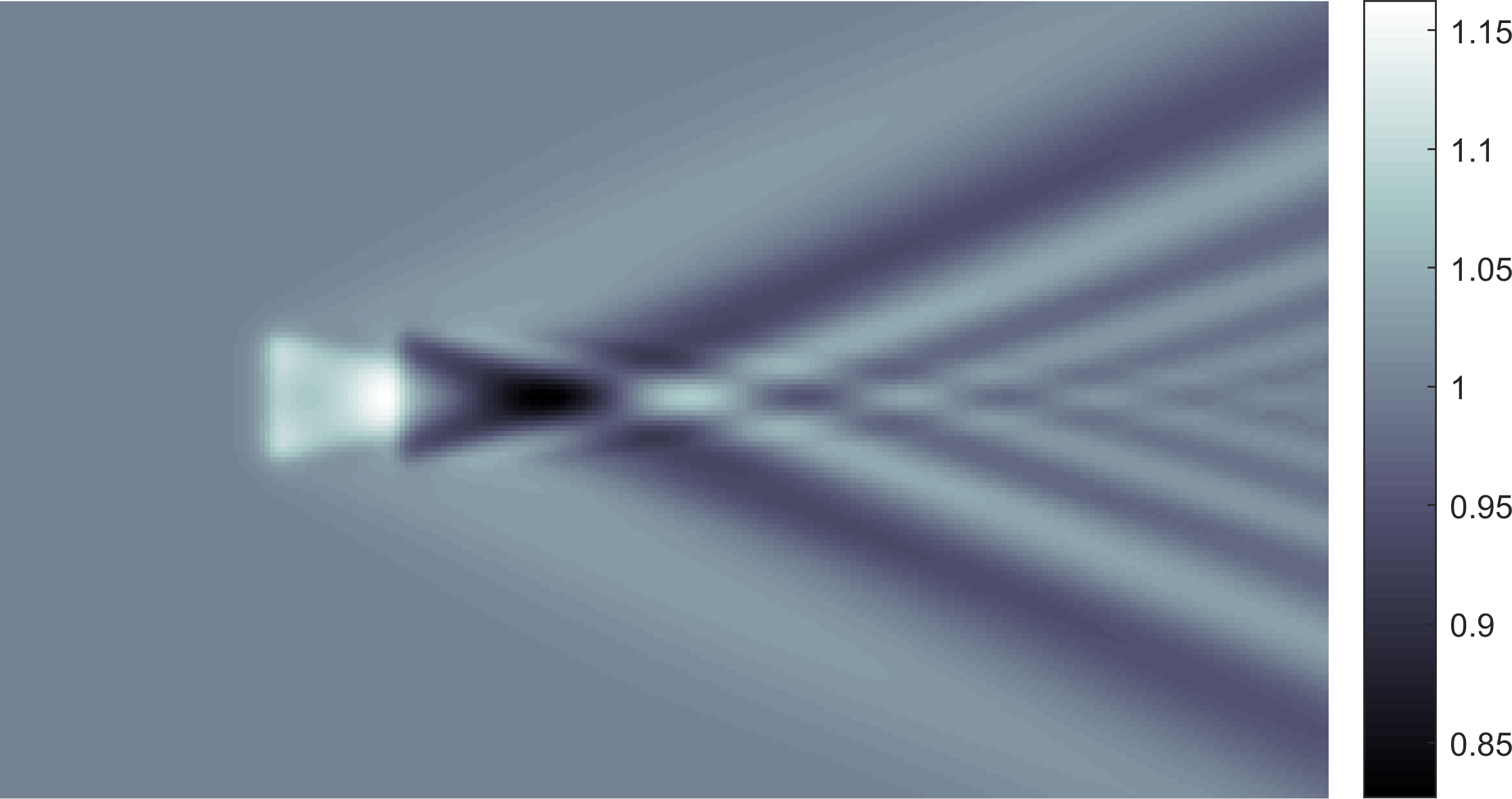}%
\label{fig:obstacle_flow_3D:supercritical:q1}
}
\caption{
\arev{
Surface elevation $\pzetas$ of supercritical flows. Left: ellipsoid. Right: box.
$\muo = 0.025$, $b\_s = 0.1$, $b = 5.0$, $\etabnil = 0.20$
}}%
\label{fig:obstacle_flow_3D:supercritical}%
\end{figure}

\arev{
Similar surface elevation plots
are presented in figure~\ref{fig:obstacle_flow_3D:supercritical} for supercritical flow conditions, i.e., the flow is too fast ($\Fr$ too high) for transverse waves to exist, leaving only diverging waves.
Only the ellipsoid and box are shown.
The current is uniform in subfigure \subref{fig:obstacle_flow_3D:supercritical:q0} and linear in subfigure \subref{fig:obstacle_flow_3D:supercritical:q1}.
Parameters are chosen such that the wake angles in the uniform and sheared currents cases are similar
(in general they will differ when shear is present, see \citet{li_2016_ship_wake_finite_depth})
and 
more of the wave pattern's far field is shown than in figure~\ref{fig:obstacle_flow_3D:wide}.
More surface profiles for uniform currents are presented in \citet{buttle_2018_river_boundary_integral_method}. }

\section{\arev{Discussion}}
\label{sec:discussion}

\arev{
The
assumption of stationary flow conditions in the presented theory excludes a range of 
the transient nonlinear  phenomena which have regularly featured in theoretical and experimental studies during the last decades. 
It is prudent to take a moment to consider what has been left out.
 
First let us mention the nonlinear features related to the free surface itself. 
The theory of Stokes waves, susceptible to modulation instability (Benjamin--Feir instability) \citep{zakharov_2009_modulation_instability} 
is a classical example and its analogue appears 
also in river flows.
Transients are often a persistent feature of the flow when sufficiently shallow relative to bathymetry obstructions  \citep{dolcetti_2016_channel_directional_spectrum}.
To wit, the free surface modulation wave resonance 
known from oceanographics \citep{Phillips_1960_mode_coupling,Zakharov_1968}
can exist also in shallow-water river systems.
Particularly, with the effect that strong boundary layer shears has on the dispersion relation, three-wave resonance becomes possible \citep{Craik_1968_resonance_critical_layer_Lommel,akselsen_ellingsen_2019}.

Another interesting nonlinear channel flow phenomenon not related to current shear is the generation solitons that radiating upstream.
These may be seen when the flow is near-critical, shallow and asymptotically two-dimensional.
The mechanism for their shedding is the local growth in amplitude within the linear regime due to lack of dispersion, followed by the subsequent `breaking free' of the soliton as increasing nonlinearities increase wave celerity \citep{wu_1987_upstream_advancing_solitons,katsis_1987_upstream_advancing_solitons}.
Amplitudes of emitted solitons are decreasing if the conditions are subcritical and constant if supercritical and below a certain cut-off Froude number \citep{akylas_1984_first_fKdV,Cole_1985_waves_forming_around_a_bump,mei_1986_solitons_off_of_slender_bodies,gourlay_2001_supercritical_bore}.
Upstream propagating solitons will slowly disperse if the flow is asymptotically three-dimensional (unrestricted by channel walls) 
\citep{katsis_1987_upstream_advancing_solitons}.
The phenomenon is inherently transient and nonlinear and, when present, not  captured in our stationary theory.
Periodic boundary conditions can however be utilised to represent a bounded channel provided the bathymetry is symmetric about the $x$-axis.

Yet more resonances are possible when the current is strongly sheared.
Critical layers (forming at depths where current advection matches wave dispersion) 
have the ability to 
transfer energy between the current and the resonating wave motion \citep{Benney_1961_O2_critical_layer_etc,Craik_1968_resonance_critical_layer_Lommel,Drivas_2016_resonance_gravity_and_vorticity_waves}.
A three-dimensional wave field scattering also occurs via interaction with critical layers \citep{zakharov1990_nth_order}.
Critical layers have been effectively excluded from the present study by considering steady states and strictly positive current profiles only. 

An instability which should be well known to all is the onset of vortex shedding behind obstacles at sufficiently high Reynolds numbers. 
Less famous is perhaps the longitudinal vortex instability,
commonly known as Craik--Leibovich CL2 instability,
possible in  sheared  three-dimensional flows.
These may grow exponentially through  three-dimensional perturbation of a two-dimensional shear current and do not
require a free surface, critical layers or inflection points to arise \citep{phillips_1996_family_of_zq_CL2_instability,phillips_1996_Langmuir_qoverz_profile}.
Such vortex generation is a fundamental mechanism for flow evolution towards three-dimensionality and increased chaos.
}


\section{Summary}
\label{sec:summary}

We have presented a perturbation solution for a free-surface shear flow following a $z^q$ velocity profile flowing over an arbitrarily but moderately sloped bed.
For various positive values of $q$ this analytical form can model bottom boundary layers, linearly sheared flows and near-surface shear layers. 
Radiation conditions are incorporated to generate the appropriate asymptotic behaviour.

Bathymetries of finite amplitude are found to introduce higher-order harmonics into the solution. 
For particular combinations of flow profile and bathymetry shapes 
these may resonate with the surface
wave modes, producing non-linear contributions to the downstream surface pattern of higher order which may become dominant. 
Resonances occurring in  flows over a sinusoidal bed manifest, in two-dimensional flows, as endless wave trains trailing an obstruction.
The expressions for criticality have been derived and 
a number of observations concerning the relationship between current vorticity, flow criticality, solution nonlinearity and the shape of the surface have been made related to the sinusoidal-bed configuration.
Recirculation is observed to occur within deep depressions in the bathymetry when shear is strong.
Further, three-dimensional bed--current vorticity interaction phenomena have been demonstrated, revealing 
for the case of flow at oblique angles to a sinusoidal bed 
both helical curving and spanwise migration of streamlines.  
Finally, examples of flows encountering localised obstructions of various shapes were presented in both two and three dimensions.

\section*{Acknowledgements}
This work was funded by the Research Council of Norway (programme FRINATEK), grant number 249740.

\bibliographystyle{jfm}
\bibliography{refs_river_clean}

\end{document}